\documentclass[12pt,a4paper]{article}
\usepackage{amsmath}
\usepackage{epsf}
\usepackage{epsfig}
\usepackage{amssymb}
\usepackage{graphicx}
\usepackage{latexsym}
\usepackage{hyperref}
\usepackage{rotating}
\usepackage{color}
\def\l{\left(}
\def\r{\right)}

\newcommand{\ba}{\begin{array}}
\newcommand{\ea}{\end{array}}
\newcommand{\be}{\begin{equation}}
\newcommand{\ee}{\end{equation}}
\newcommand{\bea}{\begin{eqnarray}}
\newcommand{\eea}{\end{eqnarray}}
\newcommand{\bg}{\begin{gather}}
\newcommand{\eg}{\end{gather}}
\newcommand{\bseq}{\begin{subequations}}
\newcommand{\eseq}{\end{subequations}}

\renewcommand{\ln}{\mathop{\rm ln}\nolimits}

\makeatletter
\def\gsim{\compoundrel>\over\sim}
\def\lsim{\compoundrel<\over\sim}
\def\compoundrel#1\over#2{\mathpalette\compoundreL{{#1}\over{#2}}}
\def\compoundreL#1#2{\compoundREL#1#2}
\def\compoundREL#1#2\over#3{\mathrel
         {\vcenter{\hbox{$\m@th\buildrel{#1#2}\over{#1#3}$}}}}
\makeatother

\numberwithin{equation}{section}

\begin{document}
\rightline{INR-TH/2014-028}

\begin{center}

    {\bf \LARGE Sgoldstino-Higgs mixing in models \\ 
      with low-scale supersymmetry breaking}
    \\
\bigskip
{\large
    K.~O.~Astapov$^{a,b,}$\footnote{{\bf e-mail}: astapov@ms2.inr.ac.ru},
    S.~V.~Demidov$^{a,b,}$\footnote{{\bf e-mail}: demidov@ms2.inr.ac.ru}} \\
    \vspace{10pt}
    $^a$\textit{Institute for Nuclear Research of the Russian Academy of
        Sciences,\\ 60th October Anniversary prospect 7a, Moscow
        117312, Russia}\\
    \vspace{5pt}
    $^b$\textit{Department of Particle Physics and Cosmology, Physics
      Faculty, \\ M.~V.~Lomonosov Moscow State University,
      \\ Vorobjevy Gory, 119991, Moscow, Russia}
  \end{center}

\begin{abstract}
    We consider a supersymmetric extension of the Standard Model with
    low-scale supersymmetry breaking. Besides usual superpartners it
    contains additional chiral goldstino supermultiplet whose scalar
    components -- sgoldstinos -- can mix with scalars from the Higgs
    sector of the 
    model. We show that this mixing can have considerable impact on
    phenomenology of the lightest Higgs boson and scalar
    sgoldstino. In particular, the latter can be a good candidate for
    explanation of $2\sigma$ LEP excess with mass around 98~GeV.
\end{abstract}

PACS numbers:
\date{\today}
\newpage

\section{Introduction}
\label{sec:1}
Discovery of a new scalar resonance at the ATLAS~\cite{Aad:2012tfa}
and CMS~\cite{Chatrchyan:2012ufa} becomes one of the most pronounced
events in the last few years. During the 1st run of the LHC experiments in
2011-2012 there was collected statistics about 5~fb at
$\sqrt{s}=7$~TeV and up to 20.6 fb at $\sqrt{s}=8$~TeV. Obtained 
results indicate that properties of the new particle are very similar
to those predicted for the Standard Model (SM) Higgs
boson~\cite{Spira:1997dg,Djouadi:2005gi} which once again confirms the
triumph of this Model. However, in spite of its beauty and capability
of explaining vast amount of experimental results in particle
physics SM has several drawbacks, e.g. zero neutrino
masses, no dark matter candidate, hierarchy problem etc.. 
We are forced to believe that SM is a part of another theory which
somehow cures its problems. Supersymmetry (SUSY) is among the most
prominent and attractive ideas for SM
extension~\cite{Haber:1984rc,Martin:1997ns}. It is interesting that
the discovery of the light Higgs-like resonance being interpreted as
the lightest Higgs boson $h$ of the Minimal Supersymmetric Standard
Model (MSSM) with mass of order 125~GeV is consistent with TeV scale
supersymmetry. It is well known that the mass of $h$ is bounded at
tree level by $Z$-boson mass and to reconcile it with the observed
value of the resonance mass requires sufficiently large quantum
corrections~\cite{Papucci:2011wy, Hall:2011aa} which implies (if 
other Higgs bosons are heavy) either heavy stop contribution or
maximal mixing in stop sector. Unobservation of light squarks at the
first run of LHC experiments indicates that this indeed may be the
case.  On the other hand it appears that the observed resonance is too  
heavy to be implemented ``naturally'' into supersymmetric
extensions~\cite{Brust:2011tb,Draper:2011aa,Baer:2012uy,Kang:2012sy,Kowalska:2013ica,Hardy:2013ywa,Baer:2013ava}. 

If supersymmetry is indeed inherent to our Nature it should be
spontaneously broken. In a particular model this may happen in some
hidden sector which does not have any renormalizable interactions with
the visible one to avoid phenomenological problems with supertrace of
squared mass matrix~\cite{Nilles:1983ge}. According to supersymmetric
analog of the Goldstone theorem~\cite{Volkov:1972jx} there should
exist a massless fermionic degree of freedom, goldstino. Being
included into supergravity framework goldstino becomes longitudinal
component of 
gravitino with mass related to the scale of supersymmetry breaking
$\sqrt{F}$ as follows $m_{3/2} = \frac{F}{\sqrt{3}M_{Pl}}$ where
$M_{Pl}$ is the Planck mass~\cite{Cremmer:1978iv}. In the simplest case
goldstino appears as a fermionic component 
of a chiral supermultiplet and interactions of this supermultiplet
with other MSSM fields are suppressed by $\sqrt{F}$. If the SUSY
breaking scale $\sqrt{F}$ is considerably higher than the 
electroweak scale than the interactions of SM particles with the
hidden sector 
are negligible. And this is the standard setup for phenomenological
consideration of supersymmetric models. For instance, for gravity
mediated SUSY breaking scenarios with soft parameters of order of
TeV-scale this implies $\sqrt{F}\gsim 10^{11}$~GeV. In the case of
gauge mediation the SUSY  breaking scale can be considerably lower,
but still its value is limited by $\sqrt{F}\gsim
50$~TeV~\cite{Giudice:1998bp}. 

However, it is phenomenologically possible (see,
e.g. Refs.~\cite{Ellis:1984kd, Gherghetta:2000qt}) to have $\sqrt{F}$
not very far from the electroweak scale, somewhere around several
TeVs. The main feature of these models is the presence of a sector
responsible for SUSY breaking, i.e. goldstino and probably  its scalar
superpartners -- sgoldstinos, in low energy spectrum. In this class of
models if R-parity 
is conserved gravitino is the lightest supersymmetric particle
(LSP) with the mass at sub-eV scale. Scalar and pseudoscalar sgoldstinos
acquire nonzero masses after integrating out particles from hidden
sector. It is phenomenologically possible to have them around
electroweak scale. If these particles are light we have an opportunity
to probe the scale of supersymmetry breaking already at present-day
experiments, in particular, at the LHC. Phenomenology of different
aspects of  low-scale supersymmetry breaking scenario have been
studied long ago. Among the most interesting signatures of these
models are gravitino pair production at particle
collisions~\cite{Dicus:1989gg,Shirai:2009kn,Mawatari:2014cja,Ellis:1996aa,Klasen:2006kb,Brignole:1998me,Brignole:1997sk,Petersson:2012dp}
and
decays~\cite{Luty:1998np,Gorbunov:2000ht,Djouadi:1997gw,Dicus:1990dy}, 
new contributions to FCNC decays of mesons, baryons, heavy
quarks and leptons with sgoldstinos in final
states~\cite{Brignole:2000wd,Gorbunov:2000th,Gorbunov:2000cz, 
  Gorbunov:2005nu, Demidov:2006pt, Demidov:2011rd}. The collider
phenomenology of sgoldstinos with masses at hundred GeV scale has
been studied in~\cite{Perazzi:2000id, Perazzi:2000ty, Gorbunov:2002er,
  Demidov:2004qt}.  

Recently, an interest to this type of models has been renewed (see,
e.g.~\cite{Antoniadis:2010hs,Petersson:2011in,Antoniadis:2012ck,Dudas:2012fa,Antoniadis:2014eta, 
  Dudas:2013mia}). One of the reasons is that these
theories allow to go beyond the setup of MSSM which presently becomes
strongly 
constrained by the LHC data. In this paper we consider possible
consequences of sgoldstino mixing with particles in the Higgs sector of MSSM
concentrating on the most intriguing 
case of mixing with the lightest Higgs boson. Interactions of
sgoldstino with the Higgs boson and some aspects of the mixing
between them have been discussed in 
Refs.~\cite{Petersson:2011in, Dudas:2012fa, Bellazzini:2012mh,
  Petersson:2012nv,Demidov:2014lda}. In particular, it has been shown that nonrenormalizable interactions
with goldstino supermultiplet result in additional contribution to the
Higgs potential and as a result to change of the Higgs selfcouplings. 
These changes can raise the value of the lightest Higgs boson mass and
on this way one try to cure naturalness
problem~\cite{Antoniadis:2014eta}. In~\cite{Bellazzini:2012mh} the
mixing of a heavy scalar sgoldstino with the lightest Higgs boson of
MSSM has been discussed to explain the excess in $h\to\gamma\gamma$
channel previously observed by ATLAS and CMS. In the present study
we discuss the case when the mixing of scalar 
sgoldstino with the lightest Higgs boson gives an additional
considerable positive contribution to the mass of the latter. 
This happens if sgoldstino mass is somewhat lower than the mass of
$h$. The most interesting consequences of this mixing  are
modifications of the lightest Higgs boson production rates and 
decays as well as presence of an additional light scalar in the low
energy spectrum. As by product we find that even small mixing can
considerably change sgoldstino signatures at
colliders\footnote{Similar well known example is the mixing of radion
  with the Higgs boson in models with extra
  dimensions~\cite{Giudice:2000av}. 
}. We perform a
scan over soft MSSM parameters in the decoupling regime, discuss
constraints from LHC and other experiments, find out acceptable
parameter space and calculate the signal strengths for the lightest
Higgs boson and scalar sgoldstino. In particular, we find that the
presence of lighter scalar sgoldstino can be consistent with small
$2\sigma$ excess observed at LEP~\cite{Barate:2003sz} in ${\rm 
  e}^+{\rm e}^{-}\to Zh$, where $h\to b\bar{b}$ with mass around
98~GeV.

The plan of the paper is the following. In Section~\ref{sec:2} we
introduce the model, describe interactions of goldstino supermultiplet
with MSSM fields and in particular with the Higgs doublets. We
calculate sgoldstino-Higgs mixing under assumption of
CP-conservation in this sector and discuss the changes in coupling
constants of new mass states. In Section~\ref{sec:3} we describe the
general strategy which we use to explore this scenario and discuss
obtained results. In section~\ref{sec:conclusions} we present our
conclusions. In Appendix~\ref{sec:4} we present several auxiliary
formulas.

\section{The low-scale SUSY breaking model}
\label{sec:2}
\subsection{The model description and sgoldstino-Higgs sector}
\label{sec:2_1}
In this section we describe a supersymmetric model within low-scale
supersymmetry breaking framework. Let us introduce goldstino chiral superfield
as $\Phi = \phi + \sqrt{2}\theta\tilde{G} + F_{\phi}\theta^2$, where
$\tilde{G}$ is goldstino, $\phi$ represents its scalar components,
sgoldstinos, and $F_{\phi}$ is auxiliary field. We suppose that due
to some dynamics in the hidden sector the auxiliary field $F_{\phi}$
acquires non-zero vacuum expectation value $\langle F_{\phi}\rangle$
and SUSY becomes spontaneously broken. Interactions of goldstino
supermultiplet with MSSM are introduced in such a way that after the
spontaneous supersymmetry breaking the standard set of soft terms
appears (see~\cite{Brignole:1996fn,Brignole:2003cm,Gorbunov:2001pd}
and references therein). Thus, we introduce the following lagrangian 
\be
\label{sgold_int}
{\cal L}_{\Phi-MSSM}={\cal L}_{K\ddot{a}hler}+{\cal
  L}_{superpotential}\;. 
\ee
Here the contribution from {\rm K$\ddot{a}$hler} potential has the
form 
\begin{eqnarray}
{\cal L}_{\rm K\ddot{a}hler}&=&\int  d^2\theta 
d^2\bar{\theta} \sum_{k}(1- {m_k^2\over
  F^2}\Phi^\dag\Phi)~\!\Phi_k^\dag  ~\!e^{g_1V_1+g_2V_2+g_3V_3}~\!\Phi_k\;,
\end{eqnarray}
where $k$ runs over all matter and Higgs supermultiplets, and the
contributions from superpotential look as
\begin{eqnarray}
\label{sup}
{\cal L}_{superpotential}&=&\int\!\!d^2\theta\left\{\epsilon_{ij}\!
\l(\mu-\frac{B}{F}\Phi)~\!H_d^iH_u^j+(Y^{L}_{ab} + {A^L_{ab}\over
  F}\Phi)~\! L_a^jE_b^cH_d^i \right.\right. \nonumber\\
&  & \left.\left. + 
(Y^{D}_{ab} + {A_{ab}^D\over F}\Phi)~\!Q_a^jD_b^cH_d^i
+(Y^{U}_{ab} + {A_{ab}^U\over
  F}\Phi)~\!Q_a^iU_b^cH_u^j\r\right. \\
& & \left.
+ \frac{1}{4}\sum_{\alpha}(1+{M_\alpha\over F}\Phi)~\!Tr 
W^\alpha W^\alpha\right\} + h.c.,\nonumber
\end{eqnarray}
where 
$\alpha$ labels all the  SM gauge fields, $\epsilon_{12}=-1$. The
physics of goldstino supermultiplet can be described by the following 
effective lagrangian
\begin{equation}
\label{gold}
{\cal L}_{\Phi} = \int{}d\theta^2d\bar{\theta}^2\left(\Phi^{+}\Phi +
\tilde{K}(\Phi^{+},\Phi)\right)-\left(\int{}d\theta^2F\Phi+h.c.\right). 
\end{equation}
Here we single out the standard kinetic term $\Phi^{+}\Phi$ from total
K$\ddot{a}$hler potential while $\tilde{K}(\Phi^{+},\Phi)$ represents
higher dimension contributions. The above lagrangian should be considered as
an effective field theory\footnote{
The lagrangian~\eqref{sgold_int} does not contain full set of
operators consistent with symmetries even to the leading order in
$1/F$ because we limit ourselves only with the simplest set of terms
which produce the MSSM soft parameters after SUSY breaking. Also here
we face with an ambiguity: the soft term $-B\epsilon_{ij}H_d^iH_u^j$
in MSSM lagrangian can be generated not only from the superpotential
as in Eq.~\eqref{sup} but also from the term
$-\frac{B}{F^2}\Phi^{\dagger}\Phi
\epsilon_{ij}H_d^iH_u^j\left|_{\theta^2\bar{\theta}^2}\right.$ in the 
K$\ddot{a}$hler potential. This is related to possibility of analytic
superfield redefinitions, discussed in~\cite{Brignole:2003cm}.
}
which is valid at energies $E\lsim \sqrt{F}$ and we consider higher
order terms in $\tilde{K}(\Phi^{+},\Phi)$ as suppressed by powers of
$F$. The linear superpotential triggers spontaneous supersymmetry
breaking  $\langle F_{\phi}\rangle=F+{\cal
  O}\left(\frac{1}{F}\right)$.  In what follows we take all soft  
parameters, $\mu$ and $F$ to be real and thus neglect possible
CP-violation. 

Let us consider the scalar sector of the model in details. By
integrating out auxiliary fields of two Higgs doublets, goldstino
supermultiplet and $D$-terms  of vector superfields we obtain the tree
level scalar potential for the sector of the Higgs fields and
sgoldstinos in the following form  
\be
\label{V}
V = V_D + V_H + V_\Phi,
\ee
where
\begin{gather}
V_D = \frac{g_{1}^{2}}{8}\left(1+\frac{M_1}{F}(\phi+\phi^*)\right)^{-1}
\left[h_d^{\dagger}h_d - h_u^{\dagger}h_u -
  \frac{\phi^*\phi}{F^2}(m_{h_d}^2h_d^{\dagger}h_d -
  m_{h_u}^2h^{\dagger}_uh_u)\right]^2\\
+\frac{g_{2}^2}{8}\left(1+\frac{M_2}{F}(\phi+\phi^*)\right)^{-1}
\left[h_d^{\dagger}\sigma^ah_d + h_u^{\dagger}\sigma^ah_u -
  \frac{\phi^*\phi}{F^2}(m_{h_d}^2h_d^{\dagger}\sigma^ah_d -
  m_{h_u}^2h^{\dagger}_u\sigma^ah_u)\right]^2\;, \nonumber
\end{gather}
\begin{gather}
V_H = \left(1-\frac{m^2_{h_u}}{F^2}\phi^*\phi\right)^{-1}
\left|\mu\epsilon_{ij}h_d^i-\frac{m_{h_u}^2}{F}\phi h_{uj}^{*}
  -\frac{B}{F}\phi\epsilon_{ij}h_d^i\right|^2
\\
+ \left(1-\frac{m^2_{h_d}}{F^2}\phi^*\phi\right)^{-1}
\left|\mu\epsilon_{ij}h_u^j-\frac{m_{h_d}^2}{F}\phi h_{di}^{*}
  -\frac{B}{F}\phi\epsilon_{ij}h_u^j\right|^2\;, \nonumber
\end{gather}
\begin{gather}
V_{\Phi} = \left(1 +
\frac{\partial^2\tilde{K}(\phi,\phi^*)}{\partial\phi\partial\phi^*} 
-\frac{m^2_{h_u}}{F^2}h_u^{\dagger}h_u -
\frac{m^2_{h_d}}{F^2}h_d^{\dagger}h_d\right)^{-1}\left|F 
+ \frac{B}{F}\epsilon_{ij}h_{d}^ih_{u}^j\right|^2
\label{Fterm_gold}
\end{gather}
We are going to investigate squared mass matrix of neutral scalars in
electroweak symmetry breaking (ESB) minimum with leading order
corrections in $1/F$. In general electroweak symmetry breaking minimum
of the scalar potential allows for non-zero value of sgoldstino field
$\phi$ because it is a singlet with respect to the SM gauge group. In
what follows we consider a case study and simplify matters by assuming
that $\langle\phi\rangle = 0$ in ESB minimum of the potential\footnote{
We note that nonzero v.e.v. of $\phi$ in particular results in
deviations of the Higgs couplings to SM fermions, see
e.g.~\cite{Dudas:2012fa}.}. This can be easily obtained by tuning
third derivatives of $\tilde{K}(\phi,\phi^*)$ as follows 
\begin{equation}
\frac{\partial^3\tilde{K}(0,0)}{\partial{}\phi^*\partial{}\phi^2}=
\end{equation}
\[
\frac{1}{F^3}\left(\mu(m_{h_u}^2+m_{h_d}^2)h^0_uh^0_d-\frac{M_2g^2+M_1g^{\prime
      2}}{8}(|h_u^0|^2-|h_d^0|^2)^2-B\mu(|h_u^0|^2+|h_d^0|^2)\right)
\]
up to higher order corrections in $1/F$.
After making this assumption we can expand scalar fields around
electroweak breaking minima as follows~\cite{Martin:1997ns}
\be
\label{exp1}
h_u^0=v_u+\frac{1}{\sqrt{2}}(h\cos{\alpha}+H\sin{\alpha})
+\frac{i}{\sqrt{2}}A\cos{\beta}, 
\ee
\be
\label{exp2}
h_d^0=v_d+\frac{1}{\sqrt{2}}(-h\sin{\alpha}
+H\cos{\alpha})+\frac{i}{\sqrt{2}}A\sin{\beta},
\ee
\be
\label{exp3}
\phi = \frac{1}{\sqrt{2}}\left(s+ ip\right)
\ee
Here $v\equiv\sqrt{v_{u}^2+v_{d}^2} = 174$~GeV and $\tan{\beta} =
\frac{v_{u}}{v_{d}}$ are introduced. The mixing angle $\alpha$ between
$h$ and $H$ is defined by the following relations 
\begin{equation}
\label{mssm_mix}
\frac{\sin{2\alpha}}{\sin{2\beta}} =
-\left(\frac{m_{H}^2+m_{h}^2}{m_{H}^2-m_{h}^2}\right),
~~~~~~~~\frac{\tan{2\alpha}}{\tan{2\beta}}
=\left(\frac{m_{A}^2+m_Z^2}{m_{A}^2-m_Z^2}\right)
\end{equation}
with standard tree level Higgs mass parameters
\be
m_{A}^2=\frac{2B}{\sin{2\beta}}=2\mu^2+m_{h_{u}}^2+m_{h_{d}}^2,
\ee
\begin{equation}
 m_{h,H}^2
 =\frac{1}{2}\left(m_{A}^2+m_Z^2\mp{}\sqrt{(m_{A}^2-m_Z^2)^2
   +4m_Z^2m_{A}^2\sin{2\beta}}\right). 
\end{equation}

In the chosen field basis~\eqref{exp1}-\eqref{exp3} the squared mass
matrices can be written in the following form
\be
\label{M_S}
{\cal M}_s^2 = 
\left(
\begin{tabular}{ccc}
$m_{H}^2$ & $0$ &2Y\\
$0$ & $m_{h}^2$ &2X\\
 2Y&2X & $m^2_{s}$
\end{tabular}
\right)
\ee
for scalars and 
\be
\label{M_P}
{\cal M}_p^2 = 
\left(
\begin{tabular}{cc}
$m_{A}^2$ & $2Z$\\
$2Z$ & $m_{p}^2$
\end{tabular}
\right)
\ee
for pseudoscalars.  The mixing terms $2X$, $2Y$ and $2Z$ are 
calculated below~\eqref{X_mixing}-~\eqref{Z-mixing}. With the assumption about zero v.e.v. of $\phi$
one finds that the only new contributions from SUSY breaking sector to
the tree level masses of the Higgs fields come from the term
$V_{\Phi}$ in the scalar potential. Another benefit of this assumption
is that mixing terms between sgoldstino and Higgses appear from linear 
in $\phi$ part of the scalar potential. The diagonal mass squared
elements for the Higgs fields read (c.f.~\cite{Petersson:2011in})
\be
\label{mh_before}
m_{h}^2 = m_{Z}^2\cos^2{2\beta} + 
\frac{v^2}{F^2}\left(B\sin{2\beta} - 2\mu^2\right)^2,
\ee 
\be
\label{mH_before}
m_{H}^2 = m_{Z}^2\sin^2{2\beta} + m_{A}^2,\;\;\;\;
m_{A}^2 = \frac{2B}{\sin{}2\beta}
+2v^2\frac{B}{F^2}\left(B-\frac{\mu^2}{\sin{}2\beta}\right).
\ee 
As compared to the MSSM case the masses get additional
contributions from new term~\cite{Antoniadis:2010hs,Petersson:2011in, Dudas:2012fa,
  Petersson:2012nv} of the fourth order in Higgs doublets  
\be
V_{F} = \frac{1}{F^2}\left|m_{h_{u}}^2h_{u}^{\dagger}h_{u} + 
m_{h_d}^2h_{d}^{\dagger}h_{d} - B\epsilon_{ij}h_{d}^ih_{u}^{j}\right|^2
\ee
which comes from the part~\eqref{Fterm_gold} of the scalar potential. 
The expressions for $m_s^2$ and $m_p^2$ can be easily obtained from
Eq.~\eqref{V} and are related to the fourth order derivatives of
the K$\ddot{a}$hler potential $\tilde{K}(\phi^{\dagger},\phi)$. 

To obtain the off-diagonal elements in the mass matrices we expand the
scalar potential to the leading order in $1/F$ and keep only the
terms which are linear in sgoldstino field $\phi$. 
For this part of the potential we find
\begin{gather}
V_{mix}
=\frac{\phi}{F}\Big(\mu(m_u^2+m_d^2)h^0_uh^0_d
-\frac{g_{1}^2M_1+g_2^2M_2}{8}(|h_u^0|^2-|h_d^0|^2)^2
\\
-B\mu(|h_u^0|^2+|h_d^0|^2)\Big)+h.c.    \nonumber
\end{gather}
and for off-diagonal terms in~\eqref{M_S} and~\eqref{M_P} we obtain
\be
\label{X_mixing}
X = 2\mu^3v\sin{2\beta} +
\frac{1}{2}v^3(g_{1}^2M_1+g_2^2M_2)\cos^2{2\beta},
\ee
\be
Y = \mu v(m_{A}^2-2\mu^2) +\frac{1}{4}(g_{1}^2M_1+g_2^2M_2)\sin{4\beta},
\ee
\be
\label{Z-mixing}
Z = -\mu v(m_{A}^2 - 2\mu^2)\cos{2\beta}.
\ee

In what follows we concentrate on the decoupling limit, i.e. $m_{A}\gg
m_{h}$. Then all the Higgs bosons except for the lightest one become
heavy. 
This limit corresponds to $\cos{\alpha}=\sin{\beta},\;
\sin{\alpha}=-\cos{\beta}$ in Eqs.~\eqref{exp1} and~\eqref{exp2}. Next,
we consider the scalar 
sgoldstino squared mass parameter $m_s^2$ to be somewhat less
than $m_{h}^2$. In this case the mixing between the two states can
give a positive contribution to the Higgs boson mass\footnote{
The case when sgoldstino mass parameter is much larger than $m_{h}$
has been studied in
Refs.~\cite{Petersson:2011in,Dudas:2012fa,Bellazzini:2012mh}.  
}. Corresponding mass states are given by the following linear
combinations  
\begin{eqnarray}
\tilde{h} = h\cos{\theta}-s\sin{\theta},\\
\tilde{s} = h\sin{\theta}+s\cos{\theta}.
\end{eqnarray}
and the expressions for their masses squared look (in the case
$m_{{h}}>m_s$) as 
\begin{equation}
\label{mh_after}
m_{\tilde{h}}^2 =\frac{1}{2}\left(m_s^2+m_{h}^2
+\sqrt{(m_s^2-m_{h}^2)^2+\left(2\frac{X}{F}\right)^2}\right), 
\end{equation}
for new Higgs-like state $\tilde{h}$ and
\be
m_{\tilde{s}}^2 =\frac{1}{2}\left(m_s^2+m_{h}^2
-\sqrt{(m_s^2-m_{h}^2)^2+\left(2\frac{X}{F}\right)^2}\right). 
\ee
for new sgoldstino-like state $\tilde{s}$. The mixing angle is given
by following relation
\begin{equation}
\label{mix_angle}
\tan{2\theta} =\frac{2X}{F(m_s^2-m_{h}^2)}.
\end{equation}
Expression for the mixing term $X$ changes if we allow for nonzero
v.e.v. of sgoldstino field. 

Also note that if other Higgs bosons are
also light the mixing pattern becomes more complicated. 
We finish this subsection by reminding that interactions of the
lightest Higgs boson with (s)quarks result in the large quantum
correction $\delta$ to its mass squared. This can be taken into
account in the expressions above by replacement $m_h^2\to m_h^2 +
\delta$.  

\subsection{Sgoldstino and Higgs boson couplings}
\label{sec:2_2}
Here we write down the couplings of new mass states $\tilde{h}$ and
$\tilde{s}$ to the SM particles. Mainly we are interested in their
couplings to the SM vector bosons and heavy fermions of the third
generation. Corresponding effective lagrangian for $h$ reads  
\begin{eqnarray} 
\label{Higgs_lagr}
{\cal L}_{h}^{eff} & = & 
\frac{2m^2_W}{\sqrt{2}v}C_WhW_{\mu}^+W^{\mu-} +
\frac{2m^2_Z}{\sqrt{2}v}C_ZhZ_{\mu}Z^{\mu} -
\frac{m_{\tau}}{\sqrt{2}v}C_\tau h\bar{\tau}\tau{} -
\frac{m_t}{\sqrt{2}v}C_th\bar{t}t 
\\
& & -\frac{m_b}{\sqrt{2}v}C_bh\bar{b}b
+g^{1-loop}_{h^{SM}\gamma\gamma}C_{\gamma\gamma}hF^{\mu\nu}F_{\mu\nu}+ 
g^{1-loop}_{h^{SM}gg}C_{gg}h\;{\rm tr}G^{\mu\nu}G_{\mu\nu} \nonumber
\end{eqnarray}
where we introduce the scaling factors $C_k$ for the couplings
relative to their SM values. Similar interaction lagrangian for the
scalar sgoldstino $s$ can be obtained from the
Eq.~\eqref{sgold_int} as follows 
\begin{eqnarray} 
\label{Sgold_lagr}
{\cal L}^{eff}_{s} & = &
-\frac{M_{2}}{\sqrt{2}F}sW^{\mu\nu *}W_{\mu\nu}
-\frac{M_{ZZ}}{2\sqrt{2}}sZ^{\mu\nu}Z_{\mu\nu}
-\frac{A^{L}_{33}v_d}{\sqrt{2}F}s\bar{\tau}\tau
-\frac{A^{U}_{33}v_u}{\sqrt{2}F}s\bar{t}t \\
& & -\frac{A^{D}_{33}v_d}{\sqrt{2}F}s\bar{b}b
-\frac{M_{\gamma\gamma}}{2\sqrt{2}}sF^{\mu\nu}F_{\mu\nu}
 - \frac{M_3}{2\sqrt{2}F}s~{\rm tr}G^{\mu\nu}G_{\mu\nu} \nonumber
\end{eqnarray}
with
\begin{eqnarray}
M_{ZZ} \equiv M_{1}\sin^{2}{\theta_{W}} +
M_{2}\cos^{2}{\theta_{W}},\;\;\;
M_{\gamma\gamma} \equiv M_{1}\cos^{2}{\theta_{W}} +
M_{2}\sin^{2}{\theta_{W}}.
\end{eqnarray}
We see that the interaction of the lightest Higgs boson $h$ and the
scalar sgoldstino $s$ with quarks and leptons have similar structure,
so the coupling constants for the Higgs-like mass state $\tilde{h}$
read 
\begin{eqnarray}
\label{htt}
g_{\tilde{h}\bar{t}t} & = & \frac{m_{t}}{v\sqrt{2}}C_t\cos{\theta} -
\frac{A_{33}^Uv\sin{\beta}}{\sqrt{2}F}\sin{\theta}, \\
\label{hbb}
g_{\tilde{h}\bar{b}b} & = & \frac{m_{b}}{v\sqrt{2}}C_b\cos{\theta} -
\frac{A_{33}^Dv\cos{\beta}}{\sqrt{2}F}\sin{\theta}, \\
\label{htau}
g_{\tilde{h}\bar{\tau}\tau} & = &
\frac{m_{\tau}}{v\sqrt{2}}C_\tau\cos{\theta} -
\frac{A_{33}^Lv\cos{\beta}}{\sqrt{2}F}\sin{\theta}. 
\end{eqnarray}
The scaling factors $C_t, C_b$ and $C_\tau$ are determined by the
mixing of $h$ and $H$ and in the decoupling limit $m_H\gg m_h$ are
close to unity, c.f.~\eqref{exp1}, ~\eqref{exp2} and~\eqref{mssm_mix}.  

The effective couplings of the SM Higgs boson with gluons and photons
result from loop contributions of quarks and $W$-bosons.  The scaling
factors $C_{\gamma\gamma}$ and $C_{gg}$ in~\eqref{Higgs_lagr} take
into account additional corrections from interactions with squarks,
charginos etc. which are typically suppressed if these superpartners
are heavy. For scalar sgoldstino the couplings to photons and gluons
appear already at tree level, see~\eqref{Sgold_lagr}, and putting them
all together one obtains for $\tilde{h}$ 
\begin{eqnarray}
g_{\tilde{h}\gamma\gamma} & = & g^{1-loop}_{h^{SM}\gamma\gamma}C_{\gamma\gamma}\cos{\theta}
+ \frac{M_{\gamma\gamma}}{2\sqrt{2}F}\sin{\theta}, \\
g_{\tilde{h}gg} & = & g^{1-loop}_{h^{SM}gg}C_{gg}\cos{\theta}
+ \frac{M_{3}}{2\sqrt{2}F}\sin{\theta}, \label{gluons}
\end{eqnarray}
where dominant SM loop contributions look as
follows~\cite{Spira:1997dg} 
\begin{eqnarray}
\label{gg_gamma}
g^{1-loop}_{h^{SM}\gamma\gamma} & = & \frac{\alpha}{4\sqrt{2}\pi{}v}\left({\cal A}_{1}(\tau_{W}) + N_cQ_t^2{\cal
  A}_{1/2}(\tau_t)\right),\\
g^{1-loop}_{h^{SM}gg} & = & \frac{3}{4}\frac{\alpha_s}{6\sqrt{2}\pi{}v}\left({\cal A}_{1/2}(\tau_t) + {\cal
  A}_{1/2}(\tau_b)\right). 
\end{eqnarray}
Here $\tau_{i}=\frac{4m_{i}^2}{m_h^2}$ and loop formfactors read 
\begin{eqnarray}
{\cal A}_{1/2} & = & 2\tau\left(1+(1-\tau)f(\tau)\right),\\
{\cal A}_1 & = & -\left(2+3\tau+3\tau(2-\tau)f(\tau)\right),
\end{eqnarray}
with
\begin{equation}
f(\tau) = \left\{
\begin{array}{ll}
\arcsin^{2}\left({1/\sqrt{\tau}}\right), \;\;\; \tau\ge 1,\\
-\frac{1}{4}\log{\frac{1+\sqrt{1-\tau}}{1-\sqrt{1-\tau}}}, \;\;\;
\tau<1
\end{array}
\right. .
\end{equation}

Interactions with $W$ and $Z$ bosons are described by different
operators for the Higgs boson and scalar sgoldstino, see 
Eqs~\eqref{Higgs_lagr} and~\eqref{Sgold_lagr}. 
Corresponding couplings for new Higgs-like mass state will have the
following form in the momentum space 
\begin{eqnarray}
\label{gZZ}
g^{\mu\nu}_{\tilde{h}ZZ} & = & g^{\mu\nu}_{hZZ}C_Z\cos{\theta}
+\frac{M_{ZZ}}{2\sqrt{2}F}2\left((k_{Z_1},k_{Z_2})\eta^{\mu\nu}
-k^{Z_2\mu}k^{Z_1\nu}\right)\sin{\theta}  \\
\label{gWW}
g^{\mu\nu}_{\tilde{h}W^+W^{-}}
& = & g^{\mu\nu}_{hW^+W^{-}}C_W\cos{\theta}
+\frac{M_2}{2\sqrt{2}F}2\left((k_{W^+},k_{W^-})\eta^{\mu\nu}
-k^{W^{-}\mu}k^{W^+\nu}\right)\sin{\theta} 
\end{eqnarray}
The scaling factors $C_W$ and $C_Z$ are again close to unity in the
decoupling regime. Effective coupling constants for sgoldstino-like
state $\tilde{s}$ can be obtained from those above by the replacement
$\cos{\theta}\to \sin{\theta}$ and $\sin{\theta}\to - \cos{\theta}$.

\section{Analysis of the model}
\label{sec:3}
\subsection{Strategy for analysis}
\label{sec:3_1}
In this Section we discuss phenomenological implications of
sgoldstino-Higgs mixing in context of the setup described above.
For a given point in parameter space of the model which is
characterized by MSSM parameters, scalar sgoldstino mass term
$m_{s}^2$ and the scale of supersymmetry breaking $\sqrt{F}$ one can
ask whether this point is compatible with experimental data and in
particular with results of LHC experiments. To explore this scenario
we perform a  scan over MSSM parameters space. In what follows we
consider two parameter sets for comparison:
\begin{itemize}
\item {\it Set 1.} $1.5 < \tan\beta < 50.0$,
  $100$~GeV~$<|\mu|<1500$~GeV, 100~GeV $< |M_1| < 
  500$~GeV, 200~GeV $<|M_2|< 500$~GeV, 1.5~TeV $<|M_3|<$ 2.0~TeV,
  $|A^{U,D,E}_{33}|<1.5$~TeV, 700~GeV $<m_{Q_{3}},m_{U_{3}},m_{D_{3}}
  <$ 1.3~TeV.  
\item {\it Set 2.} This region has higher upper borders: $100$~GeV
  $<|\mu|<2000$~GeV, 100~GeV $< |M_1| < 2000$~GeV, 200~GeV $<|M_2|<
  2000$~GeV, 1.5~TeV $<|M_3|<$ 4.0~TeV, $|A^{U,D,E}_{33}|<4$~TeV,
  700~GeV $<m_{Q_{3}},m_{U_{3}},m_{D_{3}} <$ 5~TeV.
\end{itemize}
All the MSSM parameters have been chosen at the electroweak scale.  
Other SUSY soft masses, which are not relevant for our analysis, are
taken to be sufficiently large. In particular, given that we would
like to consider decoupling regime, the Higgs pseudoscalar is also
taken also to be heavy. The main difference between the two sets
which will be important to us is that without additional contribution
only 
very small fraction of models within {\it Set 1} provides the
lightest Higgs boson with the mass higher than about 123~GeV. On the
contrary {\it Set 2} includes rather large values of trilinear
couplings $A^U_{33}$ and stop mass parameters $m_{Q_{3}}, m_{U_{3}}$
and larger values of $m_h$ (up to 128~GeV) can be obtained. For
supersymmetry breaking scale we fix the value $\sqrt{F}=10$~TeV; later
on we comment about this choice. For calculation of MSSM spectra and
the lightest Higgs boson coupling constants without contribution of
sgoldstino sector we use package NMSSMTools~\cite{Ellwanger:2004xm} in
the MSSM regime. We remind reader that in the scenario of low-scale
supersymmetry breaking gravitino is LSP. By default NMSSMTools package
in the regime of general NMSSM excludes models where neutralino is not
LSP, so we turn this option off in the program. We scan over the
chosen parameter spaces and exclude unphysical models 
by checks for absence of unphysical global minimum of the scalar
potential in Higgs sector. On this stage we use a set of experimental
constraints implemented in NMSSMTools, including constraints from
measurements of ${\rm Br}(b\to s\gamma)$  and ${\rm Br}(B_s\to 
\mu^+\mu^-)$~\cite{Chen:2009cw}. Note, that we do not impose the
condition that the SUSY contribution to the anomalous magnetic moment
of muon should explain the present $3\sigma$ difference between SM
prediction and BNL result. The result of the scan is the spectrum of
superpartners, the value $m_{h}^2$ for the squared mass of the
lightest Higgs boson including MSSM quantum corrections and coupling
constants of $h$ to photons, gluons, quarks and leptons which we use
in the following analysis. 

Then we turn on mixing with sgoldstino as follows. We randomly scan
over sgoldstino mass parameter $m_{s}$ in the interval $\left(m_{h}-x,
m_{h}\right)$ where $x = 35$~GeV. Such narrow interval was taken to
enhance the mixing angle~\eqref{mix_angle}. We accept the model if
resulting mass of the Higgs-like resonance $\tilde{h}$ falls in the
range 123~GeV$ < m_{\tilde{h}}<$127~GeV. 

Now let us discuss collider constraints which are relevant for 
our study. We start with the LHC data. Detailed determination of the
limits on the masses of superpartners for the low-scale supersymmetry
breaking scenario lies beyond the scope of this study. Still we impose
a set of constraints on masses of superpartners to omit obviously
excluded points in parameter space. 
For chosen value $\sqrt{F}=10$~TeV all superpartners firstly decay
into SM partners and next-to lightest supersymmetric particle (NLSP)
which finally decays into gravitino.  
With gravitino LSP LHC signatures
from the searches for superpartners will be the same as for general
gauge mediation models~\cite{Kats:2011qh,Melzer-Pellmann:2014eta}. 
Below we impose a set of constrains depending on the type of NLSP
which can be in our case the lightest neutralino ${\chi}^0_1$ or 3rd
generation squark, ${\tilde{t}}_1$ and $ {\tilde{b}}_1$. We do not
take into account an exotic case of ${\chi}^{\pm}_1$, which has been
studied in~\cite{chargino_NLSP}. Finally, only very small number of
models in our scan have gluino NLSP and we neglect them completely for 
simplicity. 

If NLSP is bino-like neutralino it decays mainly as
$\tilde{\chi}^0_{1}\to\gamma\tilde{G}$. Corresponding signal events 
have (multi)photon and missing $E_T$
signatures~\cite{Ruderman:2011vv}. This type of searches at ATLAS and 
CMS results in rather stringent limits on masses of superpartners: for
squarks and gluinos from 1.4 to
2~TeV~\cite{atlas1,CMS:2014koa,CMS:2014ets}. However in their analysis
it has been assumed that all squarks have the same mass and they decay
directly to bino-like NLSP, so for our sets of parameters the
constraints should be considerably weaker and we use here conservative 
bound 1.4~TeV on squarks masses. Limits on masses of the
lightest wino-like chargino and $\chi_2^0$ (if they are degenerate)
are about $600-700$~GeV~\cite{atlas1} independently of $\chi_1^0$ mass
and we used in this case the strongest constraint. 
For the case of wino-like or higgsino NLSP neutralino it decays mainly
into $Z$ and/or $h$.  Searches for a diphoton, $Z+\gamma$, $W+\gamma$
and/or jets and $E_T^{miss}$ signatures~\cite{CMS:2014koa} result in
the limits $900-2000$~GeV for gluino and squark masses. Again here
only a simplified case of degenerate squarks has been considered. The
limit on mass of NLSP neutralino $\chi_1^0$ in this case depends on
branchings of $\chi_1^0$ decay into $Z\tilde{G}$ and $h\tilde{G}$ and 
varies~\cite{Khachatryan:2014mma,Khachatryan:2014qwa} from 380~GeV for
${\rm Br}(\chi\to Z\tilde{G})=1$ to zero for ${\rm Br}(\chi\to  
h\tilde{G})=1$. Here we impose the strongest constraint by assuming
that NLSP decays to $Z$ boson pair with 100\% branching ratio. 
 When a squark is NNLSP and wino-like
neutralino is NLSP we take into account constraints from cascade
production of NLSP- lightest neutralino via stop
$m_{{\tilde{t}}_1}>560$~GeV~\cite{stop_cascad} and sbottom 
$m_{{\tilde{b}}_1}>470$~GeV~\cite{sbottom_cascad} squarks. 
In the case of squark ($\tilde{t}_1$ or $\tilde{b}_1$) NLSP we impose
the following bounds from searches for direct pair production of
sbottom $m_{{\tilde{t}}_1}>650$~GeV~\cite{sbottom_dir} and stop
$m_{{\tilde{t}}_1}>760$~GeV~\cite{stop_dir} squarks. 
Somewhat arbitrarily we impose a stringent bound on gluino mass 
$M_3>1.5$~TeV for all the cases. We comment about influence of this
constraint below.

For each chosen model we calculate predicted signal strengths $R =
\sigma/\sigma_{SM}(m_{h^{SM}}=m_{\tilde{h},\tilde{s}})$, i.e. the
ratio of a signal cross section for new Higgs or sgoldstino resonances
to the cross section of the same process in the SM with the Higgs
boson of the same mass. In the narrow width approximation for a 
final state $f$ the signal strength can be written as
\begin{equation}
\label{R}
R_{f}=\frac{\sigma(pp\to \tilde{h}(\tilde{s}))Br(\tilde{h}(\tilde{s})\to f)}
{\sigma(pp\to h^{SM})Br(h^{SM}\to f)},
\end{equation}
where $\sigma(pp\to \tilde{h}(\tilde{s}))$ is the total production rate
of the Higgs-like (sgoldstino-like) state given by sum over different
production mechanisms, $Br(\tilde{h}(\tilde{s})\to f)$ is the
branching ratio of the decay of $\tilde{h}$ ($\tilde{s}$) into final
state $f$, while $\sigma(pp\to \tilde{h}^{SM})$ and
$Br(\tilde{h}^{SM}\to f)$ are similar quantities for the SM Higgs
boson with the same mass. In what follows we consider the following
final states  $\gamma\gamma$, $ZZ$, $WW$, $b\bar{b}$ and
$\tau^{+}\tau^{-}$ which are most relevant for current LHC searches.  
Further, we distinguish between several dominant production
mechanisms, namely gluon-gluon fusion ($ggF$) and vector boson fusion
along with associated production with $W$ and $Z$ ($VBF$ and $VH$) as
they provide with different signatures. The signal strength~\eqref{R}
can be approximated by 
\begin{equation}
\label{R_gg}
R^{ggF}_{f} =\frac{\Gamma(\tilde{h}\to
  gg)Br(\tilde{h}\to f)} {\Gamma(h^{SM}\to gg)Br(h^{SM}\to f)}
\end{equation}
for the case of $ggF$ and as
\begin{equation}
\label{R_VBF}
R^{VBF/VH}_{f} =\frac{\Gamma(\tilde{h}\to
  WW, ZZ)Br(\tilde{h}\to f)} {\Gamma(h^{SM}\to WW, ZZ)Br(h^{SM}\to f)} 
\end{equation}
for $VBF$ or $VH$ production mechanisms. Similar expressions are used
for the case of sgoldstino-like state $\tilde{s}$. Here we 
should note that interactions of sgoldstino with massive vector bosons
are governed by operator which has different structure than that 
for the Higgs boson. But considering kinematics of the processes of
Higgs production via VBF or VH-strahlung mechanisms it is easy to
convince yourself that the momentum-dependent parts of~\eqref{gWW}
and~\eqref{gZZ} give negligible contribution for 
parameters in the {\it Sets 1} and {\it 2} in comparison with the SM 
parts of the couplings unless $\cos{\theta}$ is not too small. The
widths of the decays 
entering~\eqref{R_gg} and~\eqref{R_VBF} are calculated using formulas
in Ref.~\cite{Spira:1997dg} and replacing corresponding coupling
constants with those presented above. The only exception is decays
into pair of massive vector bosons. In this case for the calculation
of partial widths we use results of Ref.~\cite{Romao:1998sr} and
present corresponding formulas in Appendix~\ref{sec:4} for
completeness. 
Experimental constraints on signal strengths from ATLAS and CMS
results will be discussed in the next Section.

As it has been already noted gluon-gluon fusion is the most important 
production mechanism for $\gamma\gamma$, $ZZ$ and $WW$ channels. At
the same time as we observed above the coupling of the Higgs boson
$\tilde{h}$ to the gluons receives tree level
contribution~\eqref{gluons} due to the mixing with sgoldstino. Let us
require that this contribution should not dominate over the SM 
part. It can be suppressed either by small mixing angle or by
sufficiently large $\sqrt{F}$. Considering the case of non-negligible
mixing the sgoldstino coupling is smaller than 1-loop SM contribution
when $\sqrt{F}\gsim \left(\frac{3\pi M_3 v}
{\alpha_s}\right)^{1/2}$. Given chosen limit $M_{3}\gsim 1.5$~TeV    
from the direct searches for gluinos at the LHC one finds
$\sqrt{F}\gsim 7$~TeV. This explains our choice of sufficiently large
value of supersymmetry breaking scale $\sqrt{F}=10$~TeV. We note in
passing that the real constraint on $M_3$ in a given model can be
considerably lower with current ATLAS and CMS data. Thus, smaller
values of $\sqrt{F}$ are possible along with large sgoldstino-Higgs
mixing. 

Now let us turn to the discussion of sgoldstino-like state which is
somewhat lighter than the Higgs-like resonance. Here we impose
additional constraints from LEP~\cite{Schael:2006cr} and
TeVatron~\cite{Abazov:2013gmz}.  
Particularly strong limits come from LEP results on Higgs boson
searches~\cite{Barate:2003sz,Schael:2006cr,LEP_gamma} in $e^+e^-\to
Zh$ with $h\to b\bar{b},\; \tau^+\tau^-$ and $\gamma\gamma$. We remind
reader that a small, about $2\sigma$, excess 
has been observed at LEP in this channel around invariant mass 98~GeV
of $b\bar{b}$ pair. In what follows we would like to explore
interesting possibility that this sgoldstino-like state with mass
around 98~GeV could be source of this excess. For such models we
additionally require that the mass of $\tilde{s}$ should be in the
range $95-101$~GeV  and additionally
$0.1<R_{b\bar{b}}^{VBF/VH}(\tilde{s})<0.25$, see
Ref.~\cite{Barate:2003sz}. Alternative explanations of this excess
have been proposed within Non-minimal Supersymmetric Standard Model
(NMSSM) in papers~\cite{Belanger:2012tt,Bhattacherjee:2013vga}.    

\subsection{Results and discussions}
\label{sec:3_2}

Here we present results of the scan over MSSM parameter space with two 
sets of parameters introduced in Section~\ref{sec:3_1}. In the
figures below we show the different physical quantities for
phenomenologically acceptable models. 
\begin{figure}[htb!]
\begin{picture}(300,160)(0,160)
\put(230,150){\includegraphics[angle=0,width=0.45\textwidth]{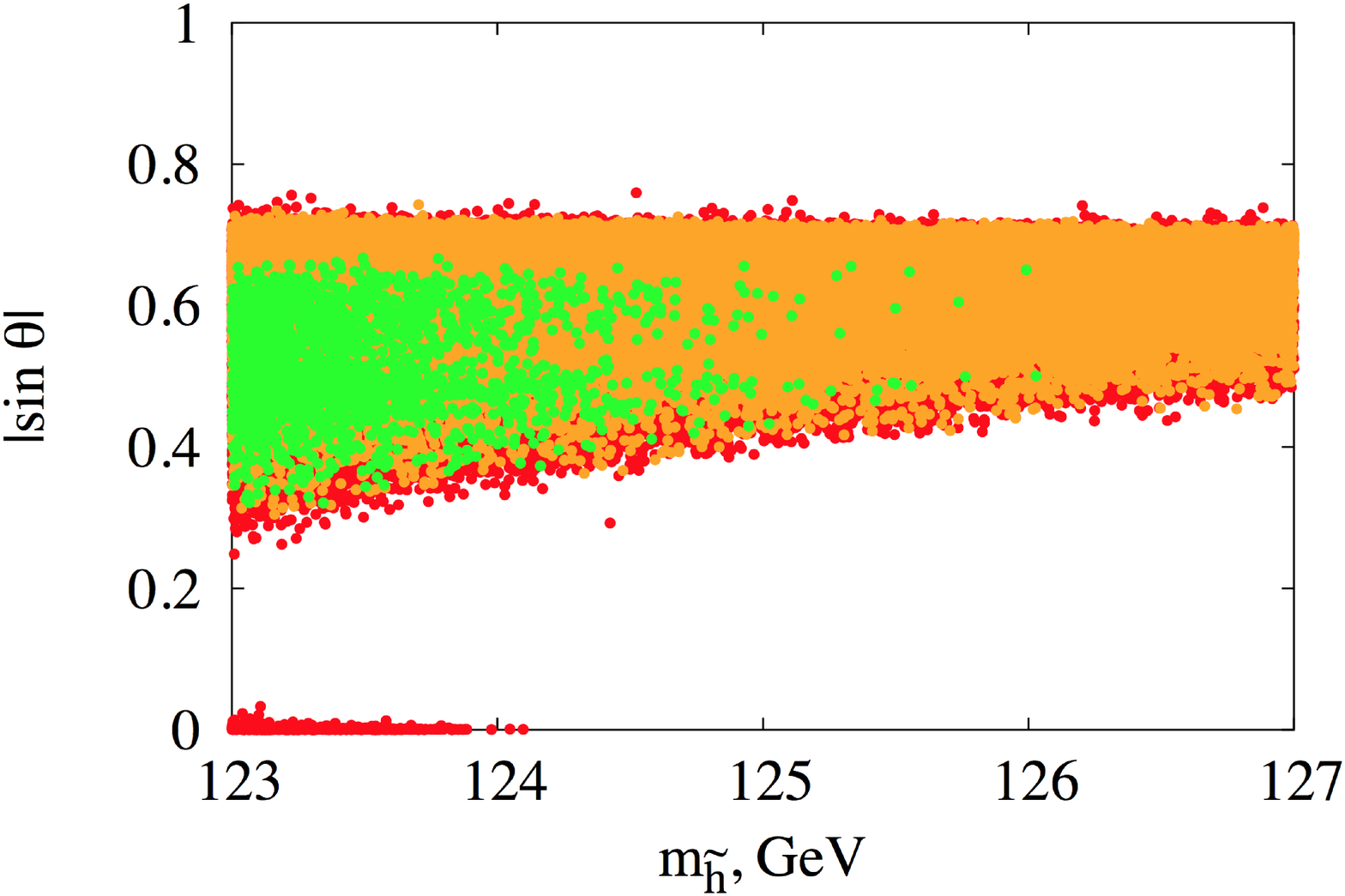}}
\put(0,150){\includegraphics[angle=0,width=0.45\textwidth]{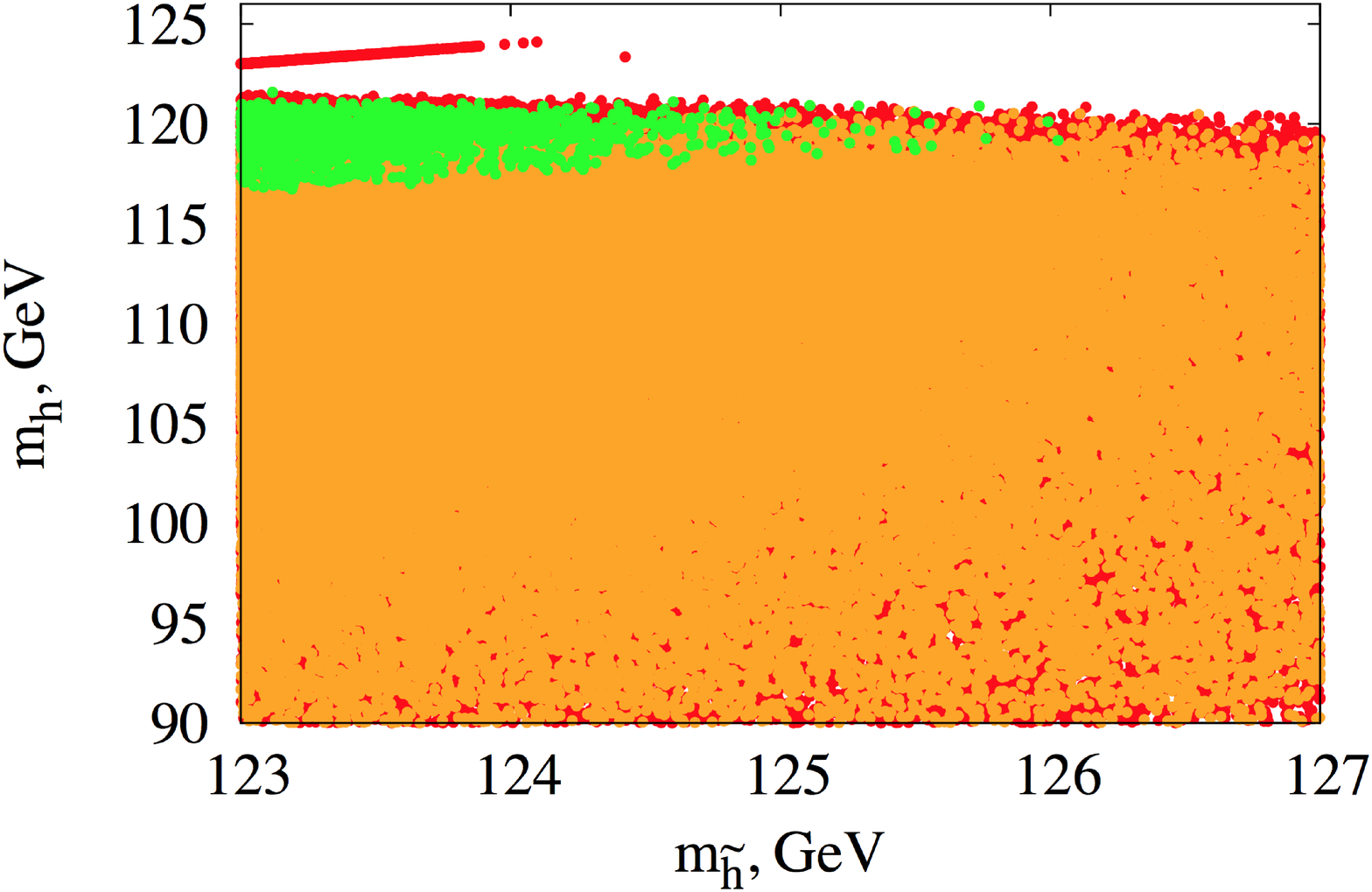}}
\end{picture}
\caption{\label{mh_mixing1} {Scatter plots in
    $m_{\tilde{h}}-m_{h}$ (left panel) and $m_{\tilde{h}}-|\sin\theta|$
    (right panel) planes for {\it Set 1 }. Models excluded by
    ATLAS and CMS searches for superpartners are tagged by  red
    color. Orange points correspond to models that satisfy LHC
    constraints but do not satisfy constraints from LEP experiment. 
    Other models are shown in  green.}}   
\end{figure}
Red points mark models which do not satisfy chosen bounds on
masses of superpartners. By orange points we show models which
are excluded by LEP constraints on sgoldstino-like state production in 
$e^+e^-\to Z\tilde{s}$ discussed above. Models which pass all these
constraints are shown in green.        

We start with {\it Set 1} of parameters. In Fig.~\ref{mh_mixing1}
(left panel) we show distributions of models over the mass of
the Higgs 
resonance before and after mixing. We see that without the mixing 
mass $m_{h}$ is always below $123$~GeV except for very
limited number of models. The mixing with sgoldstino can increase the
mass of Higgs-like state $\tilde{h}$ till observed value. 
\begin{figure}[htb!]
\begin{center}
{\includegraphics[angle=0,width=0.50\textwidth]{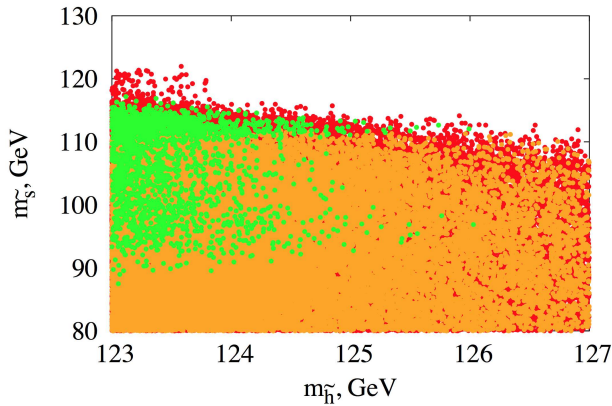}}
\end{center}
\caption{\label{mh_mixing2} {Scatter plot in $m_{\tilde{h}}-m_{\tilde{s}}$
 plane of Set 1 data. Color notations are the same as in
 Fig.~\ref{mh_mixing1}.}}    
\end{figure}
However, the number of
acceptable models considerably decreases with increase of
$m_{\tilde{h}}$. In Fig.~\ref{mh_mixing1} (right panel) we show mixing angle
versus mass $m_{\tilde{h}}$. We see that for the parameter space given
\begin{figure}[htb!]
\begin{picture}(300,300)(0,0)
\put(0,150){\includegraphics[angle=0,width=0.45\textwidth]{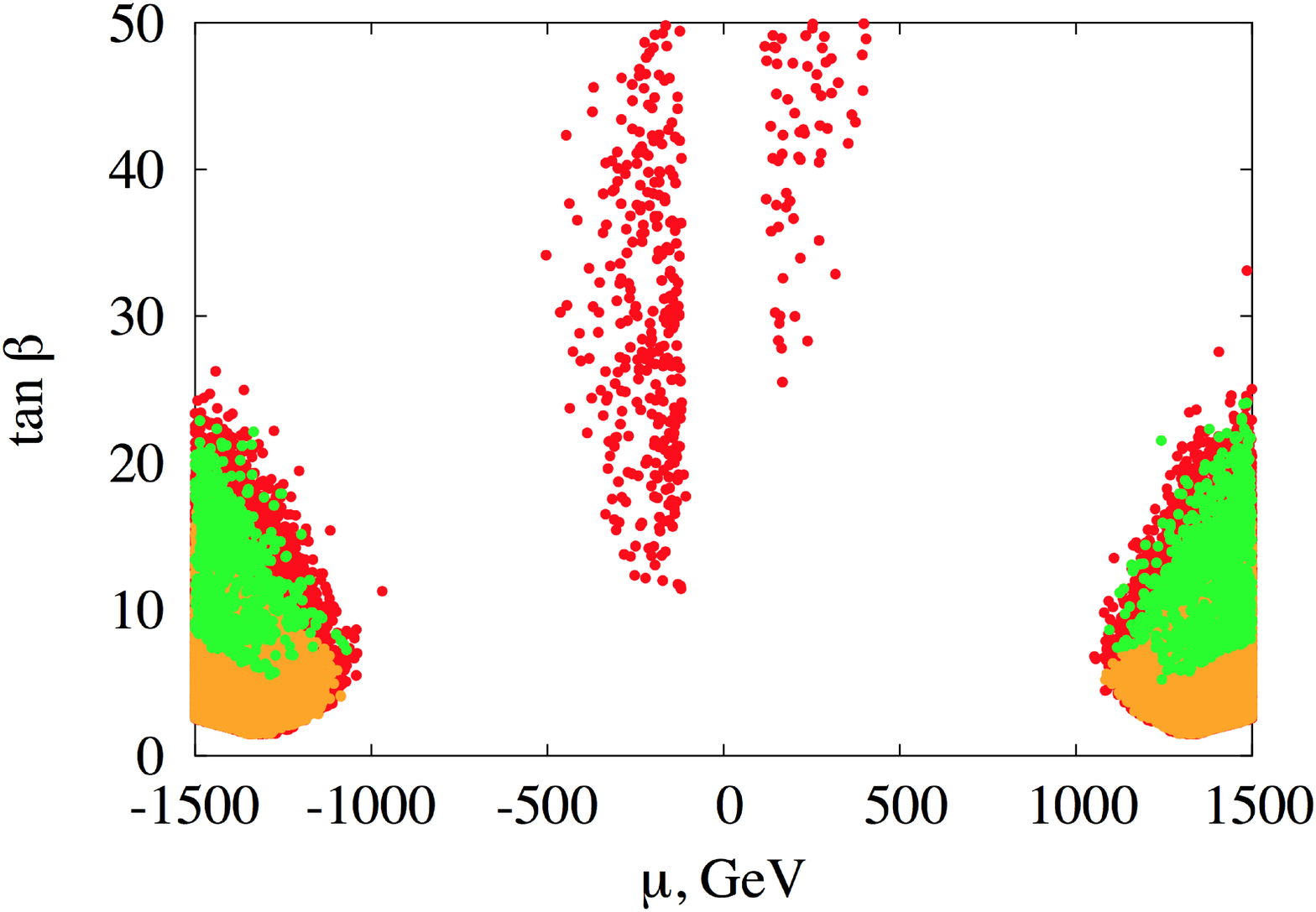}}
\put(210,150){\includegraphics[angle=0,width=0.45\textwidth]{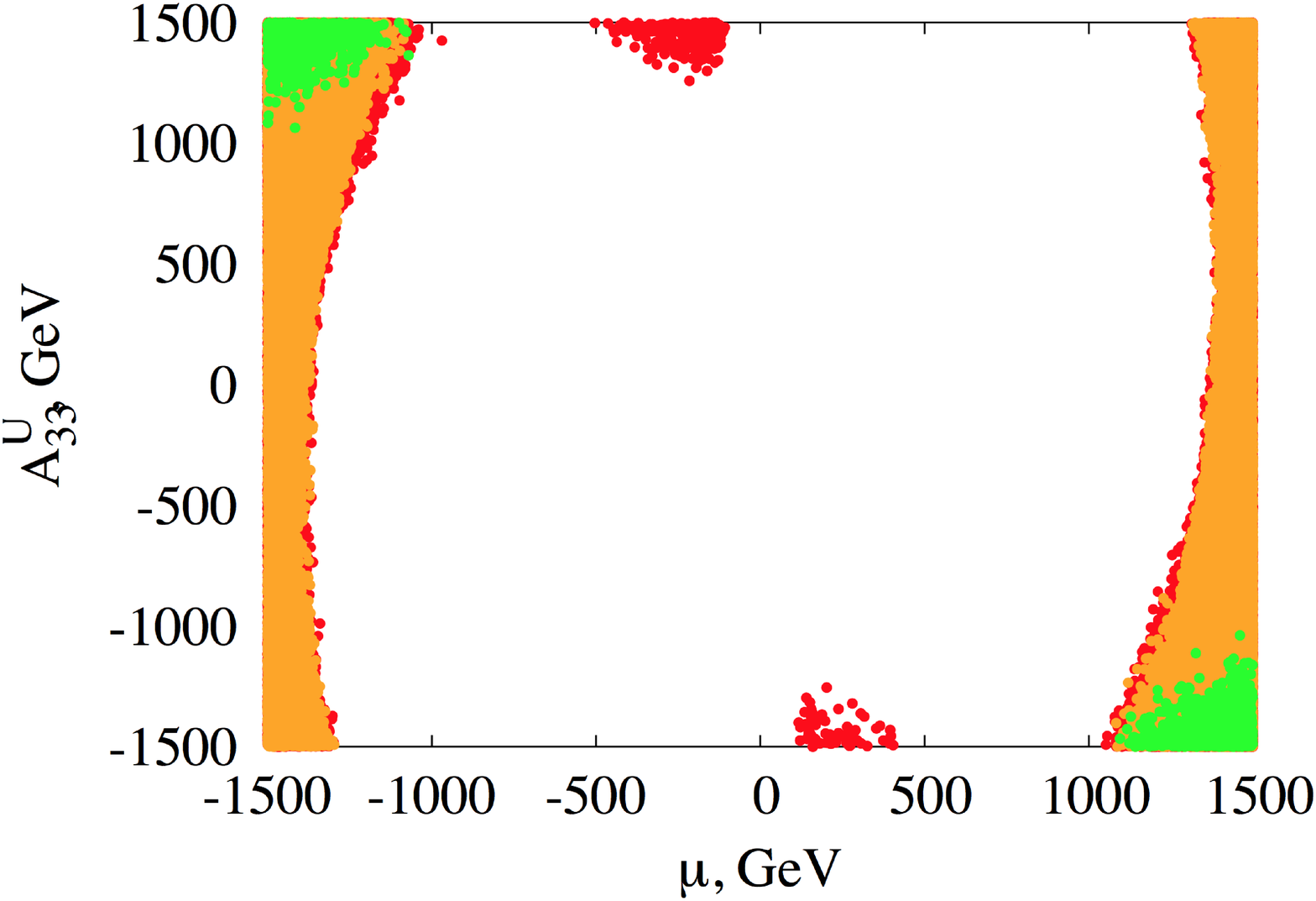}}
\put(210,0){\includegraphics[angle=0,width=0.45\textwidth]{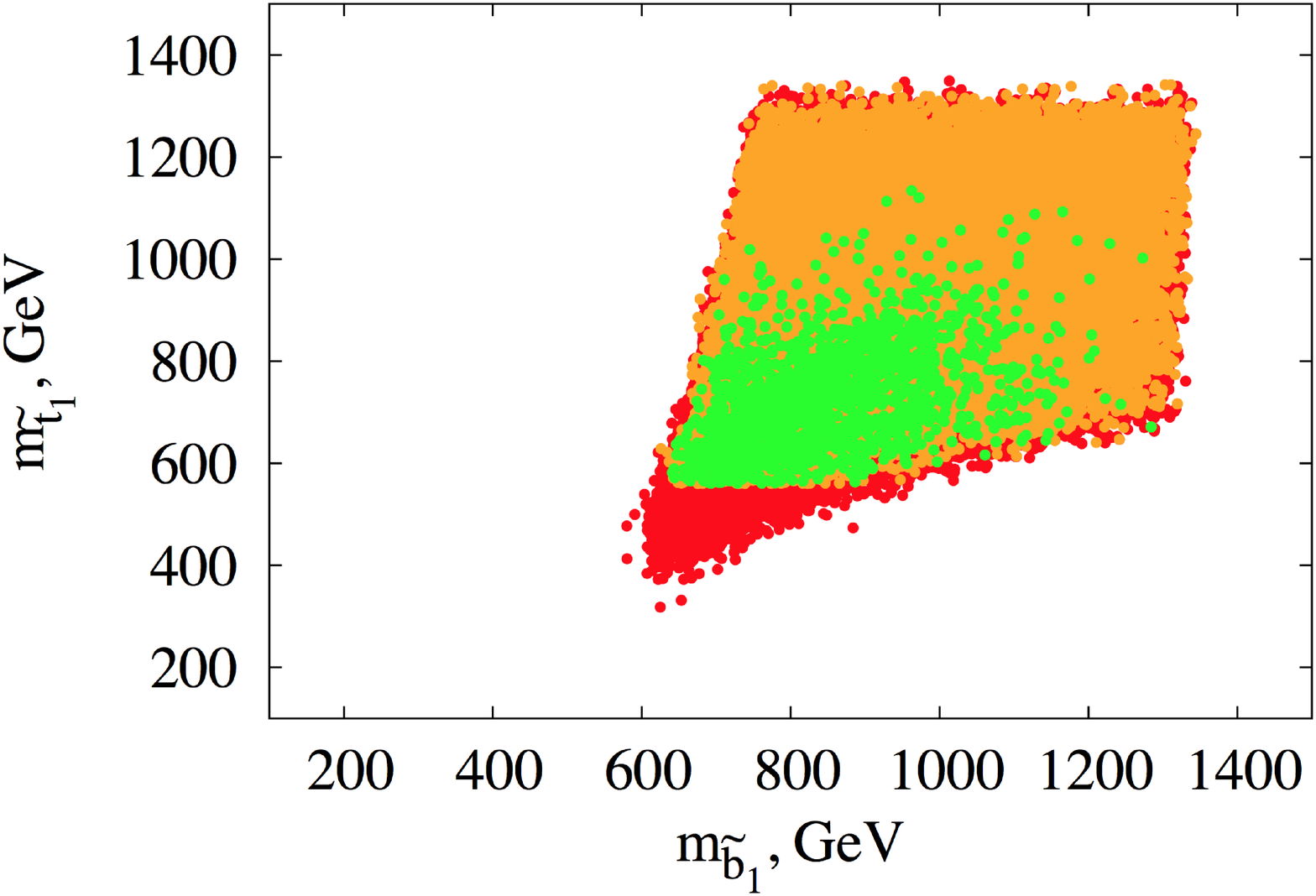}}
\put(0,0){\includegraphics[angle=0,width=0.45\textwidth]{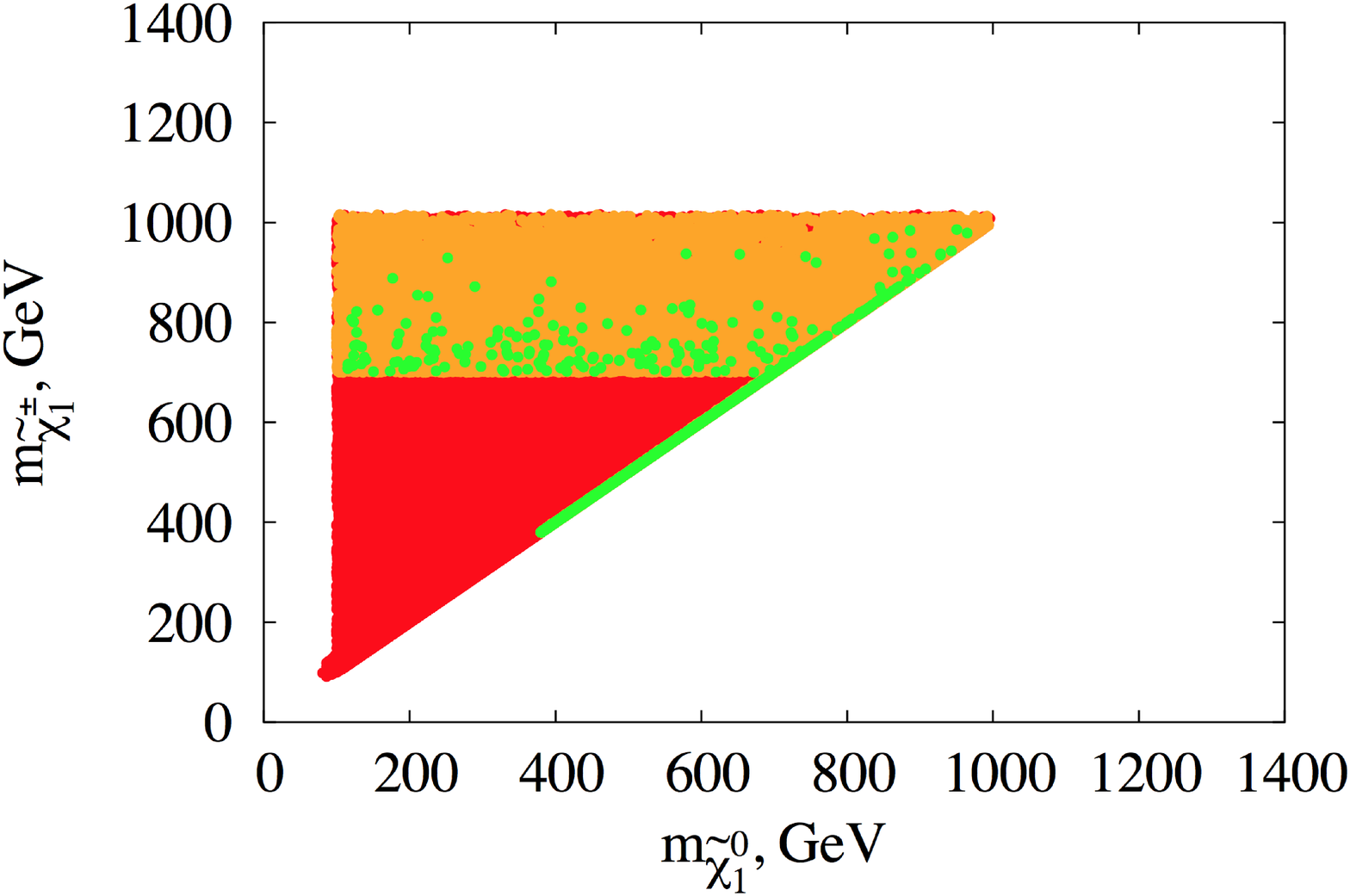}}
\end{picture}
\caption{\label{A_t} {Scatter plots in $\mu-\tan\beta$ (upper
    left panel), $\mu - A_{33}^{U}$ (upper right panel),
    $m_{{\tilde{\chi}}^0_1} - m_{{\tilde{\chi}}^{\pm}_1}$ (lower
    left panel) and $m_{{\tilde{b}}_1}-m_{{\tilde{t}}_1}$ (lower
      right panel) planes for {\it Set 1}.  Color notations are the
    same as in Fig.~\ref{mh_mixing1}. }}  
\end{figure}
by {\it Set~1}, the Higgs-like resonance should have considerable
admixture of sgoldstino with $|\sin{\theta}|\sim 0.4-0.6$ to get
observable value for its mass. Thus, in the most of the acceptable
models the Higgs mass reaches its observed value without large masses
of stops and mixing in their sector. The models with negligible mixing
with sgoldstino on the right plot correspond to those models
on the left plot in which mass $m_{h}$ exceeds 123~GeV. These models 
appear to be closed by searches for superpartners. In
Fig.~\ref{mh_mixing2} 
we show the masses of Higgs-like and sgoldstino-like resonances and
see that in the described setup sgoldstino with masses larger than
85~GeV is favorable. In Fig.~\ref{A_t} we show distributions of models
on different combinations of several MSSM parameters and masses of
superpartners for {\it Set 1} of parameters. 
One can see from the distribution in the parameters $\mu-\tan\beta$
(upper left panel) that large values of $\mu$ and moderate $\tan{\beta}$ are
preferable. This can be explained from the expression for mixing in
Eq.~\eqref{X_mixing} and the Higgs boson mass~\eqref{mh_before}
and~\eqref{mh_after}: small $\mu$ and large $\tan{\beta}$ result in
suppression in mixing parameter $X$. Smaller values of $\tan{\beta}$ are not
favorable because tree level value of the Higgs boson mass becomes
additionally suppressed, see Eq.~\eqref{mh_before}. In the upper right plot in
this Figure we show values of $A_{33}^{U}$ versus $\mu$ and see that
phenomenologically acceptable models have $A_{33}^{U}$ near its
largest value for {\it Set 1}. The reason is that such values of
$A_{33}^{U}$ increase $X_t=A^U_{33}-\mu\cot\beta$ and as a result
increase  1-loop correction to Higgs mass~\cite{Hall:2011aa}
\begin{equation}
\delta=\frac{3}{(4\pi)^2}\frac{m_t^4}{v^2}\left[\ln\frac{m_{\tilde{t}_1}m_{\tilde{t}_2}}{m_t^2}+\frac{X_t^2}{m_{\tilde{t}_1}m_{\tilde{t}_2}}\left(1-\frac{X_t^2}{12m_{\tilde{t}_1}m_{\tilde{t}_2}}\right)\right]
\end{equation}
The masses of the lightest neutralino and chargino are shown in the
lower left panel in Fig.~\ref{A_t}.   
In the lower right panel we show the masses of lightest stop and
sbottom squarks. We see that there are plenty of models in which these
masses can be as light as $500-700$~GeV what can be explored in the
future LHC runs. Scatter plots similar to those in
Figs.~\ref{mh_mixing1}--\ref{A_t} can be obtained for the {\it Set 2}
of parameters which is considerably wider. But in this case they are
not so informative as corresponding models admit arbitrary mixing
between the lightest Higgs boson and scalar sgoldstino. 

Now we turn to the discussion of LHC signal strengths for the 
Higgs-like resonance $\tilde{h}$.  On the plots below we drop all the
models excluded by the LEP constraints or LHC bounds on masses of
superpartners and for remaining models we introduce constraints for
signal strengths obtained by ATLAS and CMS experiments in their
searches for the Higgs boson~\cite{ATLAS+CMS,CMS:ril}. 
Although for $\gamma\gamma$ and $ZZ$ ($WW$) channels the dominating
production mechanism is $ggF$ while for
$\tau\tau$ and $b\bar{b}$ channels this is $VBF/VH$
still we conservatively impose the following constraints (obtained by
unification of ATLAS and CMS results) independently of the Higgs
production mechanism 
\begin{gather}
0.51<R^{ggF,\; VBF/VH}_{\gamma\gamma}(\tilde{h})<1.9,\;\;\;
0.66<R^{ggF,\; VBF/VH}_{ZZ}(\tilde{h})<1.84, \;\;\; \nonumber \\
\label{higgs_bounds}
0.53<R^{ggF,\; VBF/VH}_{WW}(\tilde{h})<1.32, \;\;\; 
0.51<R^{ggF,\; VBF/VH}_{\tau\tau}(\tilde{h})<1.9, \;\;\; \\
0<R^{ggF,\; VBF/VH}_{bb}(\tilde{h})<1.5. \nonumber
\end{gather}

In the figures below we show in magenta the models which satisfy
the bounds~\eqref{higgs_bounds}. Also we mark in 
  blue color the models in which additionally sgoldstino-like
resonance can explain 98~GeV LEP excess.  

We show signal strengths for the Higgs-like resonance 
\begin{figure}[htb!]
\begin{picture}(300,300)(0,0)
\put(0,150){\includegraphics[angle=0,width=0.45\textwidth]{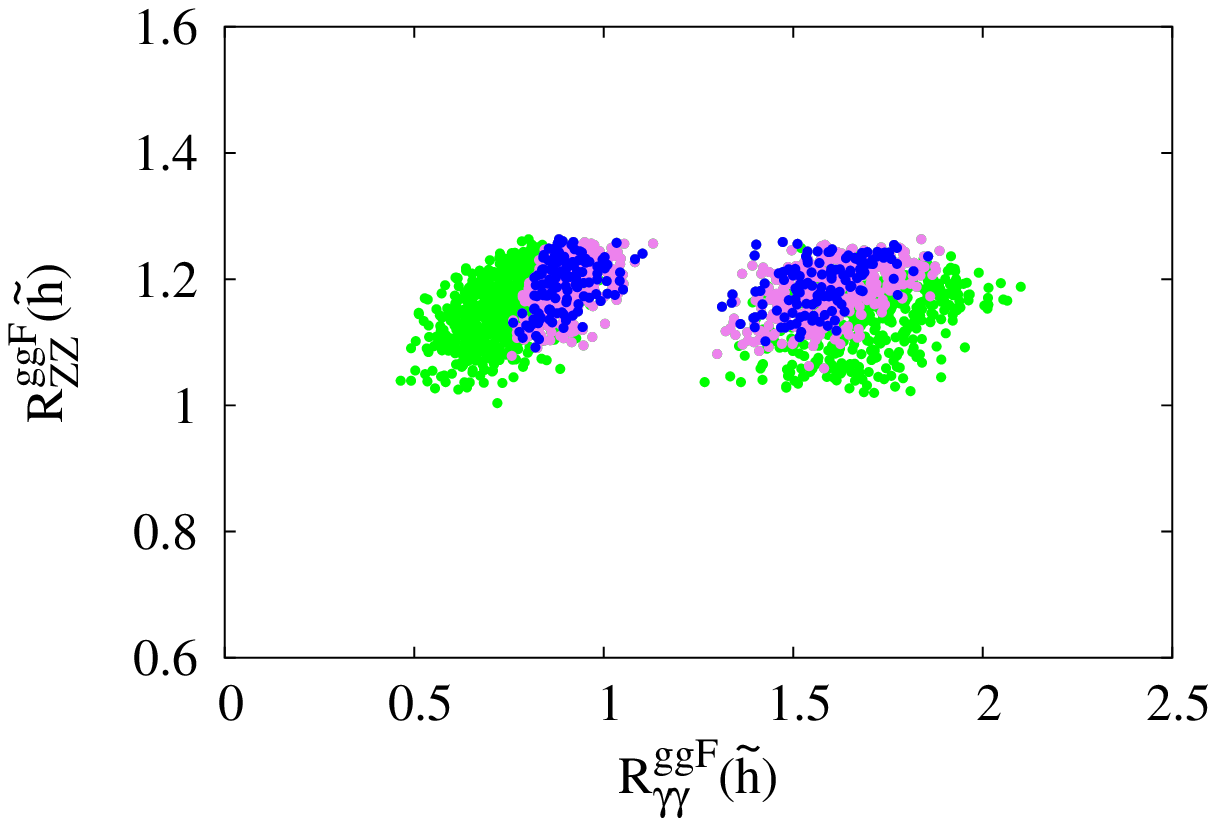}}
\put(210,150){\includegraphics[angle=0,width=0.45\textwidth]{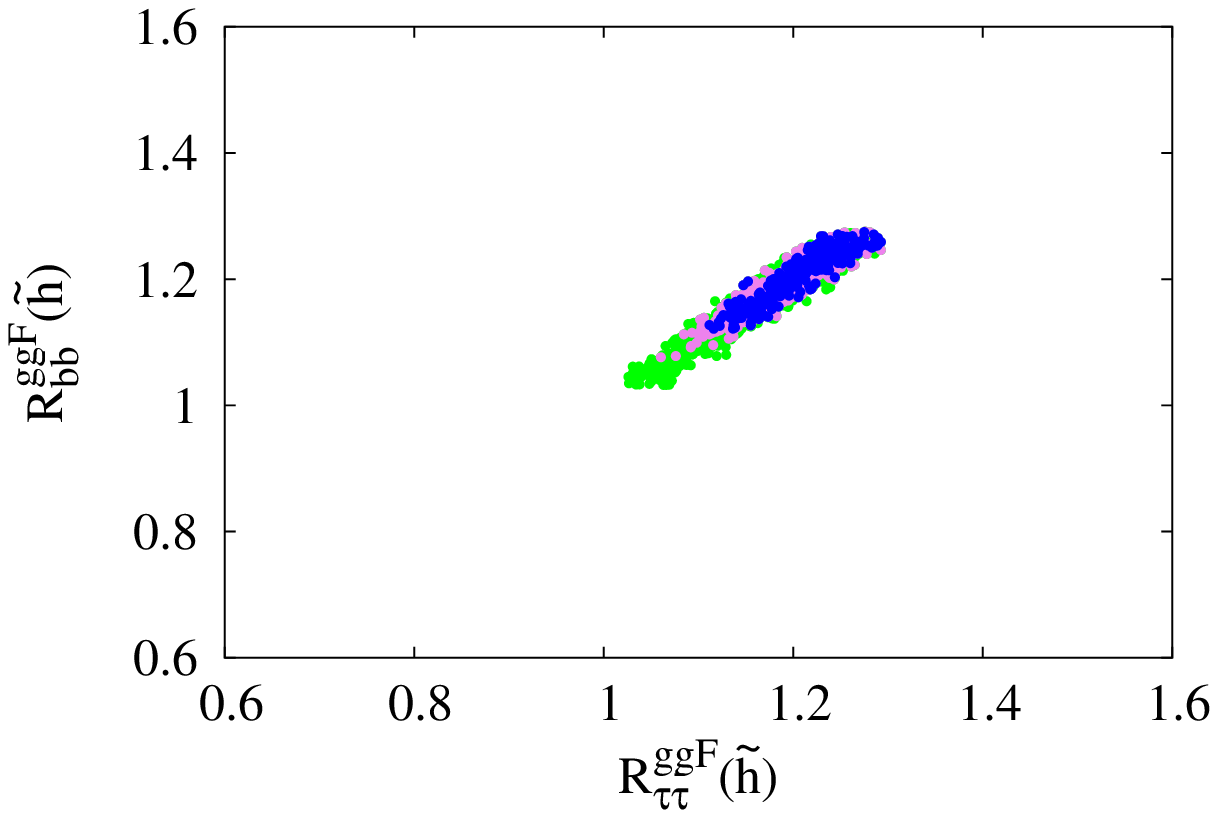}}
\put(0,0){\includegraphics[angle=0,width=0.45\textwidth]{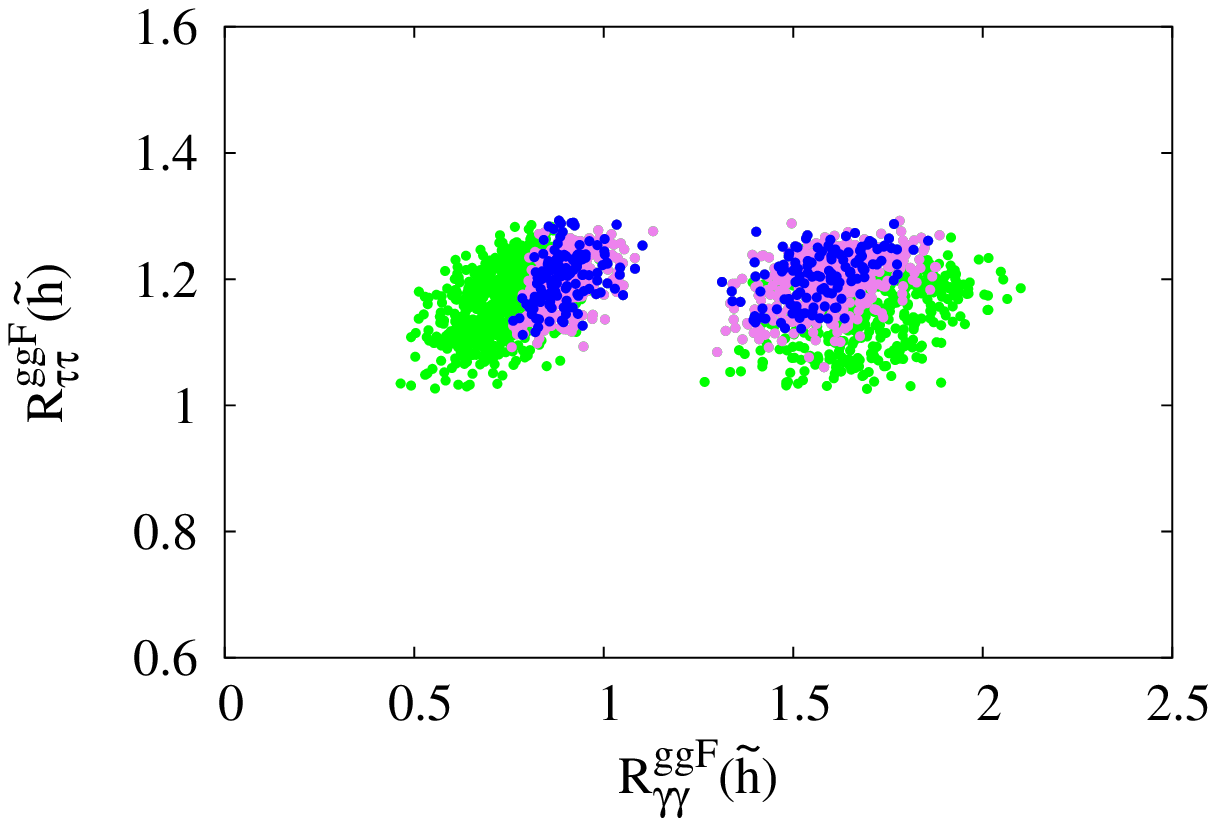}}
\put(210,0){\includegraphics[angle=0,width=0.45\textwidth]{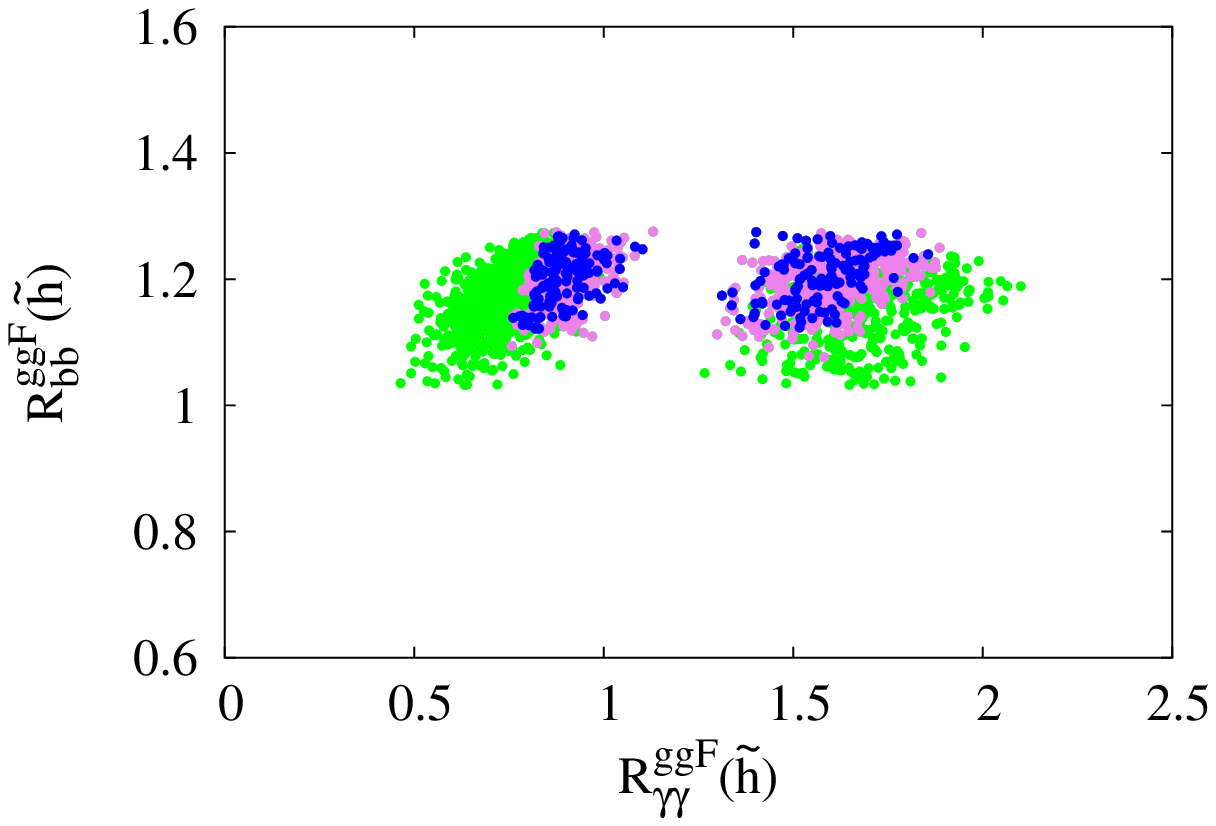}}
\end{picture}
\caption{\label{mh_Rgg} {Scatter plots in
    $R_{\gamma\gamma}^{ggF}(\tilde{h})-R_{ZZ}^{ggF}(\tilde{h})$ (upper
     left panel),
    $R_{\tau\tilde{\tau}}^{ggF}(\tilde{h})-R_{b\tilde{b}}^{ggF}(\tilde{h})$
    (upper right panel), 
    $R_{\gamma\gamma}^{ggF}(\tilde{h})-R_{\tau\tilde{\tau}}^{ggF}(\tilde{h})$
    (lower  left panel) and 
    $R_{\gamma\gamma}^{ggF}(\tilde{h})-R_{b\tilde{b}}^{ggF}(\tilde{h})$
    (lower right panel) planes for {\it Set 1}. All the models satisfy both
    LEP constraints and LHC bounds on masses of superpartners. 
    Models which satisfy constraints Higgs signal
    strength~\eqref{higgs_bounds} are marked by magenta. If in
    addition the model contains sgoldstino-like resonance capable of
    explaining 98~GeV LEP excess it is shown in blue.}} 
\end{figure}
in gluon-gluon fusion production channel in different combinations in
Fig.~\ref{mh_Rgg} for the {\it Set 1} and in Fig.~\ref{FIG10}
\begin{figure}[htb!]
\begin{picture}(300,300)(0,0)
\put(0,150){\includegraphics[angle=0,width=0.45\textwidth]{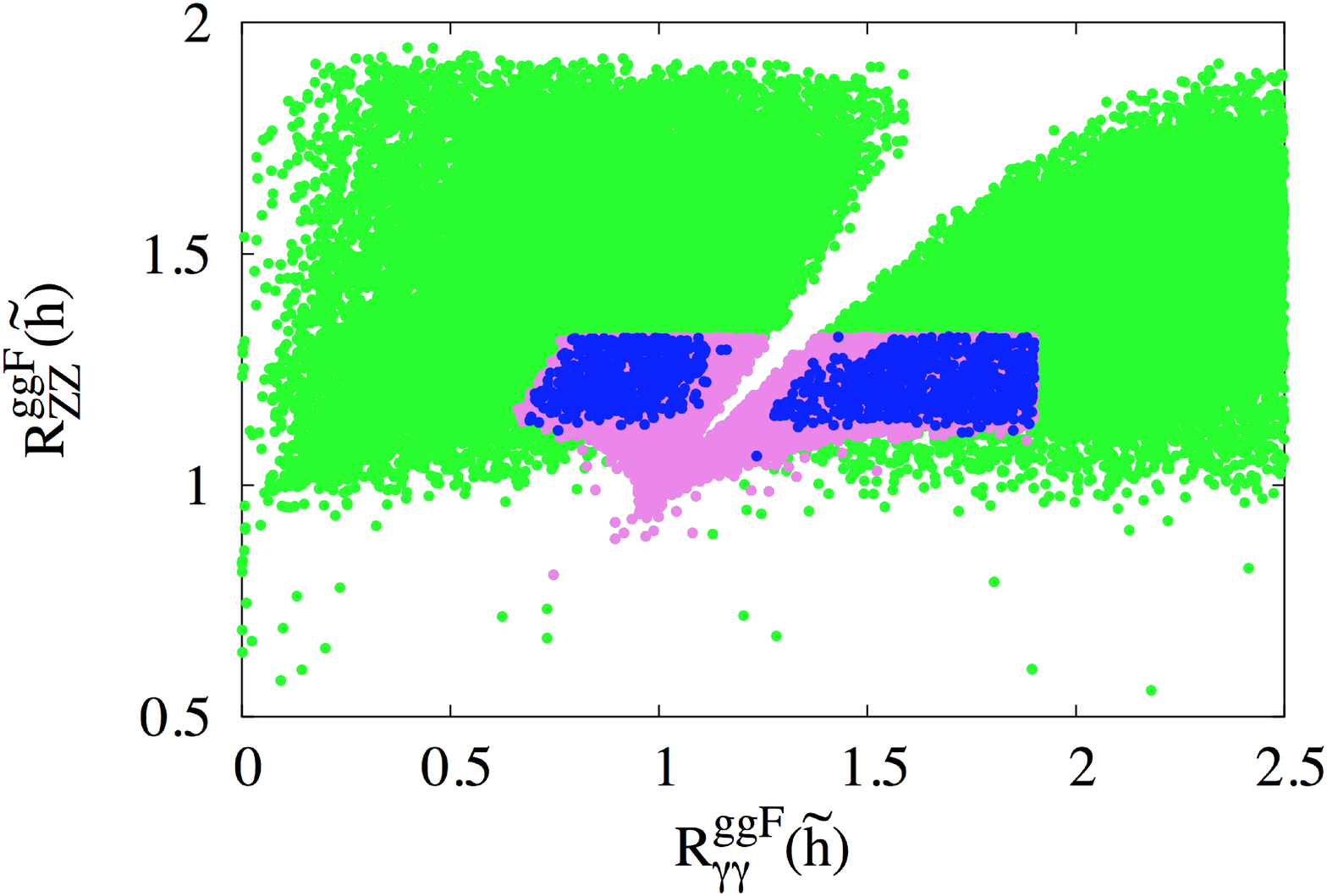}}
\put(210,150){\includegraphics[angle=0,width=0.45\textwidth]{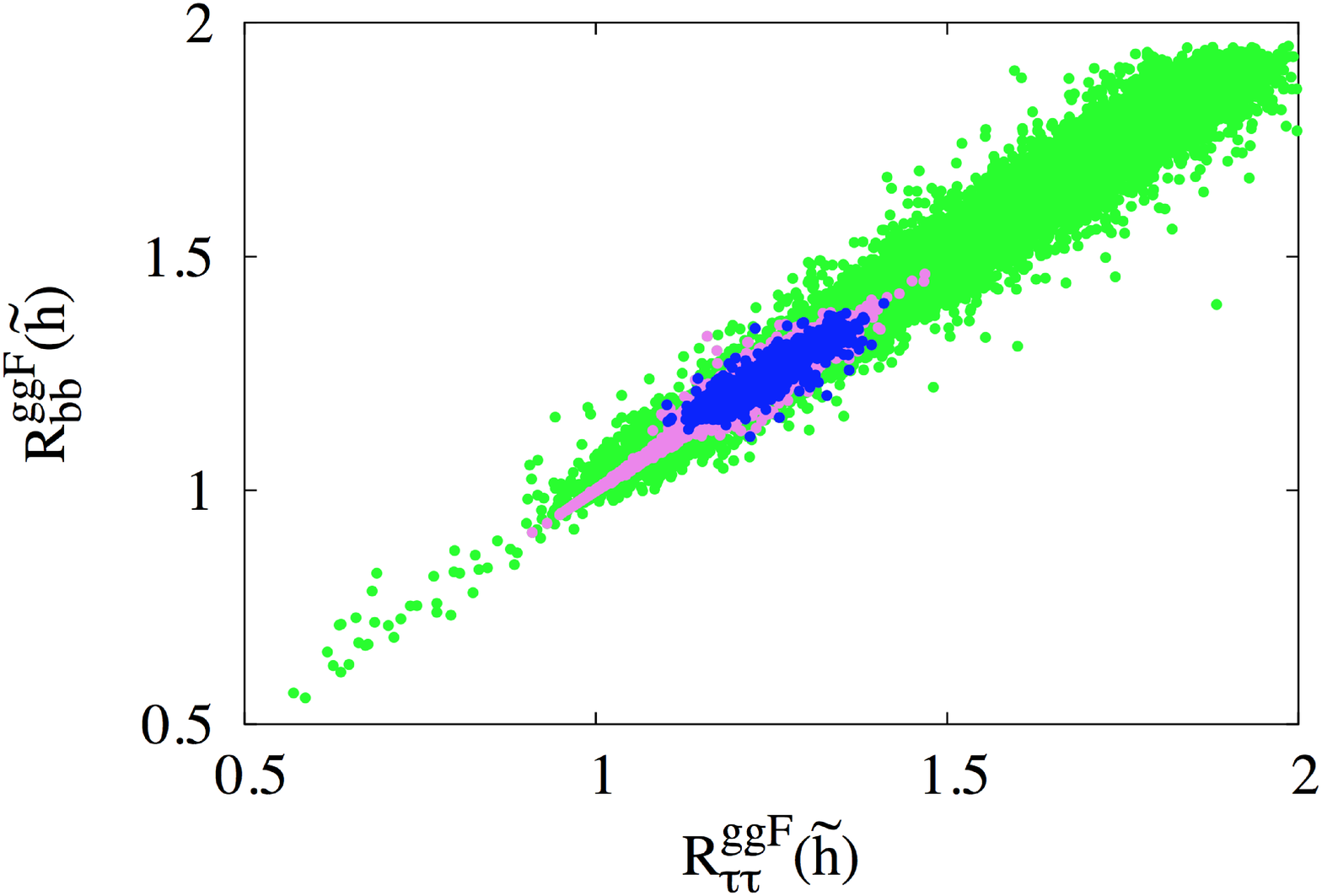}}
\put(0,0){\includegraphics[angle=0,width=0.45\textwidth]{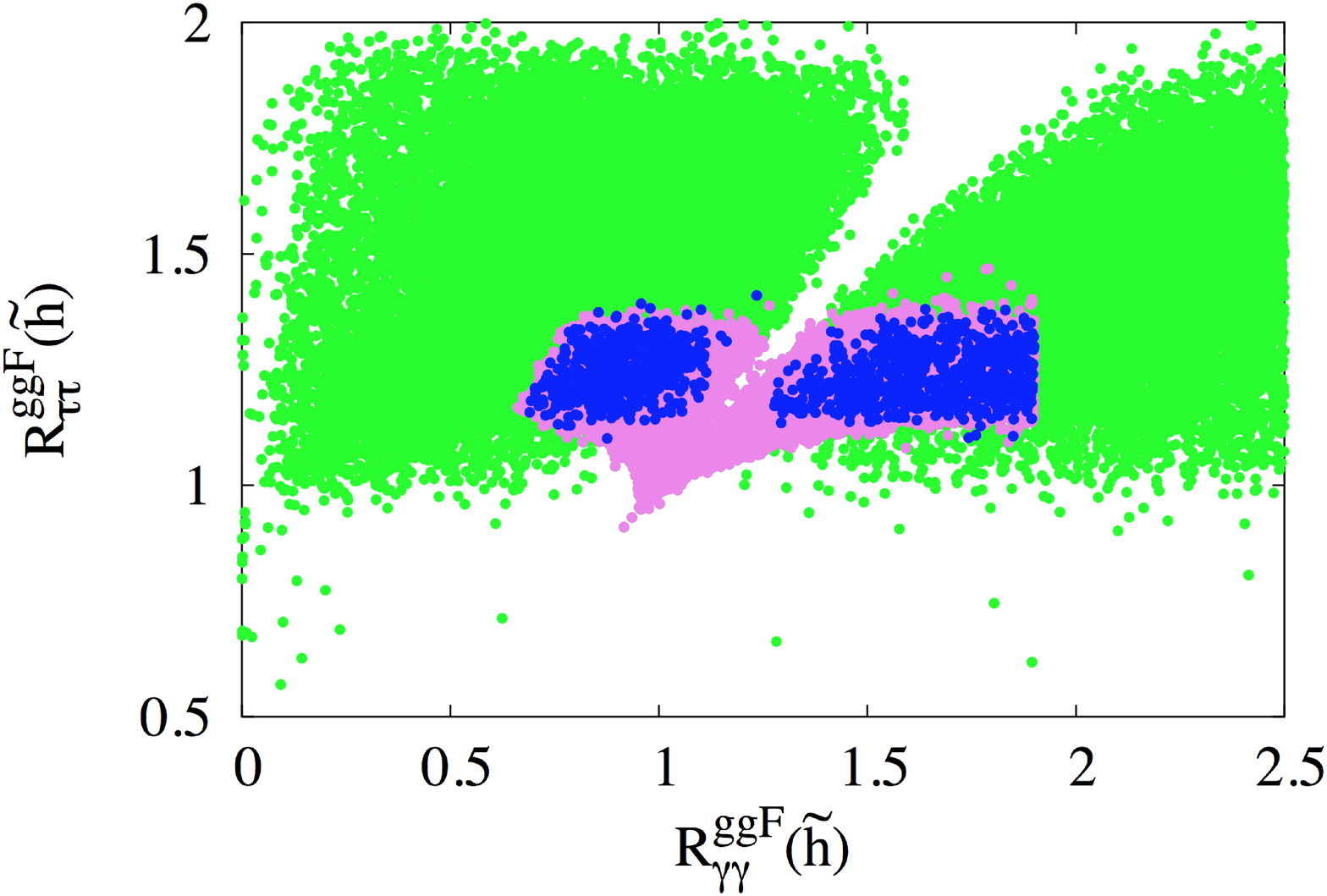}}
\put(210,0){\includegraphics[angle=0,width=0.45\textwidth]{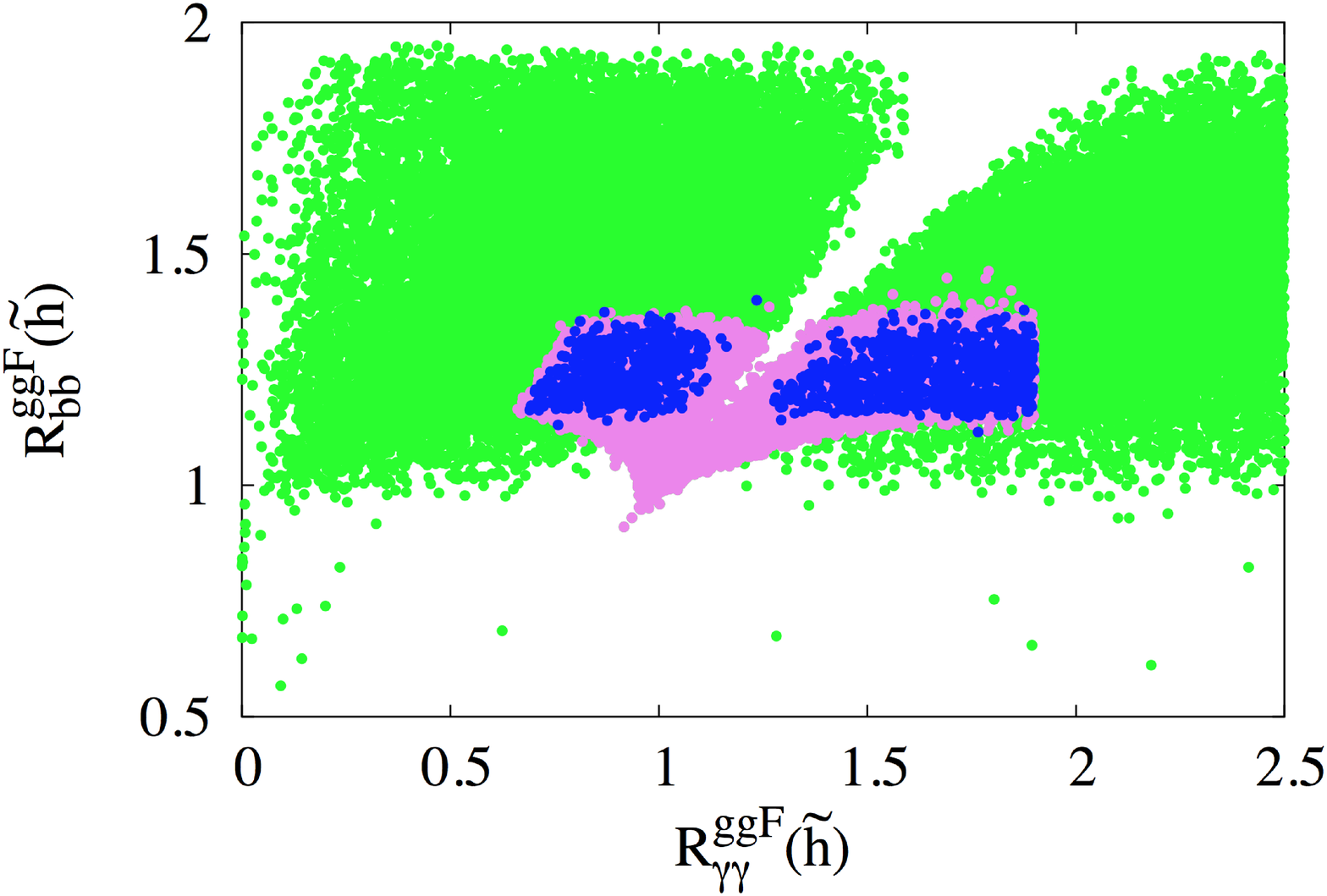}}
\end{picture}
\caption{\label{FIG10} Scatter plots in
  $R_{\gamma\gamma}^{ggF}(\tilde{h})-R_{ZZ}^{ggF}(\tilde{h})$
  (upper left panel), 
  $R_{\tau\tilde{\tau}}^{ggF}(\tilde{h})-R_{b\tilde{b}}^{ggF}(\tilde{h})$
  (upper right panel), 
  $R_{\gamma\gamma}^{ggF}(\tilde{h})-R_{\tau\tilde{\tau}}^{ggF}(\tilde{h})$
  (lower left panel), 
  $R_{\gamma\gamma}^{ggF}(\tilde{h})-R_{b\tilde{b}}^{ggF}(\tilde{h})$ (lower
  right panel) planes for {\it Set 2}.  The color notations are the
  same as in Fig.~\ref{mh_Rgg}.} 
\end{figure}
for {\it Set 2}.
From the plots in Figs.~\ref{mh_Rgg} and~\ref{FIG10} we see that
all the signal strengths except for $R_{\gamma\gamma}^{ggF}$ are
somewhat larger than unity for phenomenologically acceptable models,
while for $\gamma\gamma$ channel there are two regions with higher and
lower values of the signal strengths. 
Since sgoldstino $s$ has tree level couplings to photons and gluons
while for the Higgs boson $h$ these couplings appear only at loop
level, in general one expects large sensitivity of the couplings of
Higgs-like state $\tilde{h}$ to sgoldstino admixture and to
corresponding parameters which govern these couplings, namely
$M_3$ and $M_{\gamma\gamma}$. Depending on relative signs between the mixing
angle (which is determined by the sign of $\mu$) and soft gaugino
mass parameters $M_{1,2,3}$ the couplings to gluons and photons can
either increase or decrease with respect to their values without the
mixing. We have found that $M_3$ and $\mu$ should have opposite
signs for the coupling $g_{\tilde{h}gg}$ be close to experimentally
observed value. With another choice of the signs 
the coupling of $\tilde{h}$ to gluons becomes unacceptably small; we do 
not show corresponding models in all the Figures below.  The signs of
$M_{1}$ and $M_{2}$ can be arbitrary (we choose them of the same sign)
and they correspond to two different domains for
$R^{ggF}_{\gamma\gamma}$ in Figs.~\ref{mh_Rgg} and~\ref{FIG10}.  
The increase in the signal strengths for fermionic and massive vector
boson channels is related to the fact that with our choice of
parameters and of the signs of $\mu$ and $M_3$ the coupling 
of $\tilde{h}$ to gluons appears to be somewhat larger than its value
in SM. Hence, the production cross section in $ggF$ 
increases. 
 
\begin{figure}[htb!]
\begin{picture}(300,300)(0,0)
\put(0,150){\includegraphics[angle=0,width=0.45\textwidth]{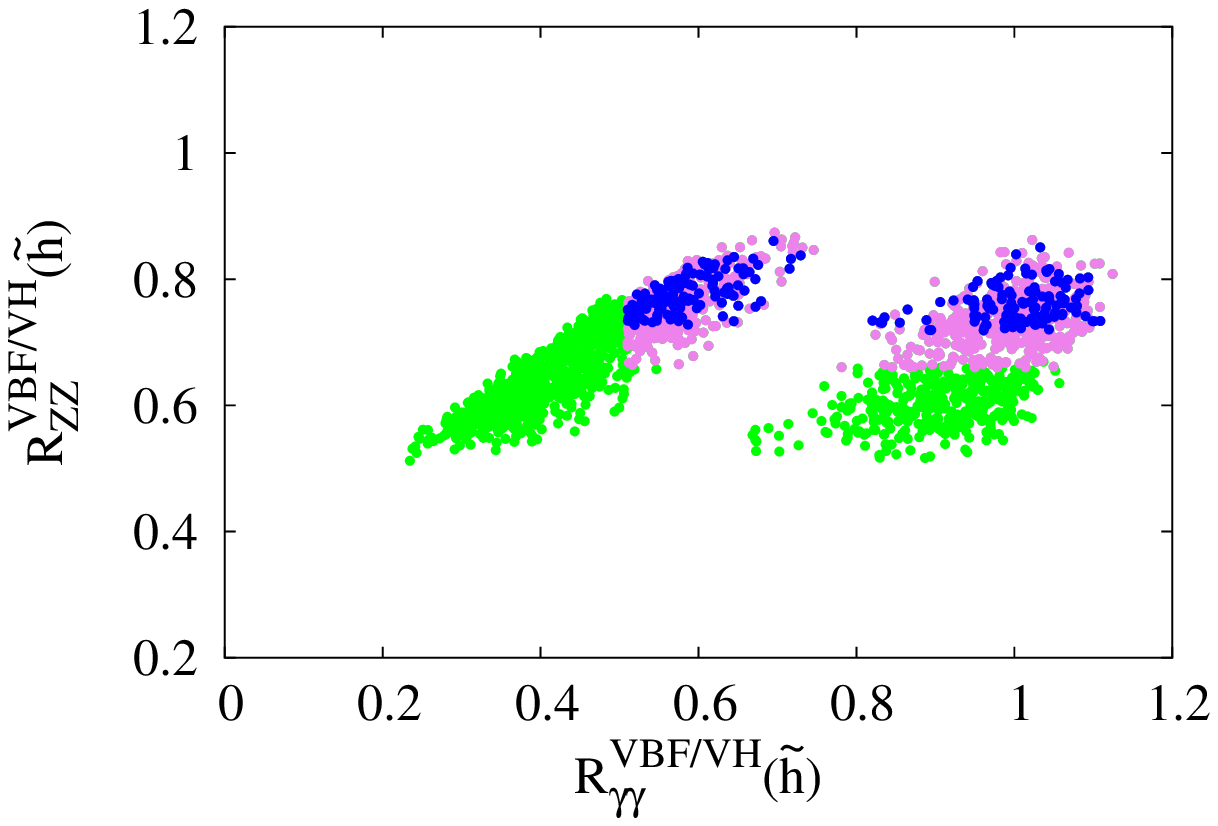}}
\put(210,150){\includegraphics[angle=0,width=0.45\textwidth]{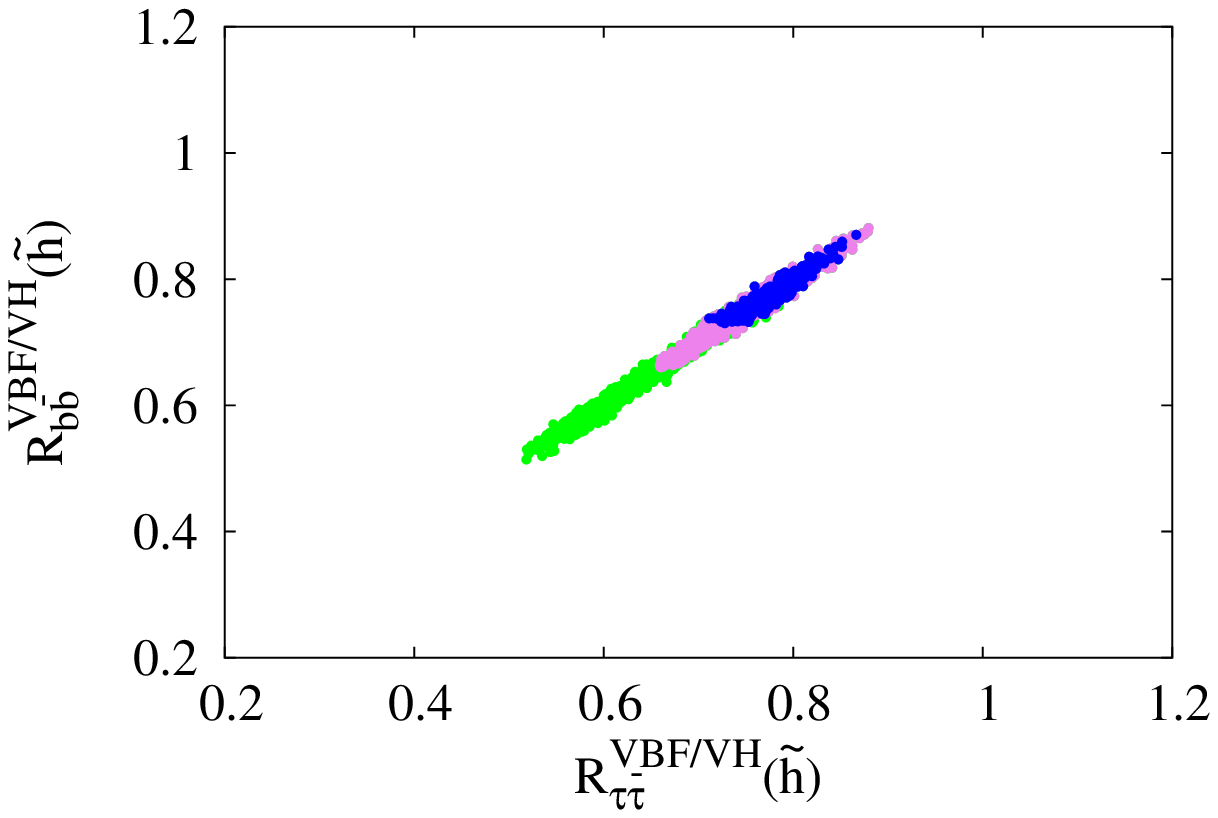}}
\put(0,0){\includegraphics[angle=0,width=0.45\textwidth]{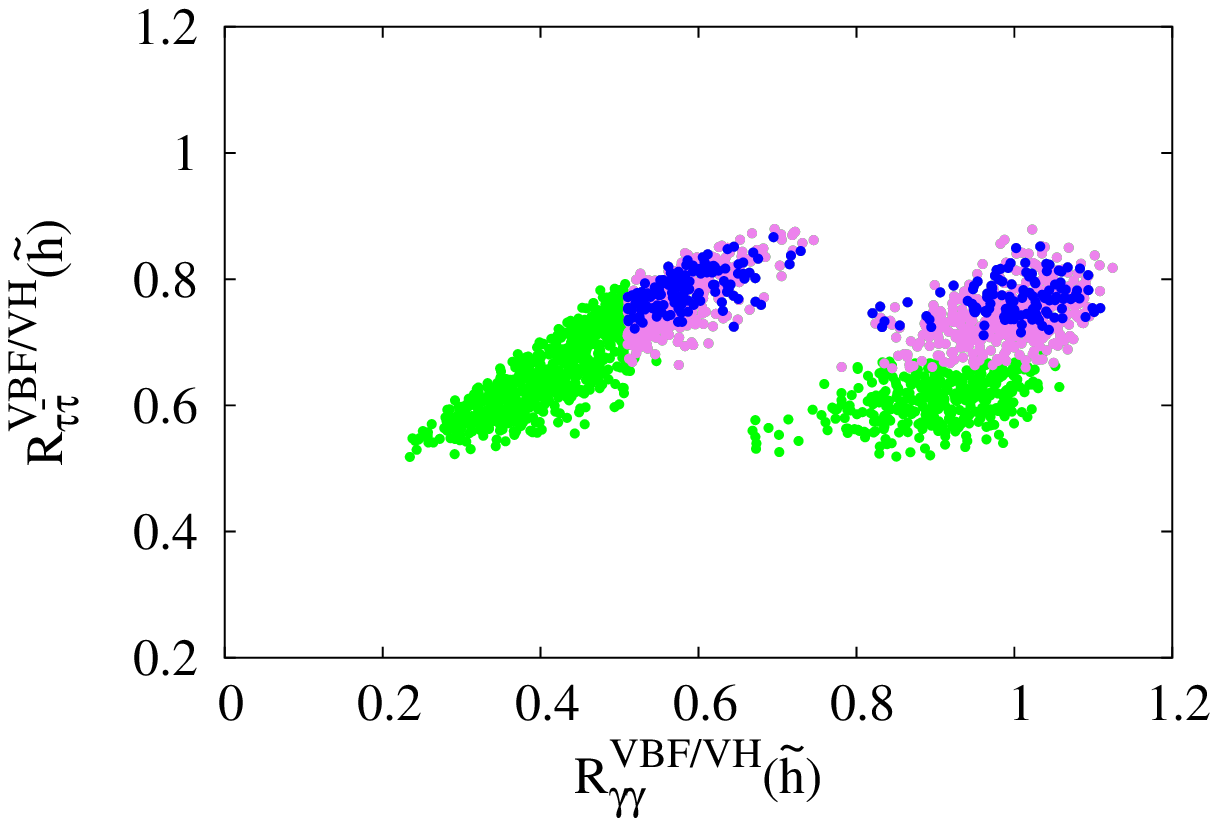}}
\put(210,0){\includegraphics[angle=0,width=0.45\textwidth]{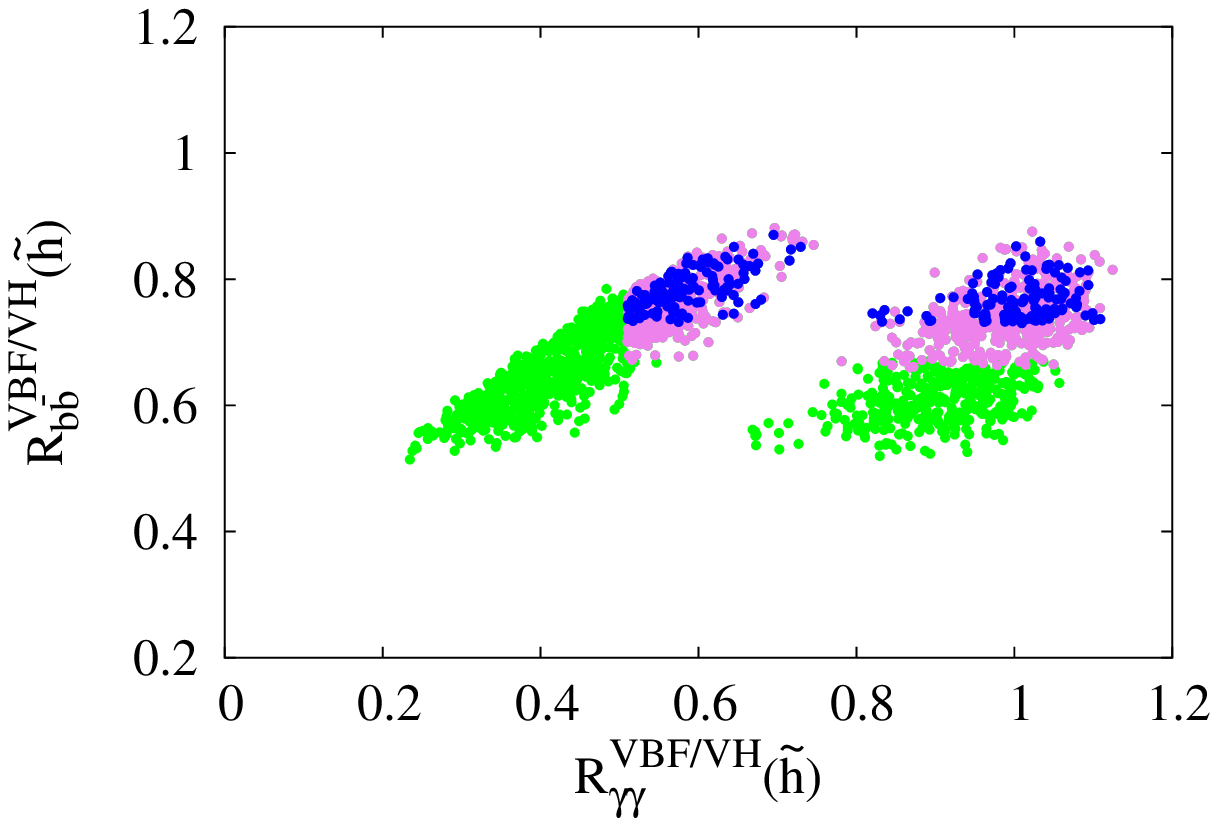}}
\end{picture}
\caption{\label{mh_RVBF} {Scatter plots in
    $R_{\gamma\gamma}^{VBF/VH}(\tilde{h})-R_{ZZ}^{VBF/VH}(\tilde{h})$
    (upper left panel), 
  $R_{\tau\tilde{\tau}}^{VBF/VH}(\tilde{h})-R_{b\tilde{b}}^{VBF/VH}(\tilde{h})$
(upper right panel),
  $R_{\gamma\gamma}^{VBF/VH}(\tilde{h})-R_{\tau\tilde{\tau}}^{VBF/VH}(\tilde{h})$
(lower left panel),
  $R_{\gamma\gamma}^{VBF/VH}(\tilde{h})-R_{b\tilde{b}}^{VBF}(\tilde{h})$
(lower right panel) planes of {\it Set 1}. The color notations are the same
    as in Fig.~\ref{mh_Rgg}.}} 
\end{figure}
\begin{figure}[htb!]
\begin{picture}(300,300)(0,0)
\put(0,150){\includegraphics[angle=0,width=0.45\textwidth]{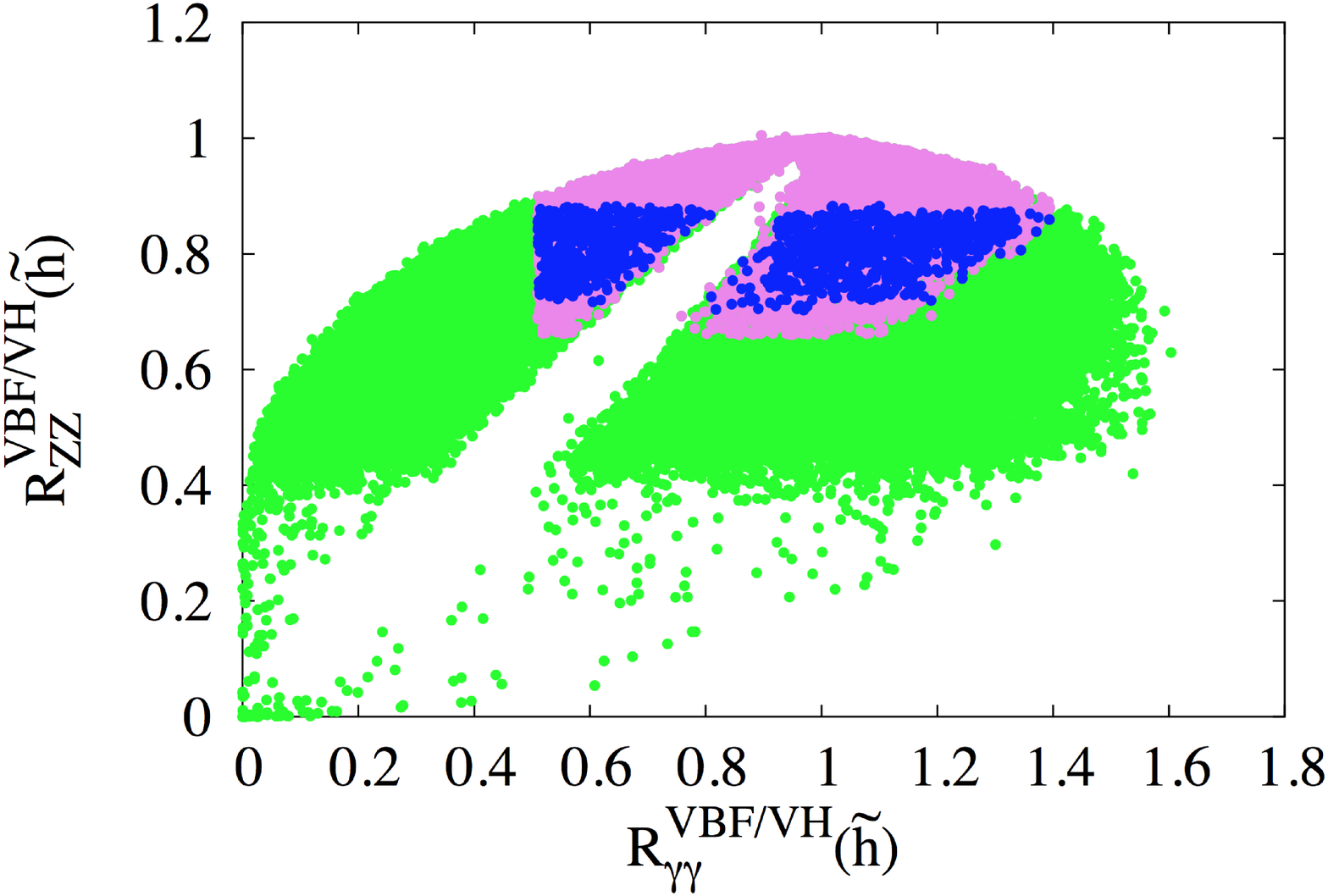}}
\put(210,150){\includegraphics[angle=0,width=0.45\textwidth]{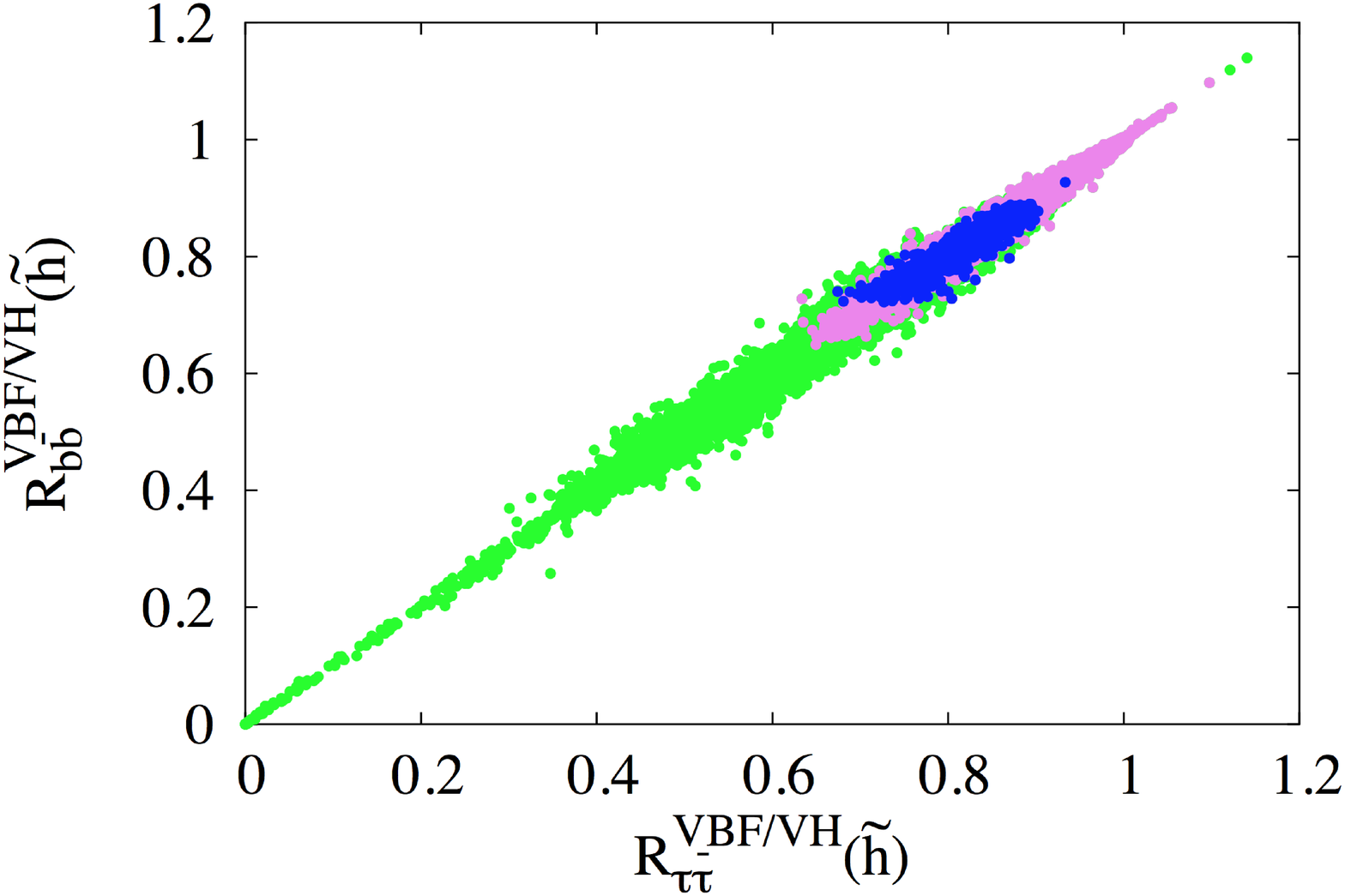}}
\put(0,0){\includegraphics[angle=0,width=0.45\textwidth]{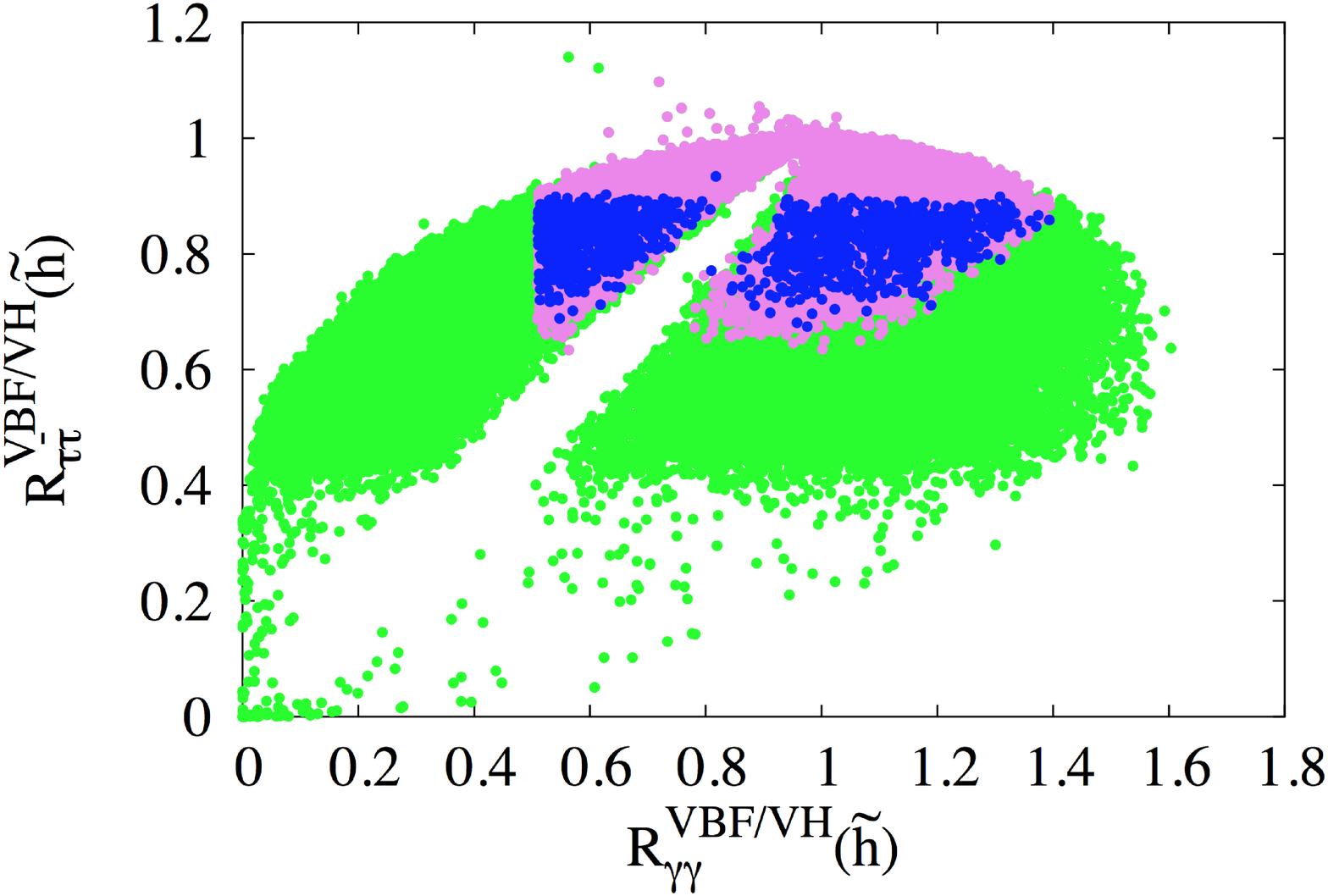}}
\put(210,0){\includegraphics[angle=0,width=0.45\textwidth]{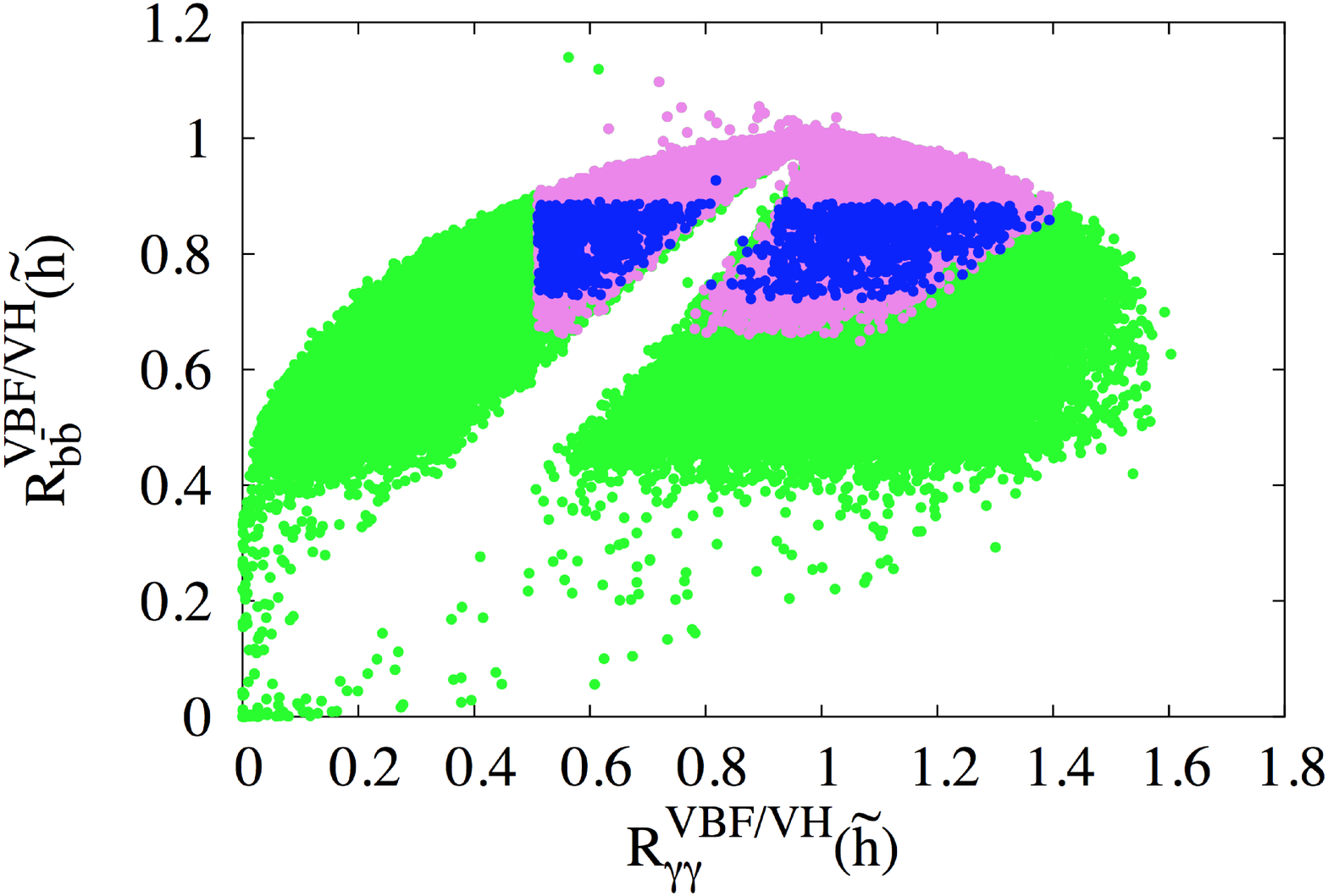}}
\end{picture}
\caption{\label{FIG11} Scatter plots in
  $R_{\gamma\gamma}^{VBF/VH}(\tilde{h})-R_{ZZ}^{VBF/VH}(\tilde{h})$ (upper
  left panel), 
  $R_{\tau\tilde{\tau}}^{VBF/VH}(\tilde{h})-R_{b\tilde{b}}^{VBF/VH}(\tilde{h})$
  (upper right panel), 
  $R_{\gamma\gamma}^{VBF/VH}(\tilde{h})-R_{\tau\tilde{\tau}}^{VBF/VH}(\tilde{h})$
  (lower left panel),
  $R_{\gamma\gamma}^{VBF/VH}(\tilde{h})-R_{b\tilde{b}}^{VBF/VH}(\tilde{h})$
  (lower right panel) planes for {\it Set 2}. The color
  notations are the same as in Fig.~\ref{mh_Rgg}. }
\end{figure}
\begin{figure}[htb!]
\begin{picture}(300,160)(0,160)
\put(0,150){\includegraphics[angle=0,width=0.45\textwidth]{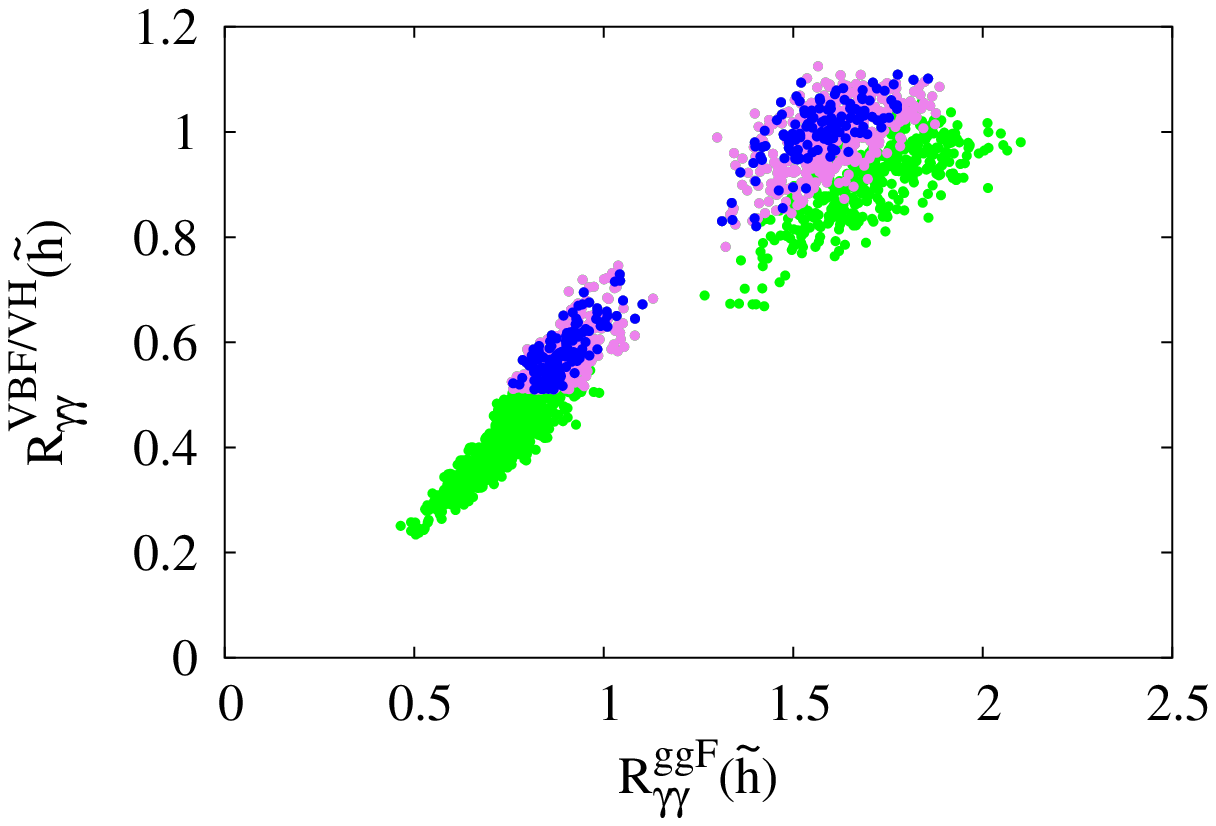}}
\put(230,150){\includegraphics[angle=0,width=0.45\textwidth]{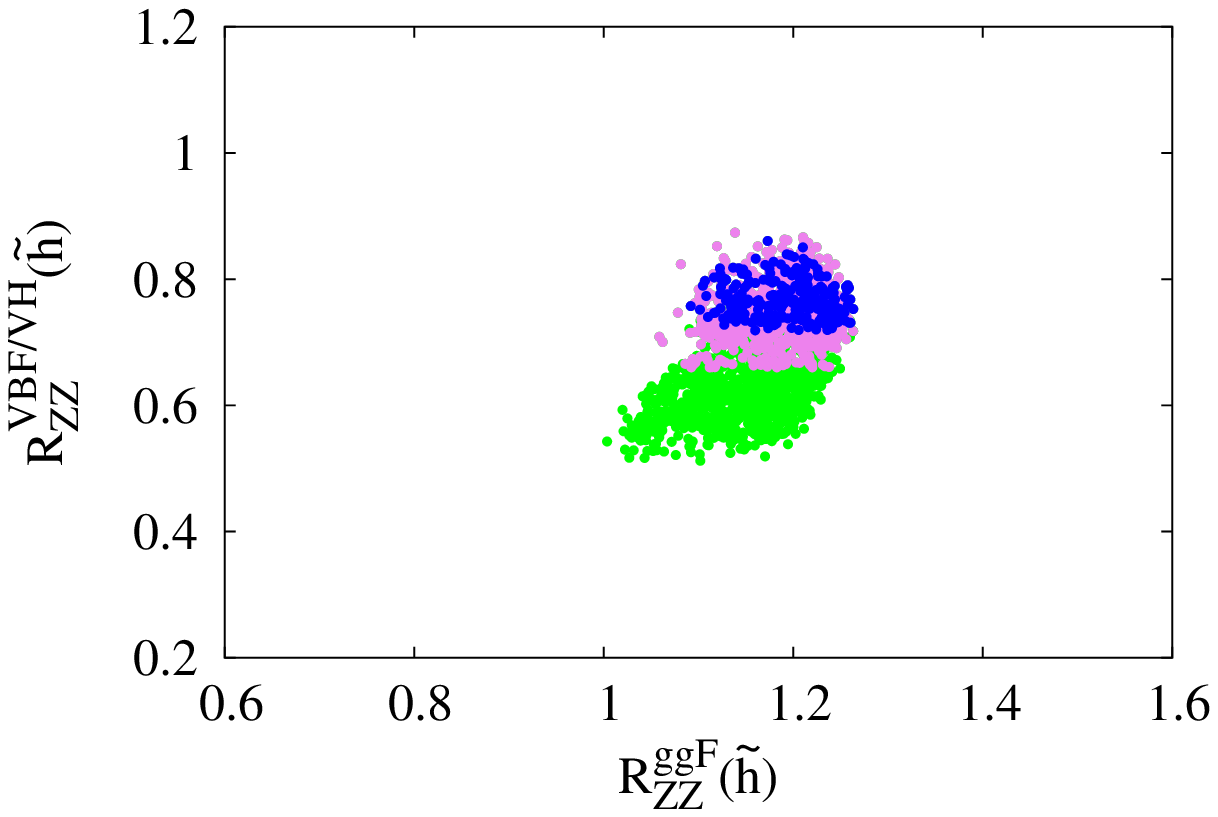}}
\end{picture}
\caption{\label{ggF_vs_VBF}  {Scatter plots in
    $R_{\gamma\gamma}^{ggF}(\tilde{h})-R_{\gamma\gamma}^{VBF/VH}(\tilde{h})$
    (left panel) and 
    $R_{ZZ}^{ggF}(\tilde{h})-R_{ZZ}^{VBF/VH}(\tilde{h})$ (right panel)
    planes for {\it Set 1}.  The color notations are the same as in
    Fig.~\ref{mh_Rgg}.}} 
\end{figure}
\begin{figure}[htb!]
\begin{picture}(300,160)(0,160)
\put(0,150){\includegraphics[angle=0,width=0.45\textwidth]{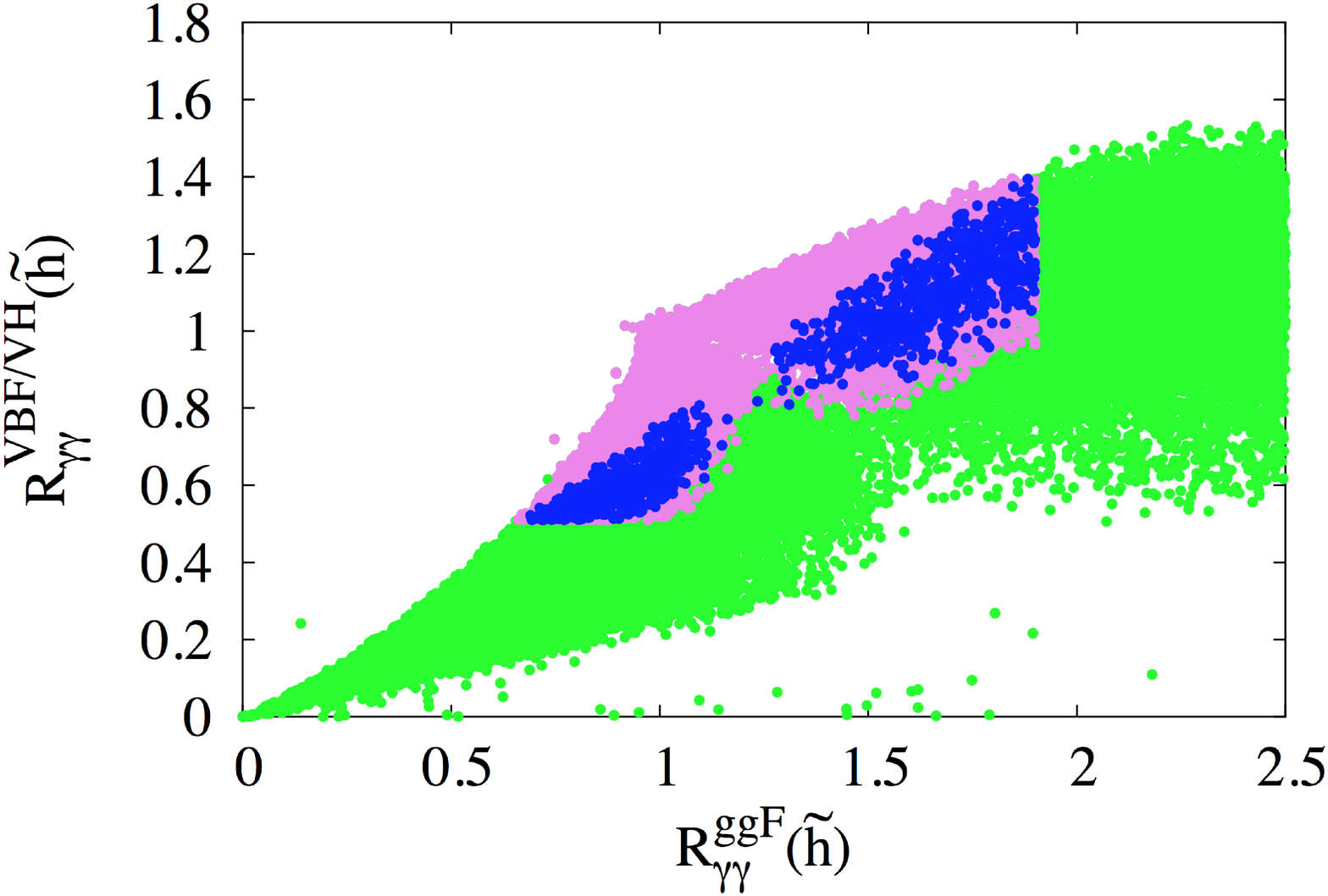}}
\put(230,150){\includegraphics[angle=0,width=0.45\textwidth]{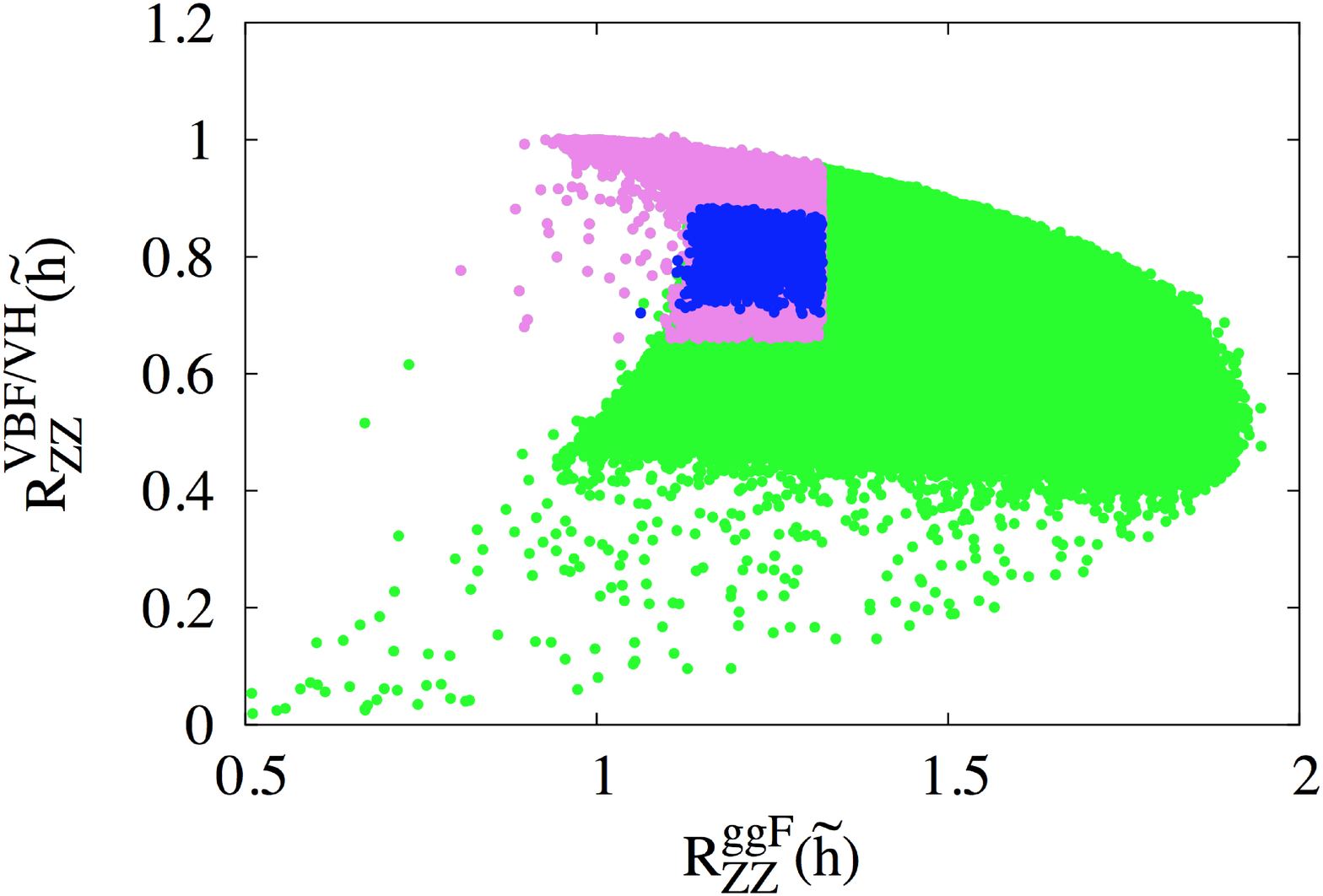}}
\end{picture}
\caption{\label{FIG12}  Scatter plots in
    $R_{\gamma\gamma}^{ggF}(\tilde{h})-R_{\gamma\gamma}^{VBF/VH}(\tilde{h})$
  (left panel),
    $R_{ZZ}^{ggF}(\tilde{h})-R_{ZZ}^{VBF/VH}(\tilde{h})$ (right panel)
  planes for {\it Set 2}.  The color notations are the same as in
  Fig.~\ref{mh_Rgg}.} 
\end{figure}

Similar plots for the case of $VBF$ and $VH$ production mechanisms
are presented in Fig.~\ref{mh_RVBF}
for {\it Set 1} and in Fig.~\ref{FIG11}
for {\it Set 2}. In this case the production cross section is
typically suppressed by the mixing as compared to the case of the SM
Higgs boson because  the contribution to the coupling with massive
vector bosons from sgoldstino is small as we discuss in
Section~\ref{sec:3_1}. Almost the same can be said about the couplings
to heavy fermions: tree level Higgs part of the couplings in
Eqs.~\eqref{htt}--\eqref{htau} are typically larger than sgoldstino 
contribution for the chosen values of parameters, in particular for
$\sqrt{F}=10$~TeV. Note that due to this 
reason we expect that the total width of the Higgs-like resonance
is suppressed by factor $\cos^2{\theta}$ with respect the SM
Higgs boson decay width. Summarizing, in Fig.~\ref{mh_RVBF} and
Fig.~\ref{FIG11} the Higgs signal strengths for fermion and massive
vector boson channels in $VBF/VH$ for most of the models become
suppressed due to the mixing with sgoldstino, in particular, for
models in which sgoldstino explains 98~GeV LEP excess. Also we show
correlations between different production mechanisms, $ggF$ and
$VBF/VH$, for $\gamma\gamma$ and $ZZ$ channels in
Fig.~\ref{ggF_vs_VBF}  
for {\it Set 1} and in Fig.~\ref{FIG12}
for {\it Set 2}.
Again for $\gamma\gamma$ we see two domains corresponding to different
signs of $M_{1,2}$.

The general conclusion from the above discussions is that mixing of
the lightest Higgs boson with a lighter sgoldstino results in an
increase of signal strengths of fermionic and massive vector boson
channels in $ggF$ and in a decrease of their values in $VBF/VH$ 
production mode and an increase in $ZZ$ channel. We do not show here
the signal strength for $WW$ channel because it is almost the
same as for $ZZ$.
Additionally, requirement that the scalar sgoldstino
explains LEP excess results in prediction of particular regions of $R$
where their values deviate from unity.  So an increase of accuracy of
measurements of the signal strength for observed Higgs-like resonance
which is expected with next LHC runs will give an opportunity to check
this scenario.

Now we turn to the discussion of sgoldstino collider phenomenology with
presented setup. It has been previously studied in 
Refs.~\cite{Gorbunov:2000ht,Perazzi:2000id,Perazzi:2000ty,Gorbunov:2002er,Demidov:2004qt}
but without including effects of its possible mixing with the Higgs
boson. As we find this mixing can be extremely important. 
Firstly, let us discuss the main decay channels and the hierarchy
between their branchings for sgoldstinos with masses at
electroweak scale. 
\begin{figure}[htb!]
\begin{picture}(300,300)(0,0)
\put(210,0){\includegraphics[angle=0,width=0.45\textwidth]{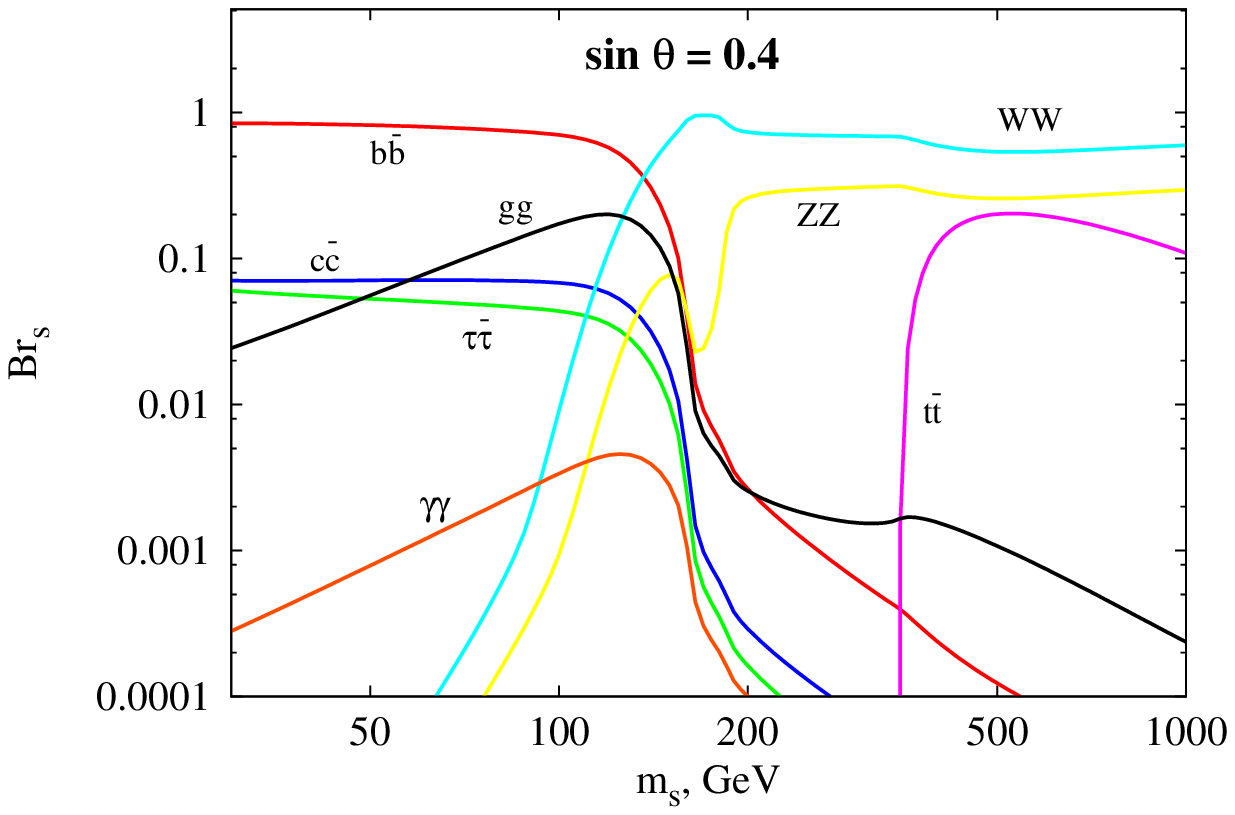}}
\put(210,150){\includegraphics[angle=0,width=0.45\textwidth]{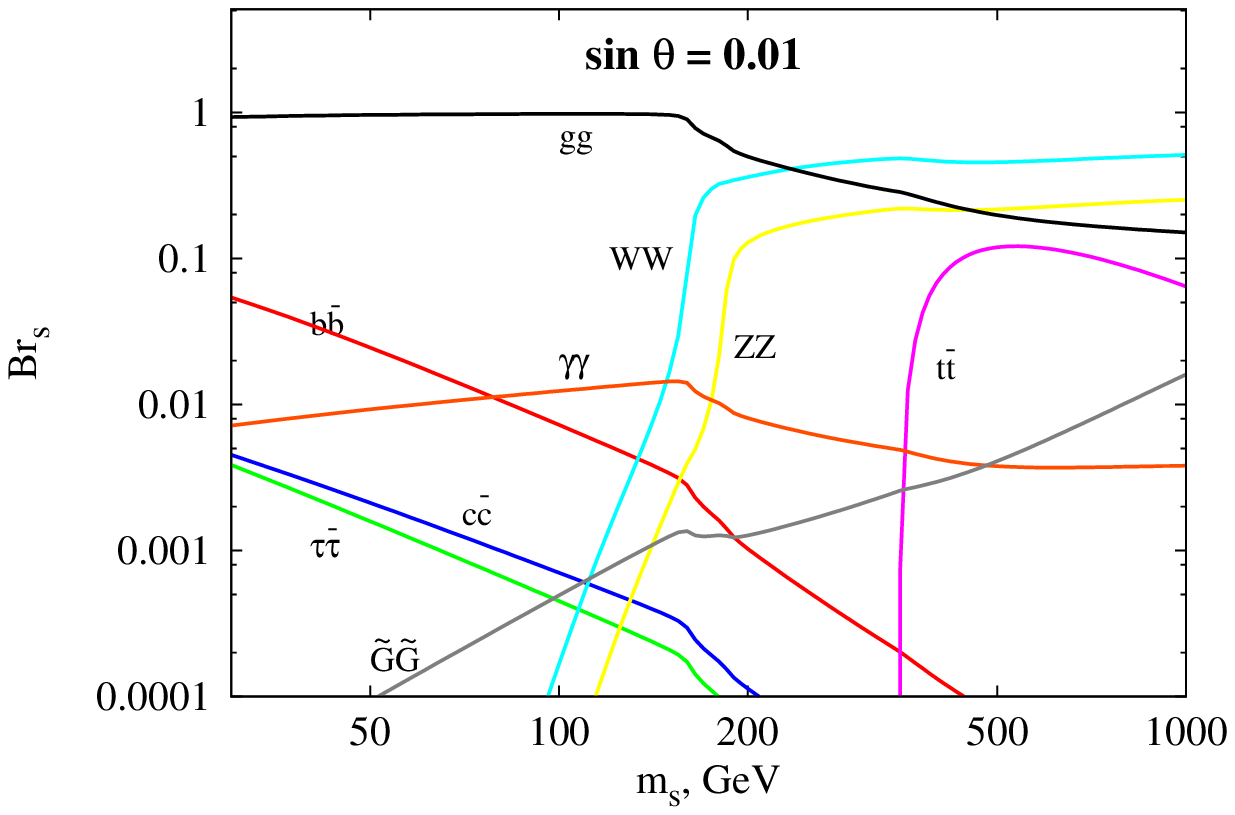}}
\put(0,0){\includegraphics[angle=0,width=0.45\textwidth]{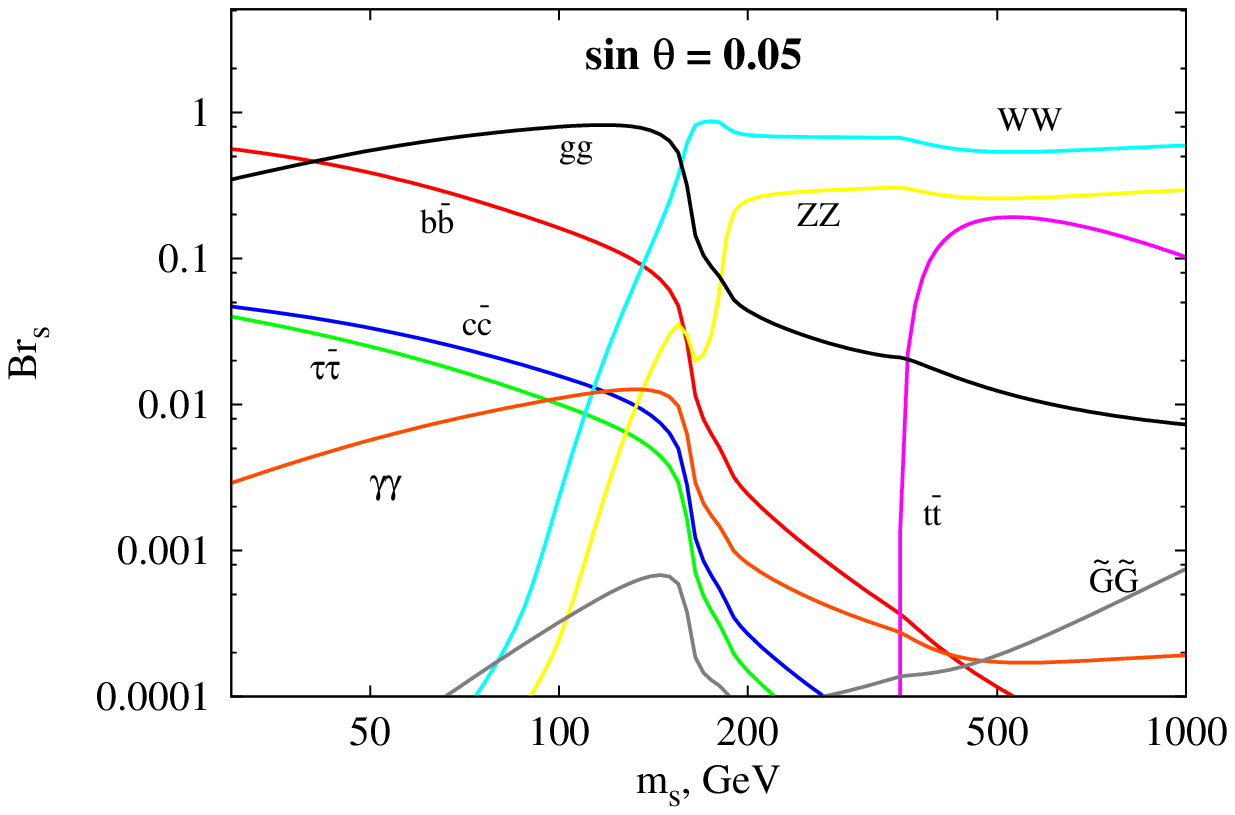}}
\put(0,150){\includegraphics[angle=0,width=0.45\textwidth]{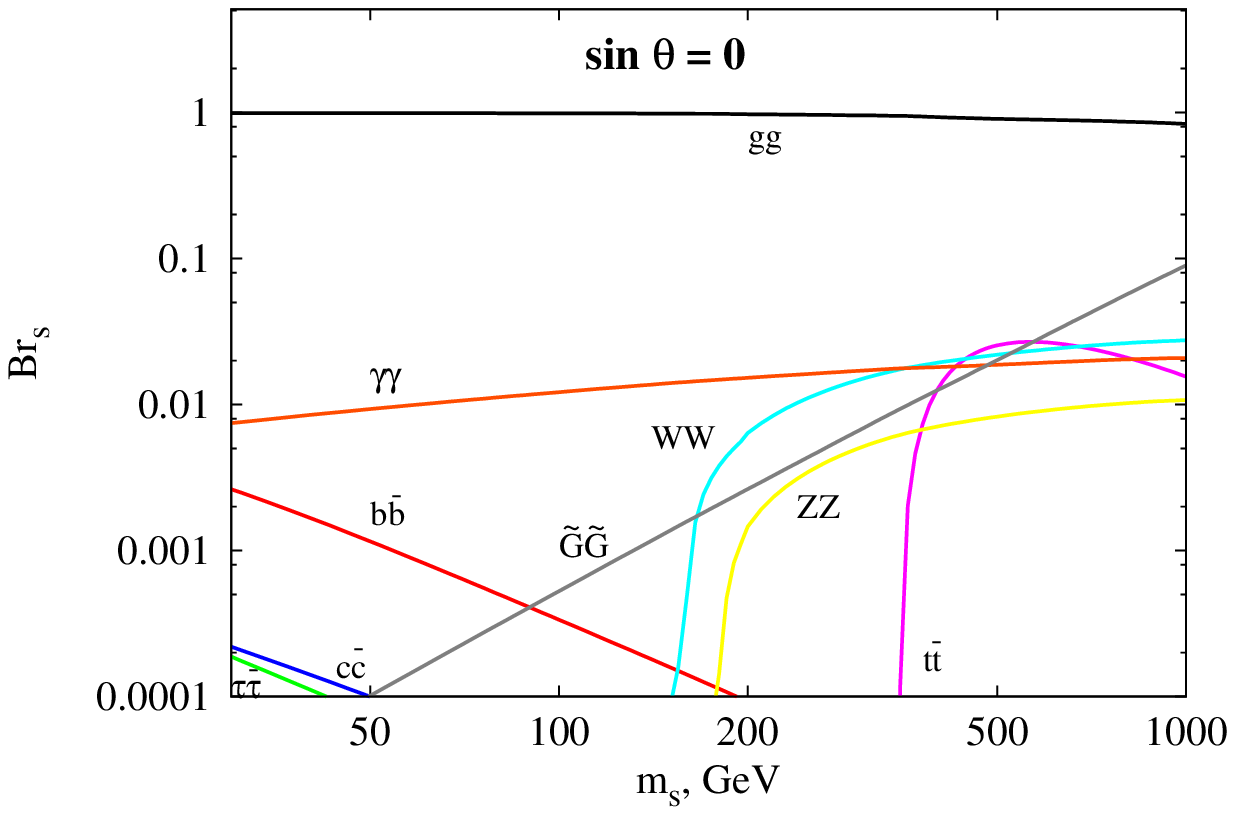}}
\end{picture}
\caption{\label{branchings} {Modification of scalar sgoldstino
    branching ratios at different values of mixing angle:
    $\sin{\theta}=0.0, 0.01, 0.05$ and 0.4. We take the following
    values for MSSM soft parameters: $\sqrt{F}=10$~TeV,
    $M_{1}=400$~GeV, $M_{2}=800$~GeV, $M_{3}=-1200$~GeV,
    $A^{U,D,E}=700$~GeV and $A^{U,D,E}_{ab} = Y^{U,D,E}_{ab}A^{U,D,E}$
    where $Y^{U,D,E}_{ab}$ are MSSM Yukawa couplings.}} 
\end{figure}
In general the interactions of scalar sgoldstino with SM particles are
similar to those of the lightest Higgs boson but the hierarchy between
the coupling constants is quite different. The main distinction is the
fact that sgoldstino couplings to gluons and photons appear already at
tree level as it have been discussed in Section~\ref{sec:2_2}. That's
why, for typical values of soft MSSM parameters pure sgoldstino with
mass around hundred GeV dominantly decays into pair of gluons and
photons which is governed by parameters $M_3$ and $M_{\gamma\gamma}$,
respectively. Then it can decay into pairs of quarks and leptons 
and corresponding decay rates are governed by corresponding trilinear
soft 
terms which enter interactions for superpartners of these quarks and
leptons in~\eqref{Sgold_lagr}. Also sgoldstinos can decay into
pair of gauge bosons and these decay widths are governed by
corresponding soft gaugino masses. And finally sgoldstinos can decay
into pair of gravitinos, which looks as invisible decay. 
The hierarchy of the branching ratios depends on hierarchy of the soft
terms in MSSM lagrangian. In Fig.~\ref{branchings}  
we show how the hierarchy of branching ratios for scalar sgoldstino
decays changes depending on mixing angle.  Again we set
$\sqrt{F}=10$~TeV and for the time being we consider here very wide
interval of sgoldstino masses. We see that even small value of mixing
angle drastically changes the hierarchy between possible decay
channels and already at mixing angle 
of 0.4 the hierarchy becomes very similar to the case of the Higgs,
except for the partial widths are now suppressed by square of sine of
mixing angle. This fact can considerably change the strategy of
sgoldstino searches at colliders~\cite{Gorbunov:2000ht,Perazzi:2000id,Perazzi:2000ty,Gorbunov:2002er,Abreu:2000ij}. 

\begin{figure}[htb!]
\begin{picture}(300,300)(0,0)
\put(210,0){\includegraphics[angle=0,width=0.45\textwidth]{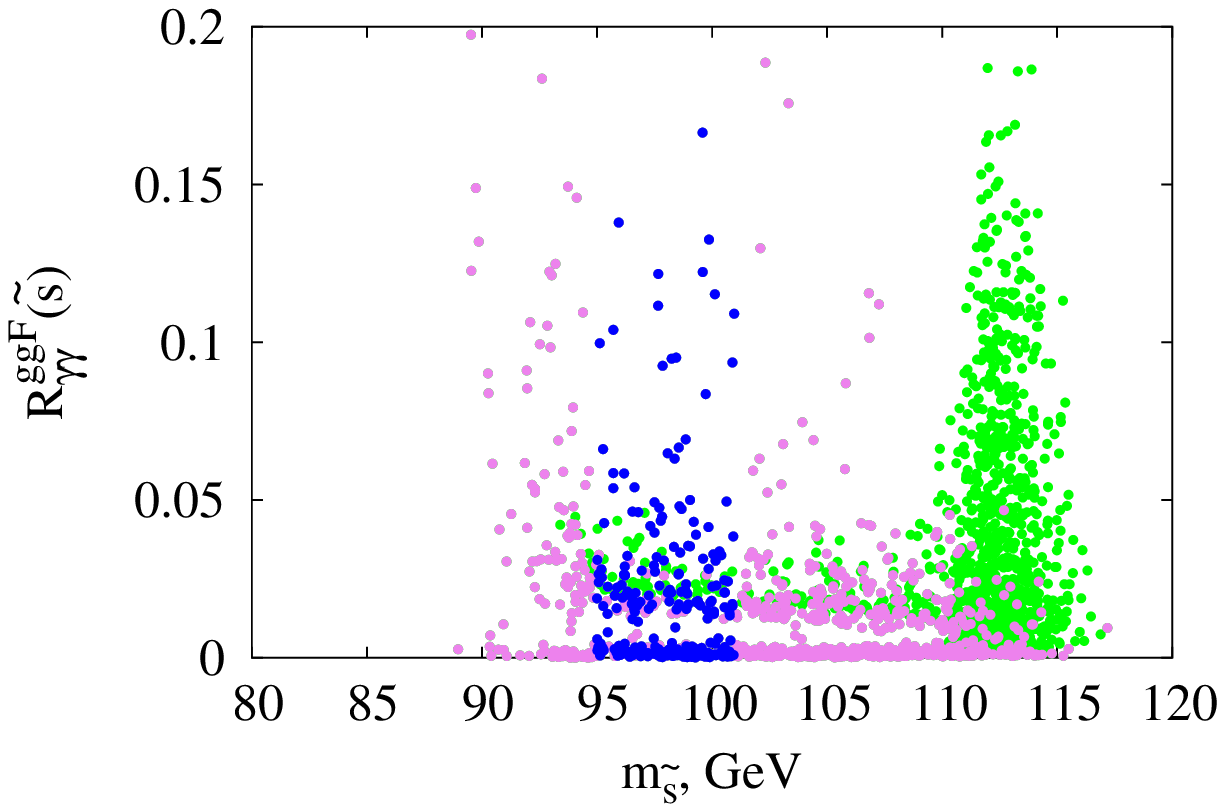}}
\put(210,150){\includegraphics[angle=0,width=0.45\textwidth]{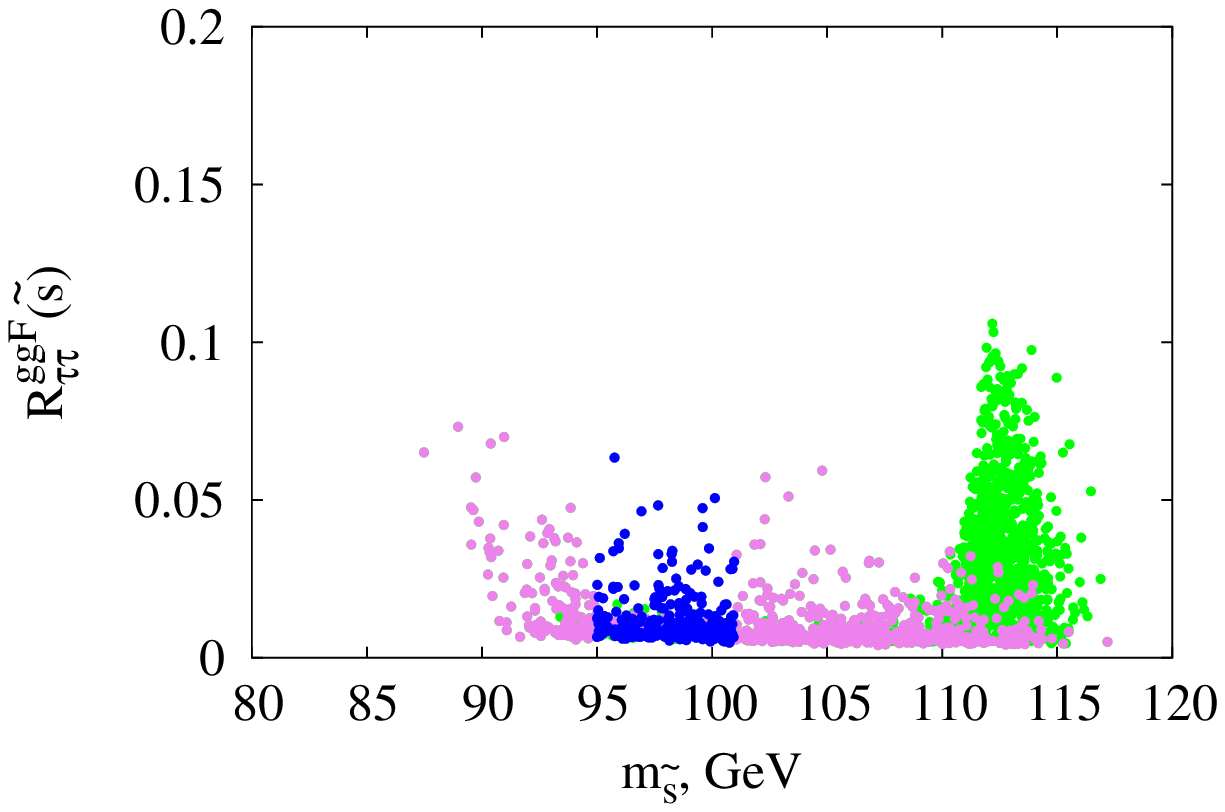}}
\put(0,0){\includegraphics[angle=0,width=0.45\textwidth]{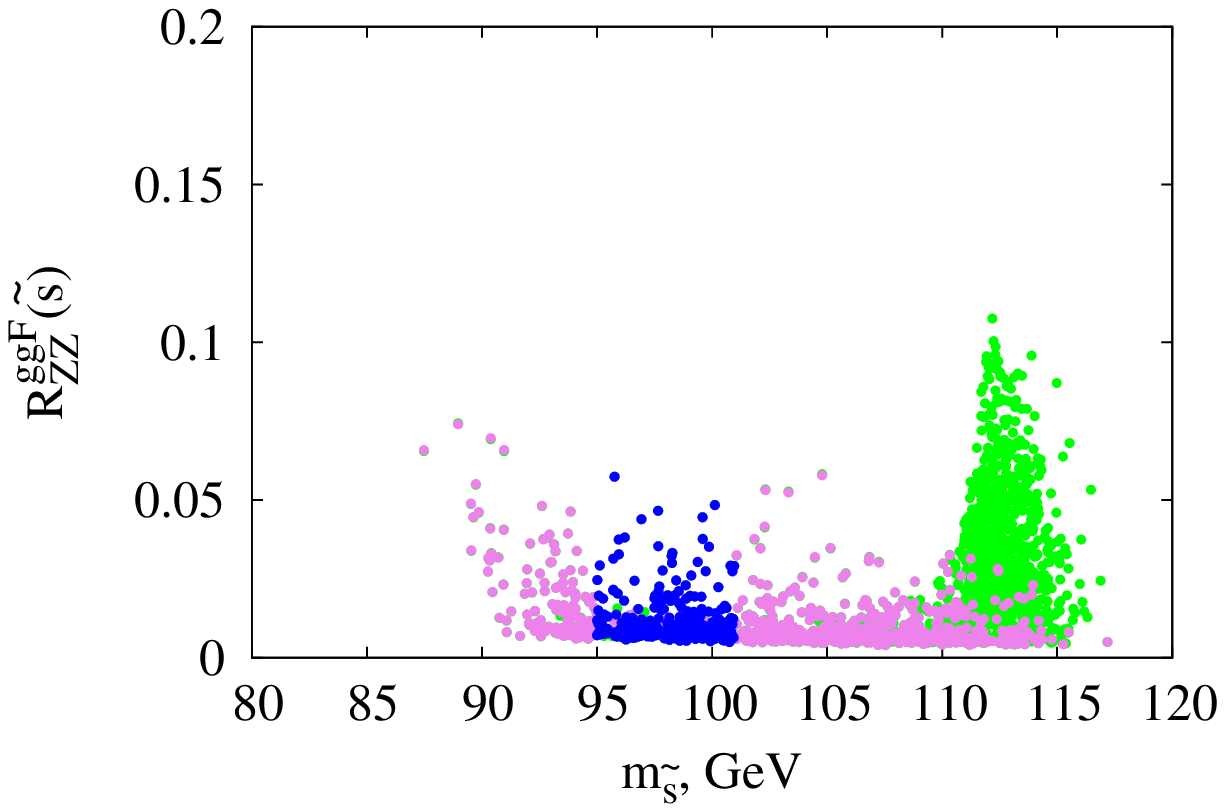}}
\put(0,150){\includegraphics[angle=0,width=0.45\textwidth]{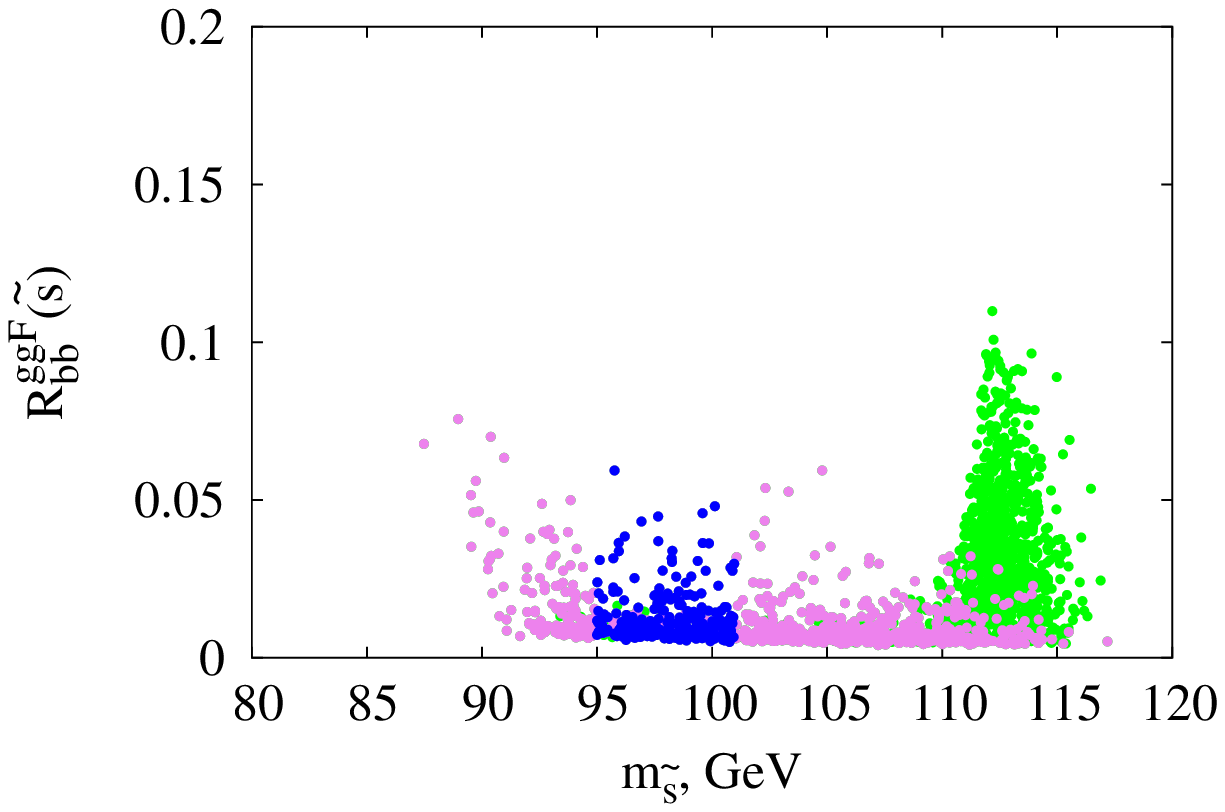}}
\end{picture}
\caption{\label{ms_R1} {Scatter plots in
    $m_{\tilde{s}}-R_{b\tilde{b}}^{ggF}(\tilde{s})$ (upper  left panel),
    $m_{\tilde{s}}-R_{\tau\tilde{\tau}}^{ggF}(\tilde{s})$ (upper 
      right panel),
    $m_{\tilde{s}}-R_{ZZ}^{ggF}(\tilde{s})$ (lower left panel),
    $m_{\tilde{s}}-R_{\gamma\gamma}^{ggF}(\tilde{s})$ (lower
      right panel)
    planes for {\it Set 1}. The color notations are the same as in
    Fig.~\ref{mh_Rgg}.}}  
\end{figure}
\begin{figure}[htb!]
\begin{picture}(300,300)(0,0)
\put(210,0){\includegraphics[angle=0,width=0.45\textwidth]{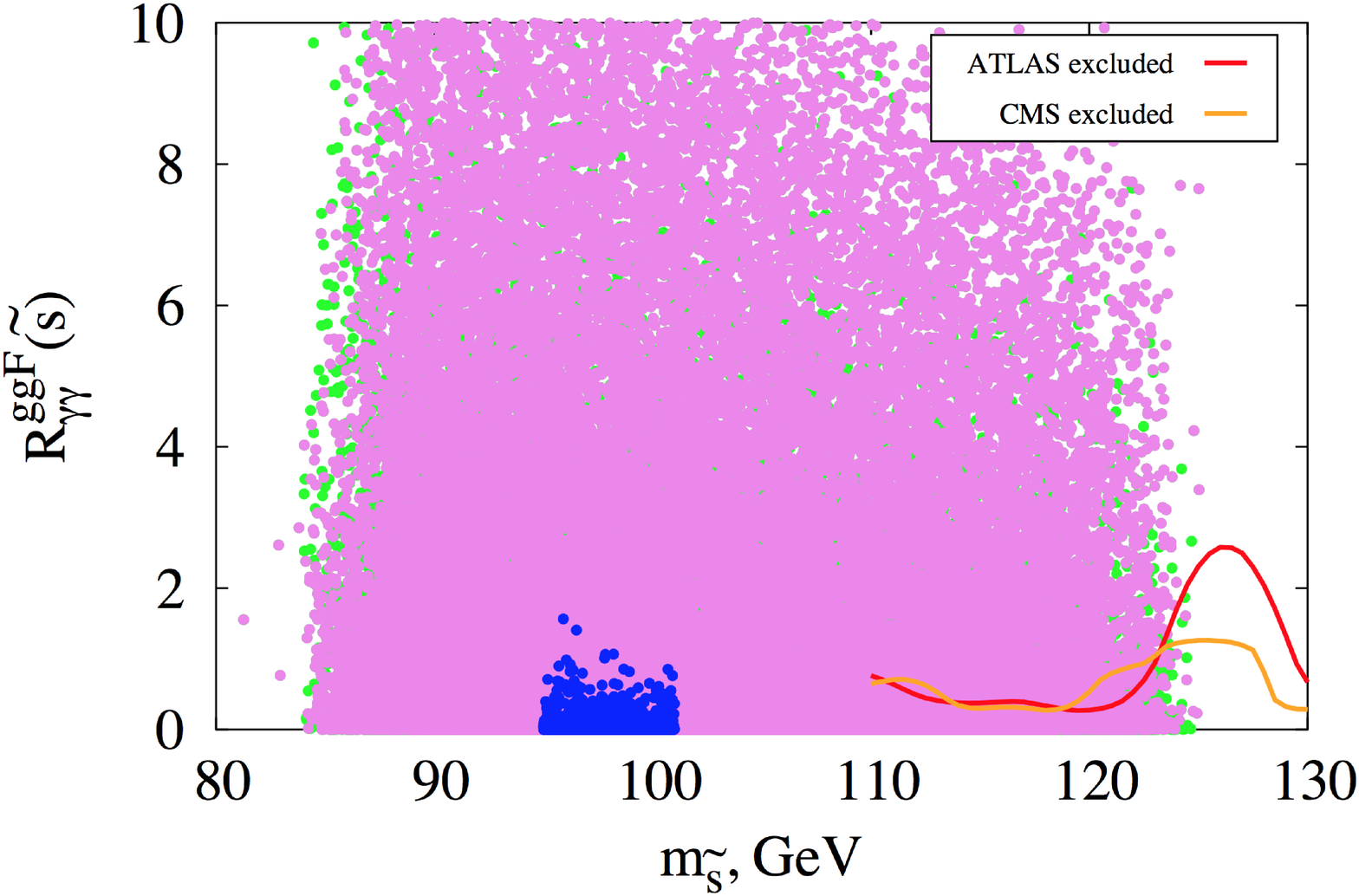}}
\put(210,150){\includegraphics[angle=0,width=0.45\textwidth]{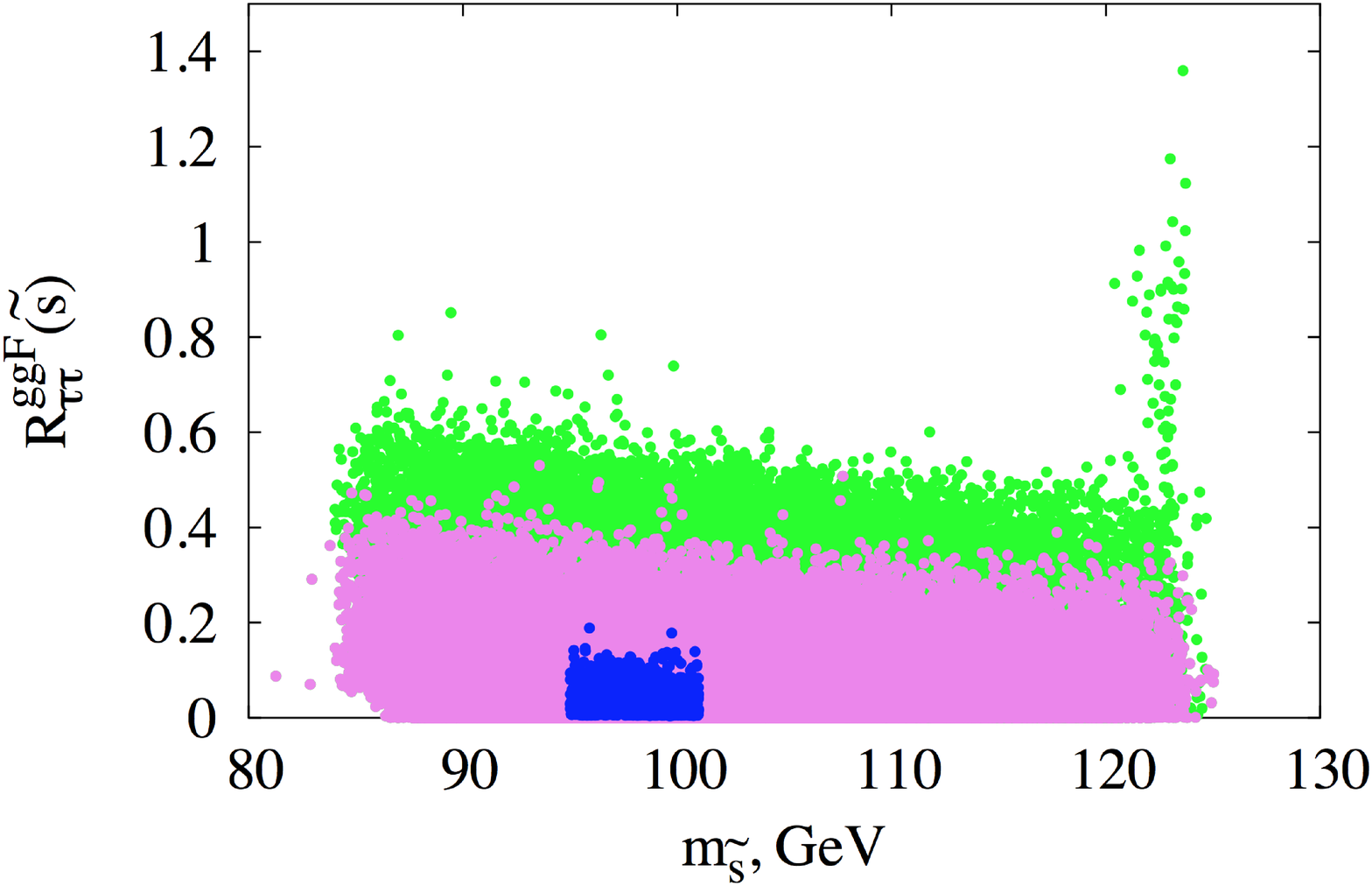}}
\put(0,0){\includegraphics[angle=0,width=0.45\textwidth]{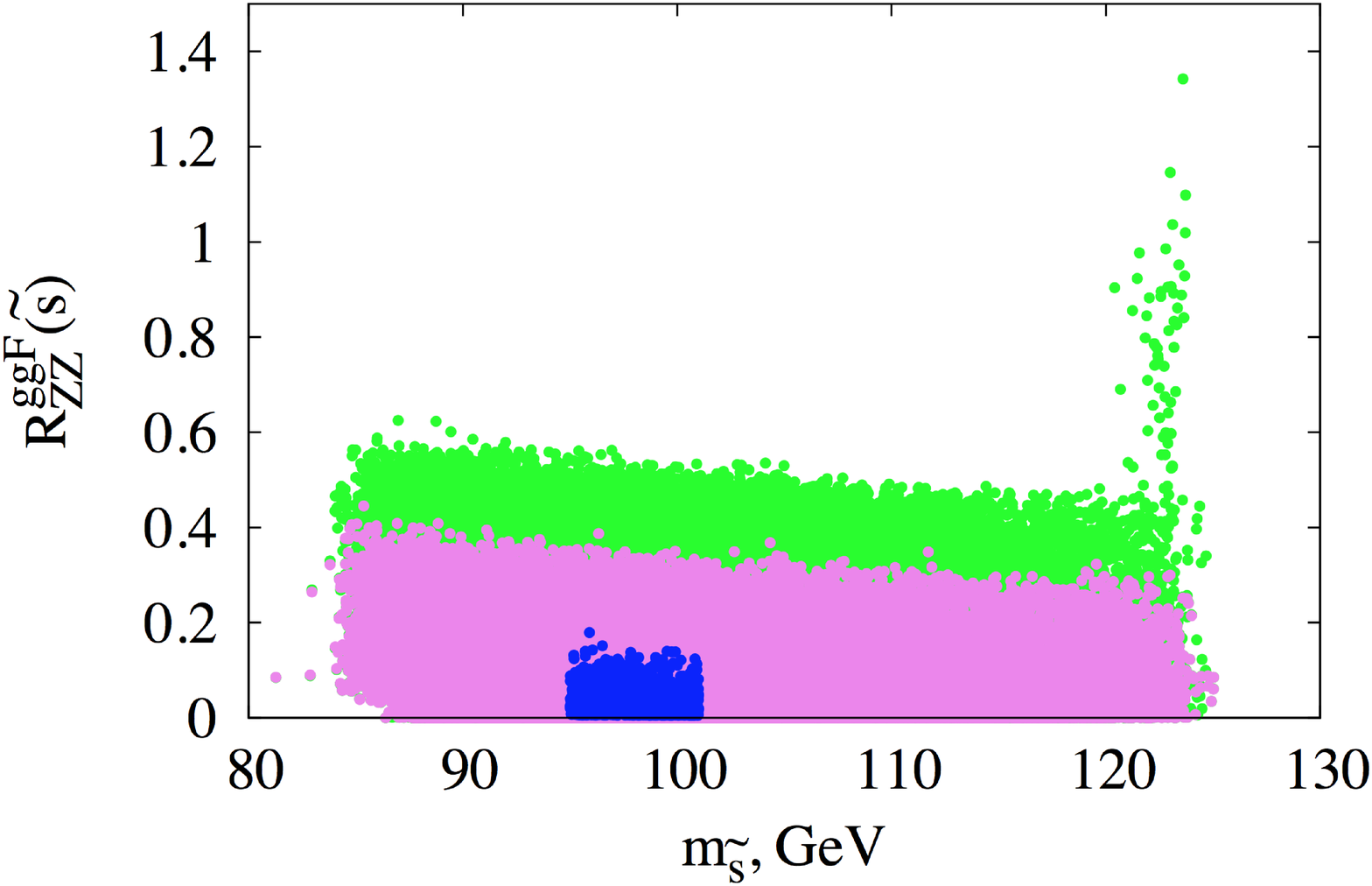}}
\put(0,150){\includegraphics[angle=0,width=0.45\textwidth]{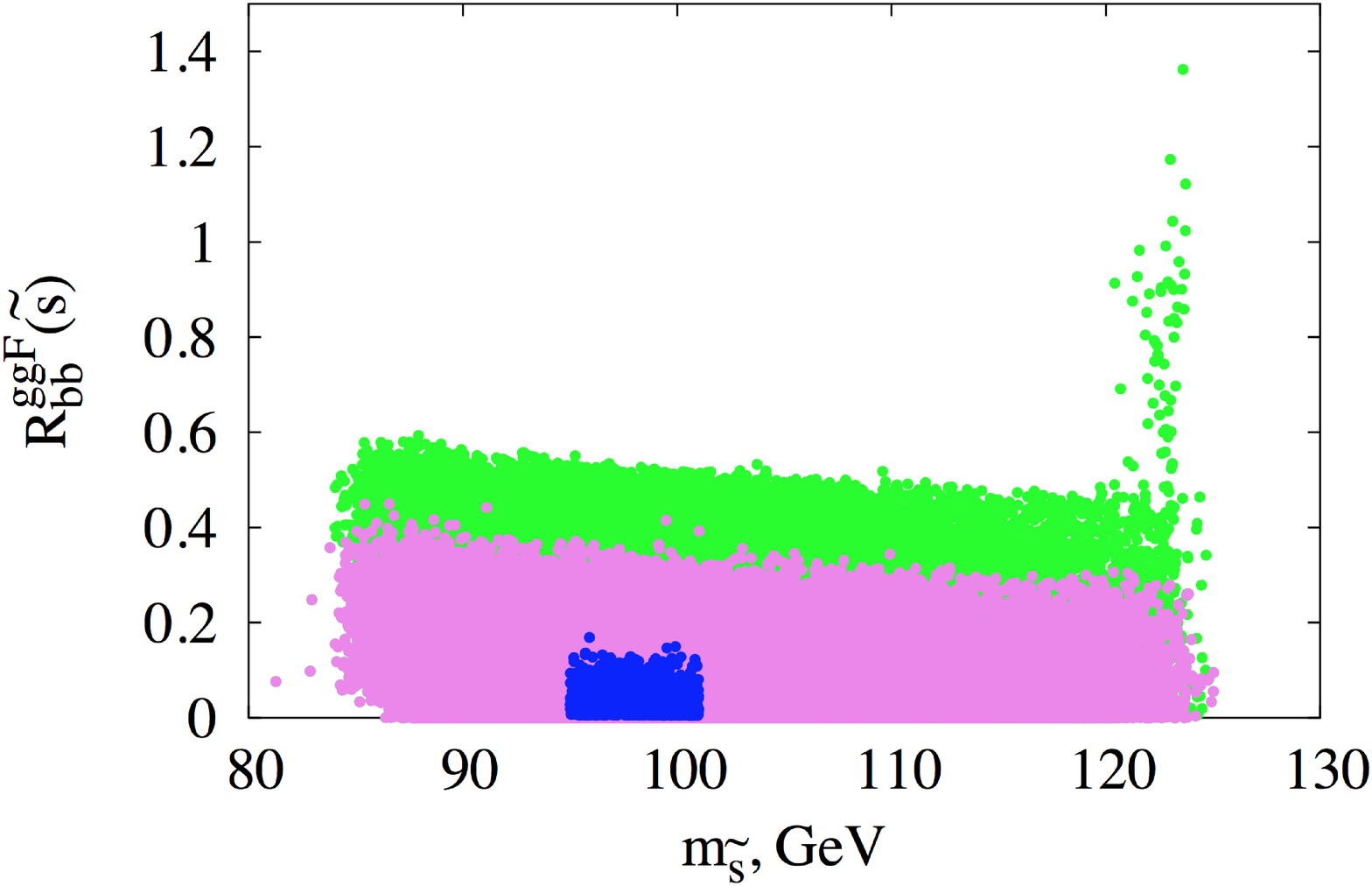}}
\end{picture}
\caption{\label{FIG13} Scatter plots in
  $m_{\tilde{s}}-R_{b\tilde{b}}^{ggF}(\tilde{s})$ (upper  left panel), 
  $m_{\tilde{s}}-R_{\tau\tilde{\tau}}^{ggF}(\tilde{s})$ (upper 
    right panel),
  $m_{\tilde{s}}-R_{ZZ}^{ggF}(\tilde{s})$ (lower  left panel), 
  $m_{\tilde{s}}-R_{\gamma\gamma}^{ggF}(\tilde{s})$ (lower  right
  panel)
  planes for {\it Set 2}.  The color notations are the same as in
  Fig.~\ref{mh_Rgg}.} 
\end{figure}

Now we return to the light sgoldstino-like state in our scenario and 
we show the signal strengths of $\tilde{s}$ for $b\tilde{b},
\tau\tilde{\tau}, \gamma\gamma$ and $ZZ$ channels in gluon-gluon
fusion production process in Fig.~\ref{ms_R1}
for {\it Set 1} and in Fig.~\ref{FIG13}
for {\it Set 2}.
We see that for $ggF$ production the sgoldstino signal strength does
not exceed 0.1 for fermionic and $ZZ$ ($WW$) channels for {\it Set 1}
and is less than $0.4-0.5$ for {\it Set 2}. While in the $\gamma\gamma$
channel $R_{\gamma\gamma}^{ggF}(\tilde{s})$ can reach values about 0.2
for {\it Set 1}  and can be quite large for some models in {\it Set
  2}. In the last case we can use results of the CMS~\cite{CMS:ril}
and ALTAS~\cite{ATLAS:2012znl} searches for Higgs boson in
$\gamma\gamma$ channel and put additional constraints on
$R_f^{ggF}(\tilde{s})$. They are shown in lower right panel in 
Fig.~\ref{FIG13} where all the models above red and orange curves
are excluded. Other searches for the Higgs boson made by LHC and
TeVatron~\cite{Abazov:2013gmz} experiments put limits which do not
introduce additional constraints.

\begin{figure}[htb!]
\begin{picture}(300,300)(0,0)
\put(210,0){\includegraphics[angle=0,width=0.45\textwidth]{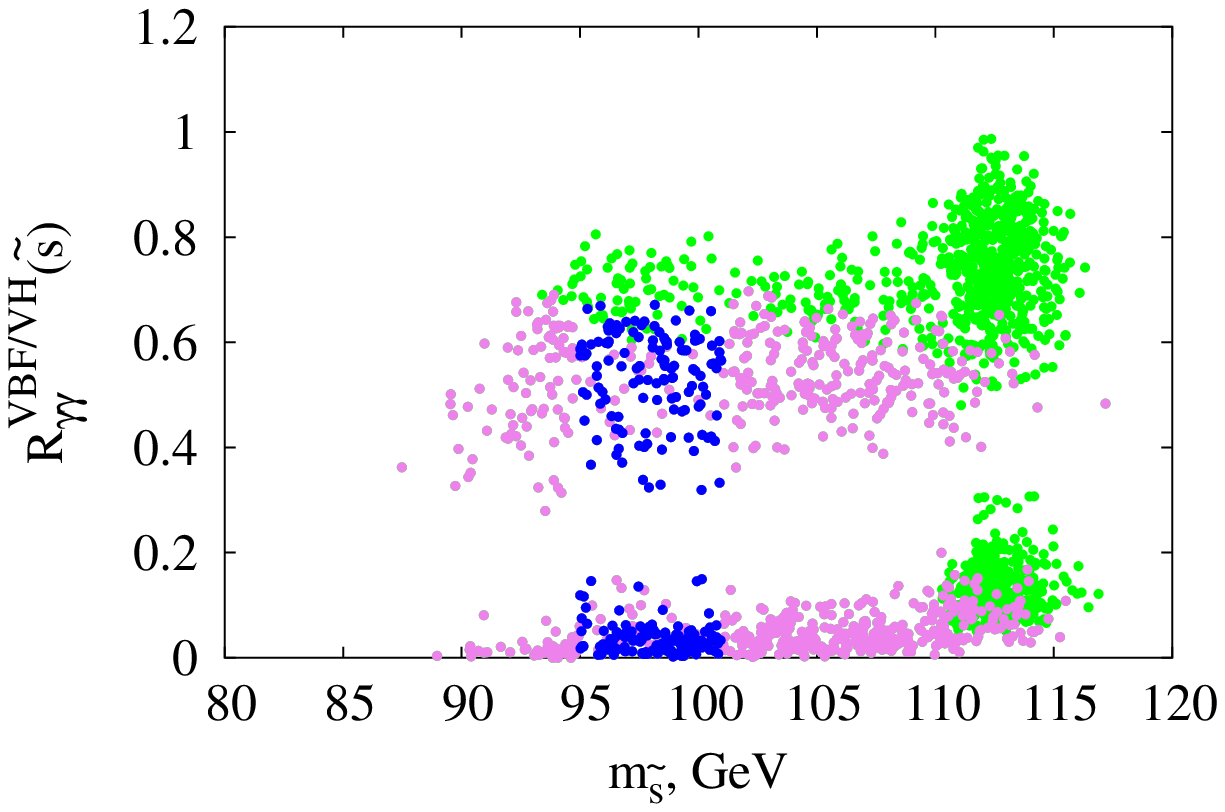}}
\put(210,150){\includegraphics[angle=0,width=0.45\textwidth]{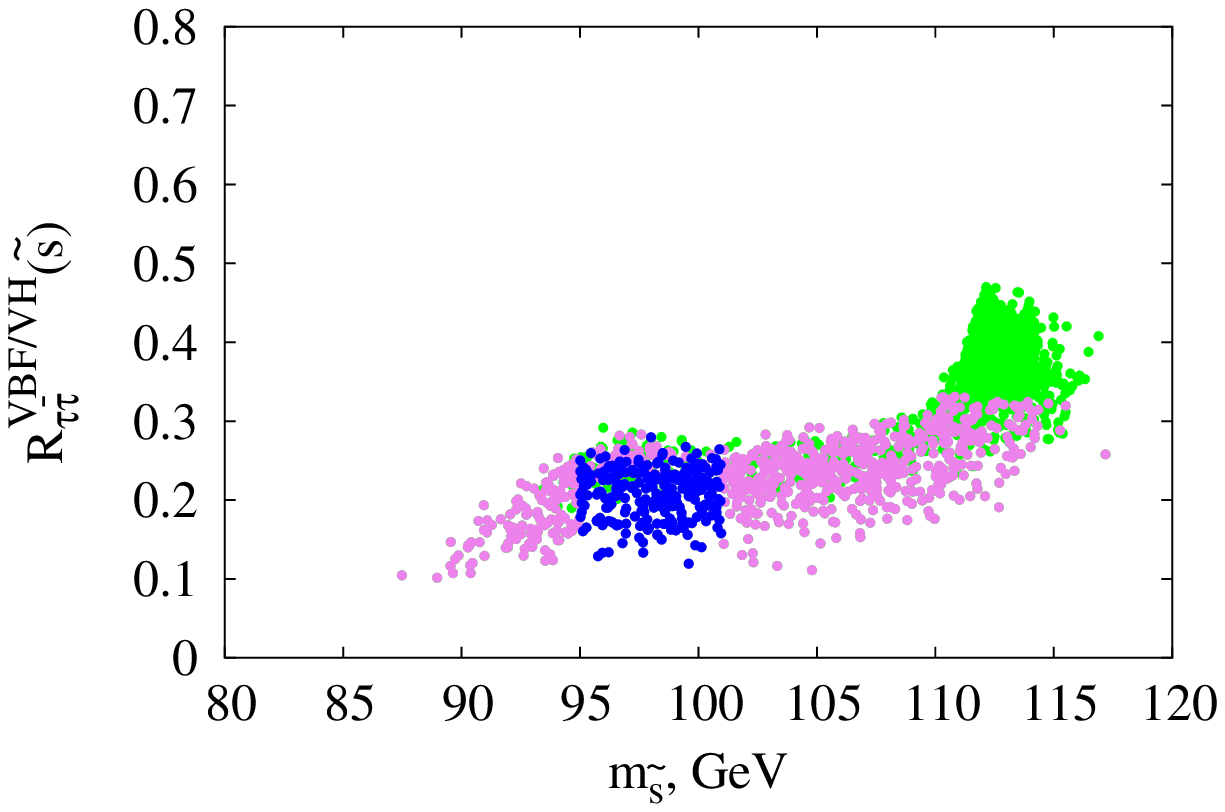}}
\put(0,0){\includegraphics[angle=0,width=0.45\textwidth]{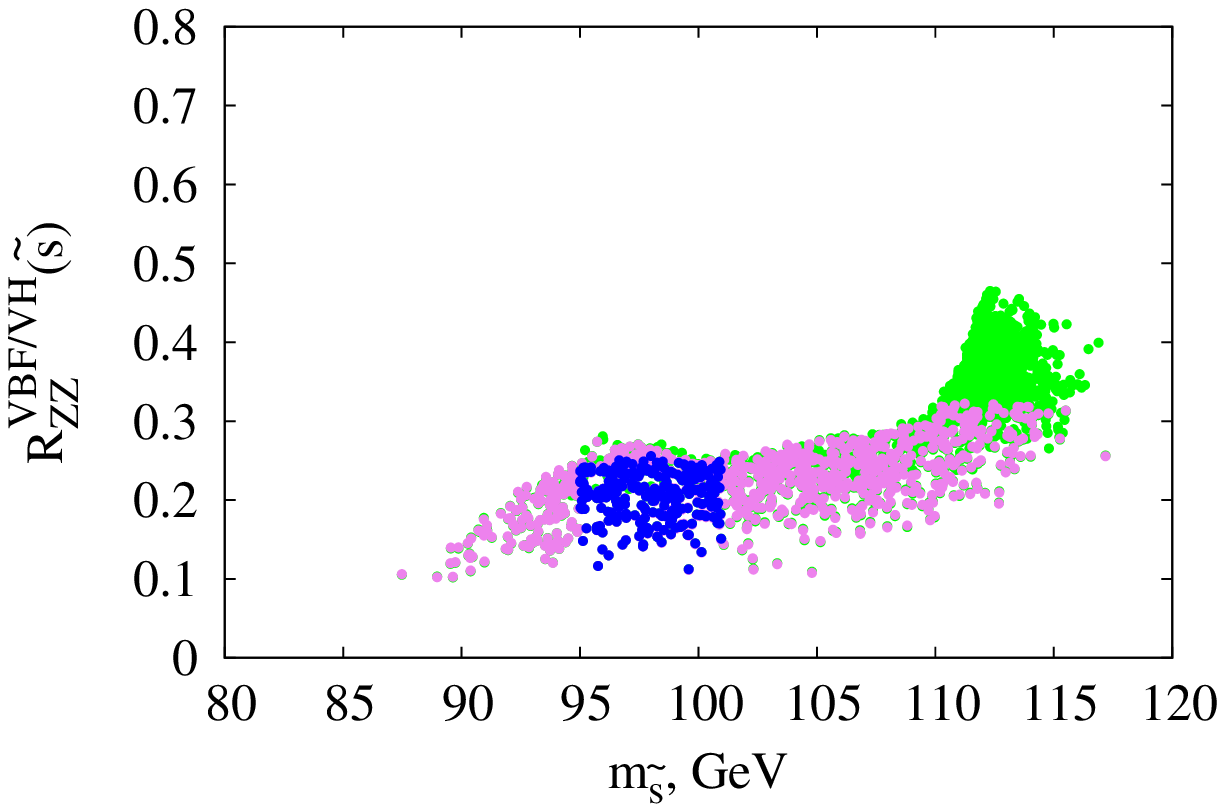}}
\put(0,150){\includegraphics[angle=0,width=0.45\textwidth]{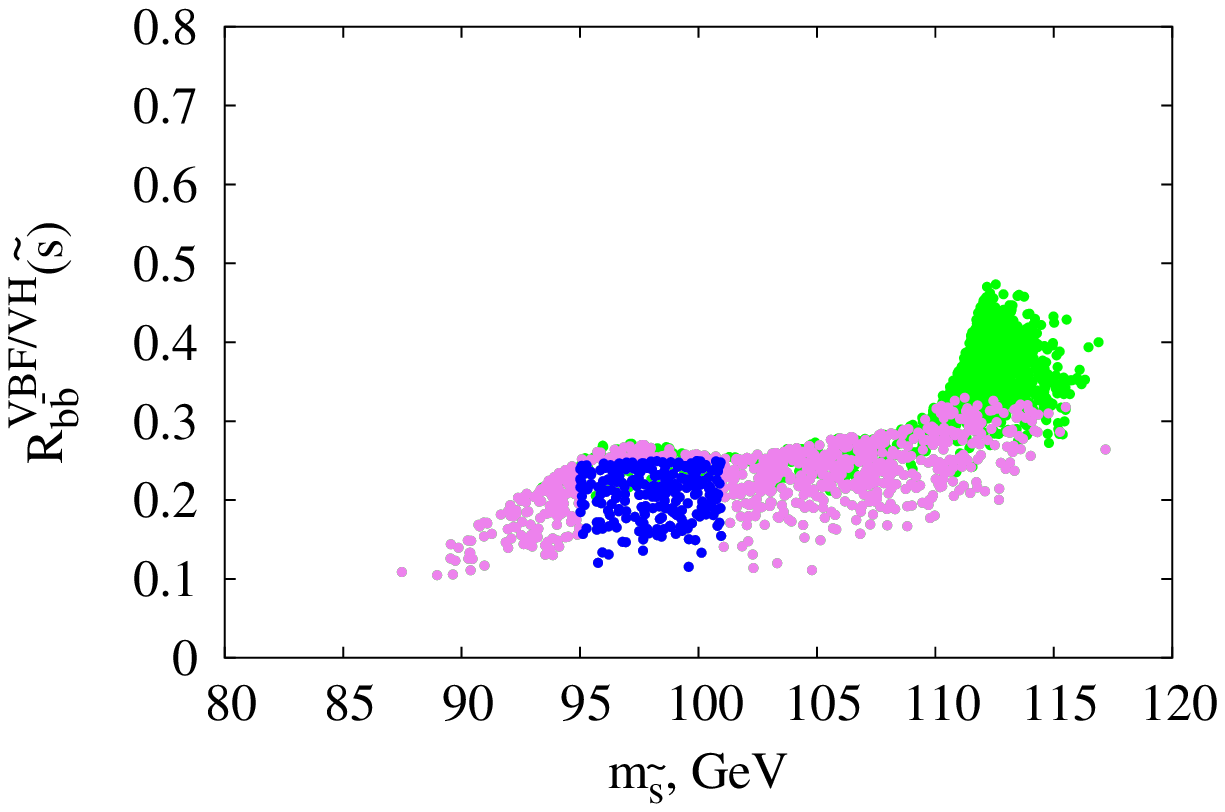}}
\end{picture}
\caption{\label{ms_R2} {Scatter plots in
    $m_{\tilde{s}}-R_{b\tilde{b}}^{VBF/VH}(\tilde{s})$ (upper 
      left panel), 
    $m_{\tilde{s}}-R_{\tau\tilde{\tau}}^{VBF/VH}(\tilde{s})$ (upper
     right panel),
    $m_{\tilde{s}}-R_{WW}^{VBF/VH}(\tilde{s})$ (lower  left panel),
    $m_{\tilde{s}}-R_{\gamma\gamma}^{VBF/VH}(\tilde{s})$ (lower 
      right panel)
    planes for {\it Set 1}.  The color notations are the same as in 
    Fig.~\ref{mh_Rgg}.}}  
\end{figure}

\begin{figure}[htb!]
\begin{picture}(300,300)(0,0)
\put(210,0){\includegraphics[angle=0,width=0.45\textwidth]{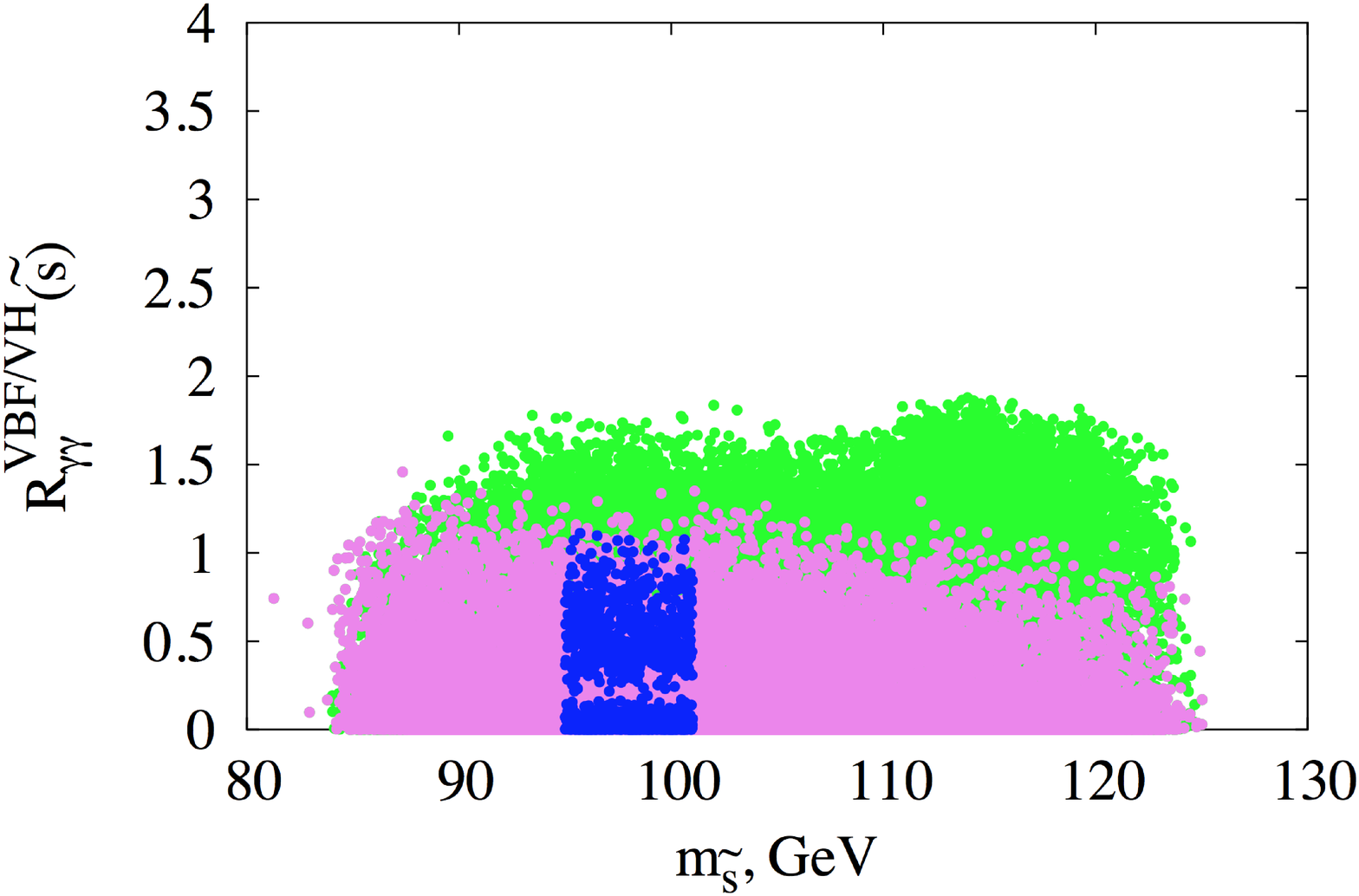}}
\put(210,150){\includegraphics[angle=0,width=0.45\textwidth]{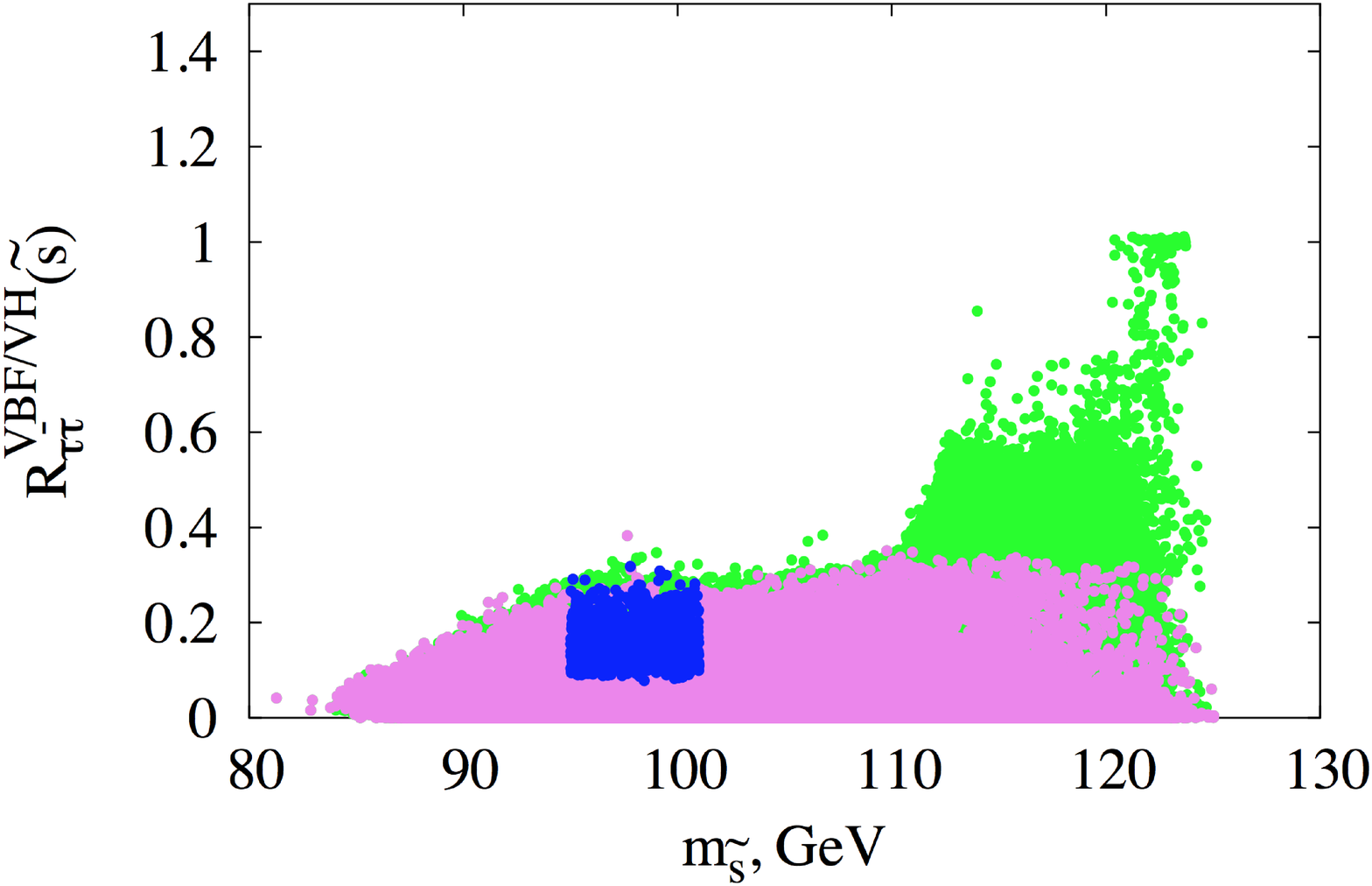}}
\put(0,0){\includegraphics[angle=0,width=0.45\textwidth]{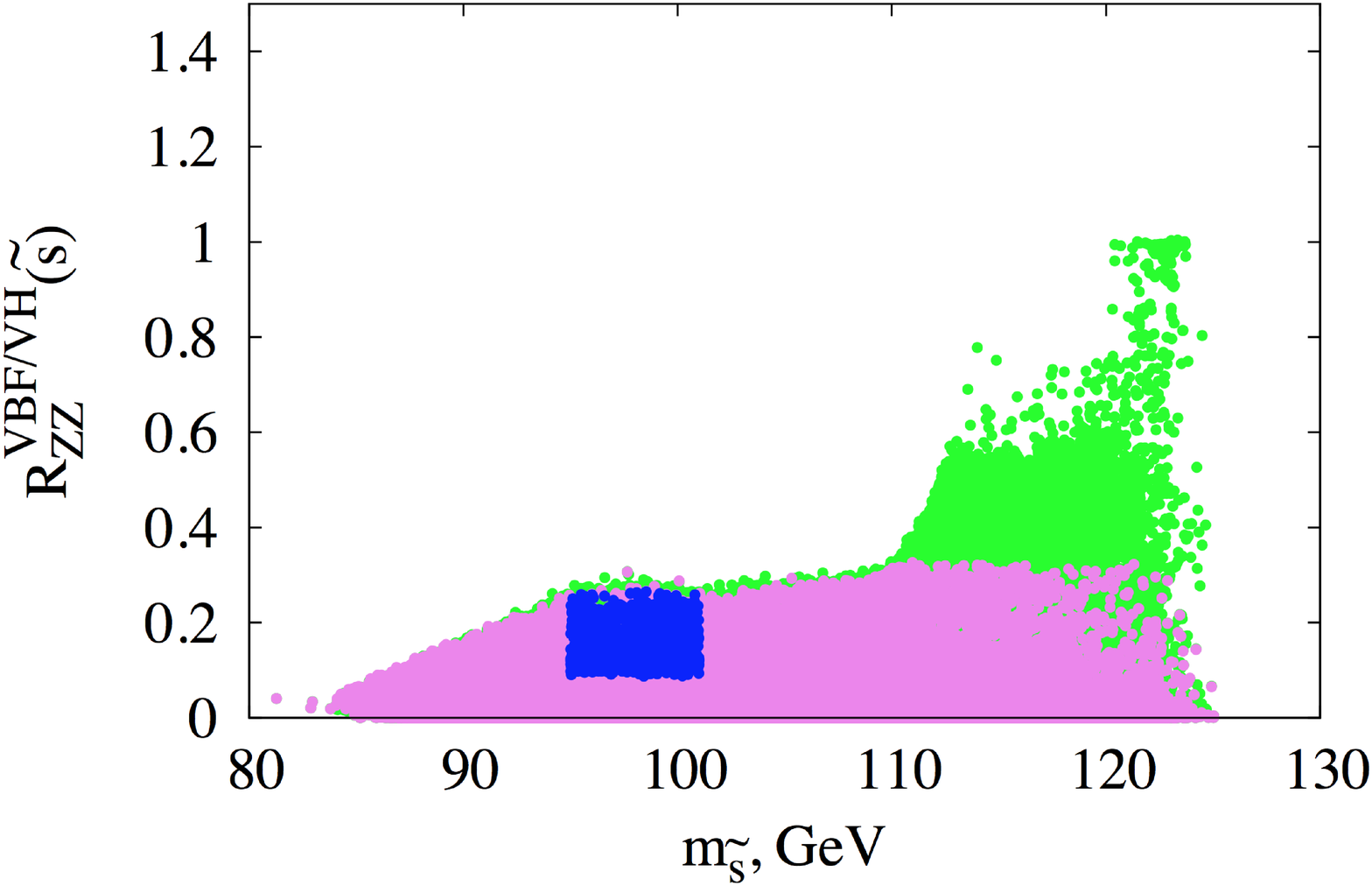}}
\put(0,150){\includegraphics[angle=0,width=0.45\textwidth]{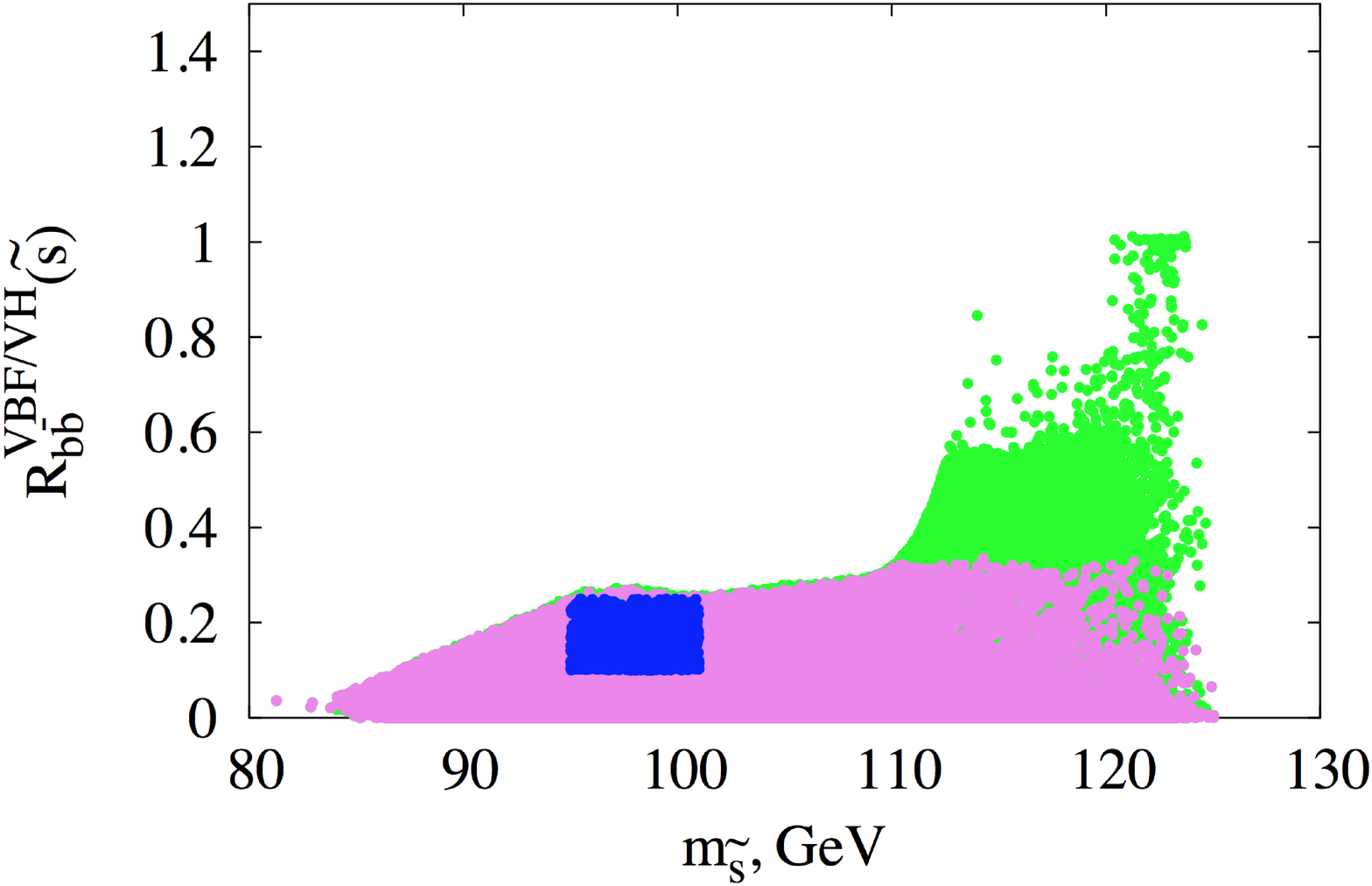}}
\end{picture}
\caption{\label{FIG14} Scatter plots in
  $m_{\tilde{s}}-R_{b\tilde{b}}^{VBF/VH}(\tilde{s})$ (upper  left
  panel),
  $m_{\tilde{s}}-R_{\tau\tilde{\tau}}^{VBF/VH}(\tilde{s})$ (upper 
    right panel),
  $m_{\tilde{s}}-R_{WW}^{VBF/VH}(\tilde{s})$ (lower  left panel), 
  $m_{\tilde{s}}-R_{\gamma\gamma}^{VBF/VH}(\tilde{s})$ (lower 
    right panel)
  planes for {\it Set 2}.  The color notations are the same as in
  Fig.~\ref{mh_Rgg}.} 
\end{figure}

Similar scatter plots for $VBF/VH$ production process are shown in
Fig.~\ref{ms_R2} 
for the case of {\it Set 1} and in Fig.~\ref{FIG14}
for the case of {\it Set 2}. We see that the signatures of $VBF/VH$
sgoldstino production look quite promising: corresponding signal
strengths can reach values up to $1.2-1.3$ for $\gamma\gamma$
and for other channels they can be as large as 0.3. This indicates
that the discussed scenario is out of reach of TeVatron experiments
but hopefully can be probed in the future LHC runs. 

\section{Concluding remarks}
\label{sec:conclusions}
To summarize, in this paper we discussed implications of the possible
mixing between the supersymmetric Higgs sector and hidden sector in
models with low-scale supersymmetry breaking. We have found that the
mixing of scalar sgoldstino $\tilde{s}$ to the lightest Higgs boson
$\tilde{h}$ can result in an additional increase of mass of the
latter. As an attractive feature of this scenario, we have found that 
new sgoldstino-like scalar state $\tilde{s}$ which is somewhat lighter
than the Higgs-like boson is present in low energy spectrum. In
particular, there is a region in the parameter space of the model 
where this state can explain $2\sigma$ LEP excess  in $e^+e^-\to
Z\tilde{s}$ with $\tilde{s}\to b\bar{b}$ having mass around
98~GeV. 

Performing a scan over parameters for $\sqrt{F}=10$~TeV and selecting
phenomenologically acceptable models we have found that the mixing
with sgoldstino results in a distinctive features in signal strengths
for the Higgs-like resonance in this scenario. In gluon-gluon fusion
the signal  strengths for fermion and massive vector boson channels
are somewhat larger than unity with values about $1.0-1.5$. On the
contrary, for vector boson fusion or associative production with
massive vector boson the signal strengths are predicted to 
be within the range about $0.7-1.0$. If sgoldstino is required to be
98~GeV LEP resonance then even more strict bounds on the signal
strength are predicted, which hopefully can be probed in the next runs
of the LHC experiments.

Note that here we have performed a simplified analysis by limiting
ourselves to the case of MSSM decoupling limit, zero vacuum
expectation value for sgoldstino field and fixed value for
supersymmetry breaking scale $\sqrt{F}=10$~TeV.  By going beyond these 
assumptions one could obtain that the life with sgoldstino-Higgs
mixing can become even more complicated. In particular, we expect
different mixing patterns due to presence of heavier Higgs boson in
spectrum (see e.g.~\cite{Han:2013mga}) and shifts in the Yukawa
couplings of the lightest Higgs boson to
fermions~\cite{Dudas:2012fa}. Among the other possible
phenomenological issues which are not covered in the present study we
mention possibility of new decays of the lightest Higgs boson in which
sgoldstino can be involved including those with flavour violation (see
also Ref.~\cite{Petersson:2012dp}). For sufficiently light sgoldstinos 
decays $\tilde{h}\to\tilde{s}\tilde{h}^*$ with subsequent
$\tilde{s}\to \gamma\gamma$ and $\tilde{h}\to b\bar{b}$ or
$\tilde{h}\to\tilde{s}\tilde{s}^*$ with $\tilde{s}\to
\gamma\gamma$ become possible resulting in new signatures in the
Higgs boson decays.
Another interesting area to explore is models of low-scale
supersymmetry breaking in which gauginos have Dirac masses (see,
e.g.~\cite{Goodsell:2014dia} and references therein). In this case
sgoldstino interactions with the SM fields can be different as
compared to the case discussed in our paper resulting in different
mixing properties and the couplings of mass states. We leave
investigations of these interesting possibilities for future work. 

\paragraph*{Acknowledgments}
We thank D.~Gorbunov for valuable discussions and careful reading the
manuscript. The work was supported by the RSCF grant 14-22-00161. The
numerical part of the work was performed on Calculational Cluster of
the Theory Division of INR RAS.

\appendix
\section{Modifications of decays $\tilde{h}\to ZZ$ and $\tilde{h}\to
  W^+W^-$. }
\label{sec:4}
In this Appendix we present formulas for partial widths of the
decay of the Higgs-like resonance $\tilde{h}$ into pair of massive
vector bosons $V$, where $V$ is $W$ or $Z$-boson.  A complication
arises due to the fact that the Higgs boson $h$ and scalar sgoldstino
$s$ have different interactions with $W^\pm$ and $Z$-bosons, see
Eqs.~\eqref{Higgs_lagr} and~\eqref{Sgold_lagr}. Using results of
Ref.~\cite{Romao:1998sr} we obtain  
\begin{eqnarray}
\Gamma(\tilde{h}\to{}VV^*) =\delta_V\frac{G_Fm^3_{\tilde{h}}}
      {16\pi^2\sqrt{2}}\int{}d(\Delta^2)
      \sqrt{\lambda(m_V^2,\Delta^2,m^2_{\tilde{h}})} 
      \frac{\Gamma_Vm_V}{|D(\Delta^2)|^2}\times \\
      \times\bigg[\lambda(m_V^2,\Delta^2,m^2_{\tilde{h}})+
        +12\frac{m_V^2\Delta^2}{m_{\tilde{h}}^4}+X(\Delta)\bigg]   \nonumber
\end{eqnarray}
where 
\begin{equation}
\lambda(x,y,x)=\left(1-\frac{x}{z}
-\frac{y}{z}\right)^2-4\frac{xy}{z^2}, \;\;\; 
D(\Delta^2)=\Delta^2-m_V^2+im_V\Gamma_V
\end{equation}
and 
\begin{eqnarray}
X(\Delta) & = & \frac{m_V^2\Delta^2f}{m_{\tilde{h}}^4}
\Big(12(-\Delta^2-m_V^2+m_h^2)+\\
& & +4f \Big[\frac{1}{2}\Delta^4
  +\frac{1}{2}(m_V^2-m_h^2)^2 +(m_V^2-m_h^2)\Delta^2
  +\Delta^2m_V^2\Big]\Big) \nonumber
\end{eqnarray}
{${\delta}_V=2(1)$ for $V=W(Z),$ $\Delta-$is 4-momenta of off-shell particle $V^*$ and  $f$ is defined by}   
\begin{equation}
f=\frac{-M_{ZZ(2)}v}{2Fm_V^2}.
\end{equation}

\bibliographystyle{asmplain}


\end{document}